\documentclass{LMCS}
\usepackage{ifpdf}
\pdffalse
\newif\ifmakeindex              
\makeindexfalse
\newif\ifdoprintindex           
\doprintindextrue
\newif\ifshowindex              
\showindexfalse
\newif\ifdraft                  
\draftfalse                     
\newif\ifmarginnotes            
\newif\ifcheckednotes           
\checkednotesfalse
\newif\ifxypicloaded            
\xypicloadedfalse
\usepackage{amsmath,amsthm,amsfonts,amssymb} 
\usepackage{calc}               
\newdimen\proofrulebreadth \proofrulebreadth=.05em
\newdimen\proofdotseparation \proofdotseparation=1.25ex
\newdimen\proofrulebaseline \proofrulebaseline=2ex
\newcount\proofdotnumber \proofdotnumber=3
\let\then\relax
\def\hfi{\hskip0pt plus.0001fil}
\mathchardef\squigto="3A3B
\newif\ifinsideprooftree\insideprooftreefalse
\newif\ifonleftofproofrule\onleftofproofrulefalse
\newif\ifproofdots\proofdotsfalse
\newif\ifdoubleproof\doubleprooffalse
\let\wereinproofbit\relax
\newdimen\shortenproofleft
\newdimen\shortenproofright
\newdimen\proofbelowshift
\newbox\proofabove
\newbox\proofbelow
\newbox\proofrulename
\def\shiftproofbelow{\let\next\relax\afterassignment\setshiftproofbelow\dimen0 }
\def\shiftproofbelowneg{\def\next{\multiply\dimen0 by-1 }%
\afterassignment\setshiftproofbelow\dimen0 }
\def\setshiftproofbelow{\next\proofbelowshift=\dimen0 }
\def\setproofrulebreadth{\proofrulebreadth}

\def\prooftree{
\ifnum  \lastpenalty=1
\then   \unpenalty
\else   \onleftofproofrulefalse
\fi
\ifonleftofproofrule
\else   \ifinsideprooftree
        \then   \hskip.5em plus1fil
        \fi
\fi
\bgroup
\setbox\proofbelow=\hbox{}\setbox\proofrulename=\hbox{}%
\let\justifies\proofover\let\leadsto\proofoverdots\let\Justifies\proofoverdbl
\let\using\proofusing\let\[\prooftree
\ifinsideprooftree\let\]\endprooftree\fi
\proofdotsfalse\doubleprooffalse
\let\thickness\setproofrulebreadth
\let\shiftright\shiftproofbelow \let\shift\shiftproofbelow
\let\shiftleft\shiftproofbelowneg
\let\ifwasinsideprooftree\ifinsideprooftree
\insideprooftreetrue
\setbox\proofabove=\hbox\bgroup$\displaystyle 
\let\wereinproofbit\prooftree
\shortenproofleft=0pt \shortenproofright=0pt \proofbelowshift=0pt
\onleftofproofruletrue\penalty1
}

\def\eproofbit{
\ifx    \wereinproofbit\prooftree
\then   \ifcase \lastpenalty
        \then   \shortenproofright=0pt  
        \or     \unpenalty\hfil         
        \or     \unpenalty\unskip       
        \else   \shortenproofright=0pt  
        \fi
\fi
\global\dimen0=\shortenproofleft
\global\dimen1=\shortenproofright
\global\dimen2=\proofrulebreadth
\global\dimen3=\proofbelowshift
\global\dimen4=\proofdotseparation
\global\count255=\proofdotnumber
$\egroup  
\shortenproofleft=\dimen0
\shortenproofright=\dimen1
\proofrulebreadth=\dimen2
\proofbelowshift=\dimen3
\proofdotseparation=\dimen4
\proofdotnumber=\count255
}

\def\proofover{
\eproofbit 
\setbox\proofbelow=\hbox\bgroup 
\let\wereinproofbit\proofover
$\displaystyle
}%
\def\proofoverdbl{
\eproofbit 
\doubleprooftrue
\setbox\proofbelow=\hbox\bgroup 
\let\wereinproofbit\proofoverdbl
$\displaystyle
}%
\def\proofoverdots{
\eproofbit 
\proofdotstrue
\setbox\proofbelow=\hbox\bgroup 
\let\wereinproofbit\proofoverdots
$\displaystyle
}%
\def\proofusing{
\eproofbit 
\setbox\proofrulename=\hbox\bgroup 
\let\wereinproofbit\proofusing
\kern0.3em$
}

\def\endprooftree{
\eproofbit 
  \dimen5 =0pt
\dimen0=\wd\proofabove \advance\dimen0-\shortenproofleft
\advance\dimen0-\shortenproofright
\dimen1=.5\dimen0 \advance\dimen1-.5\wd\proofbelow
\dimen4=\dimen1
\advance\dimen1\proofbelowshift \advance\dimen4-\proofbelowshift
\ifdim  \dimen1<0pt
\then   \advance\shortenproofleft\dimen1
        \advance\dimen0-\dimen1
        \dimen1=0pt
        \ifdim  \shortenproofleft<0pt
        \then   \setbox\proofabove=\hbox{%
                        \kern-\shortenproofleft\unhbox\proofabove}%
                \shortenproofleft=0pt
        \fi
\fi
\ifdim  \dimen4<0pt
\then   \advance\shortenproofright\dimen4
        \advance\dimen0-\dimen4
        \dimen4=0pt
\fi
\ifdim  \shortenproofright<\wd\proofrulename
\then   \shortenproofright=\wd\proofrulename
\fi
\dimen2=\shortenproofleft \advance\dimen2 by\dimen1
\dimen3=\shortenproofright\advance\dimen3 by\dimen4
\ifproofdots
\then
        \dimen6=\shortenproofleft \advance\dimen6 .5\dimen0
        \setbox1=\vbox to\proofdotseparation{\vss\hbox{$\cdot$}\vss}%
        \setbox0=\hbox{%
                \advance\dimen6-.5\wd1
                \kern\dimen6
                $\vcenter to\proofdotnumber\proofdotseparation
                        {\leaders\box1\vfill}$%
                \unhbox\proofrulename}%
\else   \dimen6=\fontdimen22\the\textfont2 
        \dimen7=\dimen6
        \advance\dimen6by.5\proofrulebreadth
        \advance\dimen7by-.5\proofrulebreadth
        \setbox0=\hbox{%
                \kern\shortenproofleft
                \ifdoubleproof
                \then   \hbox to\dimen0{%
                        $\mathsurround0pt\mathord=\mkern-6mu%
                        \cleaders\hbox{$\mkern-2mu=\mkern-2mu$}\hfill
                        \mkern-6mu\mathord=$}%
                \else   \vrule height\dimen6 depth-\dimen7 width\dimen0
                \fi
                \unhbox\proofrulename}%
        \ht0=\dimen6 \dp0=-\dimen7
\fi
\let\doll\relax
\ifwasinsideprooftree
\then   \let\VBOX\vbox
\else   \ifmmode\else$\let\doll=$\fi
        \let\VBOX\vcenter
\fi
\VBOX   {\baselineskip\proofrulebaseline \lineskip.2ex
        \expandafter\lineskiplimit\ifproofdots0ex\else-0.6ex\fi
        \hbox   spread\dimen5   {\hfi\unhbox\proofabove\hfi}%
        \hbox{\box0}%
        \hbox   {\kern\dimen2 \box\proofbelow}}\doll%
\global\dimen2=\dimen2
\global\dimen3=\dimen3
\egroup 
\ifonleftofproofrule
\then   \shortenproofleft=\dimen2
\fi
\shortenproofright=\dimen3
\onleftofproofrulefalse
\ifinsideprooftree
\then   \hskip.5em plus 1fil \penalty2
\fi
}

\ifpdf
  \usepackage[matrix,arrow,arc,tips,frame,2cell]{xy} 
  \xypicloadedtrue
\else
  \usepackage[matrix,arrow,arc,tips,frame,2cell]{xy} 
  \xypicloadedtrue
  \fi
\usepackage{hyperref}
\ifxypicloaded
\newdir{ >}{{}*!/-10pt/@{>}}
\newdir{ (}{{}*!/-5pt/@^{(}}
\newdir{ )}{{}*!/-5pt/@_{(}}
\newdir{|>}{!/4.5pt/@{|}*:(1,-.2)@{>}*:(1,+.2)@_{>}}
\fi
\newcommand{\defnfont}{\textbf}

\newcommand{\fixedcatfont}{\mathbf}

\newcommand{\fixedtermtypekindfont}{\mathsf}

\newcommand{\setsofnumbersfont}{\mathbb}

\newcommand{\opcat}[1]{{{#1}^{\mathrm{op}}}}

\newcommand{\id}{\mathit{id}}

\newcommand{\bang}{!}

\newcommand{\mor}[3]{\xymatrix@1{{#1}\ar[r]^-{#2}&{#3}}}
\newcommand{\twomor}[5]{\xymatrix@1{{#1}\ar[r]^-{#2}&{#3}\ar[r]^-{#4}&{#5}}}

\newcommand{\cover}{\mathrel{\mbox{${-}\kern-5pt\vartriangleright$}}}

\newcommand{\pair}[2]{\langle{#1},{#2}\rangle}

\newcommand{\prodmor}[5]{\xymatrix@1{{#1}&{#3}\ar[l]_-{#2}\ar[r]^-{#4}&{#5}}}

\newcommand{\coprodmor}[5]{%
  \xymatrix@1{{#1}\ar[r]^-{#2}&{#3}&\ar[l]_-{#4}{#5}}}
\newcommand{\adjdisp}[4]{
  \xymatrix{
  {#1}\ar@<1.3ex>[r]^-{#3}_-\perp & 
  {#2}\ar@<1.3ex>[l]^-{#4}
  }}

\newcommand{\pseudoeqrdisp}[4]{
  \xymatrix{
    {#1}\ar@<1.0ex>[d]^{#3} \ar@<-1.0ex>[d]_{#2} \\
    {#4}
  }}
\newcommand{\hfibp}[3]{\xymatrix@1{{#1}\ar[r]^{#2} & {#3}}}
\newcommand{\vfibp}[3]{%
 \NoCompileMatrices
 \def\objectstyle{\scriptstyle}
 \def\labelstyle{\scriptstyle}
 \SelectTips{cm}{}
 \vcenter{\xymatrix@1 @R=0.8pc{ {#1}\ar[d]^{#2} \\ {#3}}}}

\newcommand{\settwoobj}[3]{(\xymatrix@1@C-0.5pc{{#1}\ar[r]^{#2} & {#3}})}

\newcommand{\meet}{\wedge}

\newcommand{\imp}{\supset}
\newcommand{\biimp}{\supset\!\!\subset}

  {\end{align*}}

\newcommand{\emphdefn}[1]{\defnfont{#1}}
\newcommand{\ts}{\vdash}

\newcommand{\co}{\colon}
\newcommand{\ld}{\mathpunct{.}}

\newcommand{\verystrong}%
  {\left(\stackrel{\text{\small very}}{\text{\small strong}}\right)}

\newcommand{\Ctx}{\fixedtermtypekindfont{Ctx}}

\newcommand{\Prop}{\fixedtermtypekindfont{Prop}}
\newcommand{\Type}{\fixedtermtypekindfont{Type}}

  {\begin{displaymath}%
   \begin{array}{c}}
  {\end{array}\end{displaymath}\noindent}
\def\hypquad{\quad\qquad}  

\newbox\tempa
\newbox\tempb
\newdimen\tempc
\newbox\tempd

\def\mud#1{\hfil $\displaystyle{#1}$\hfil}
\def\rig#1{\hfil $\displaystyle{#1}$}

\def\inruleanhelp#1#2#3{\setbox\tempa=\hbox{$\displaystyle{\mathstrut #2}$}%
                        \setbox\tempd=\hbox{$\; #3$}%
                        \setbox\tempb=\vbox{\halign{##\cr
        \mud{#1}\cr
        \noalign{\vskip\the\lineskip}%
        \noalign{\hrule height 0pt}%
        \rig{\vbox to 0pt{\vss\hbox to 0pt{\copy\tempd \hss}\vss}}\cr
        \noalign{\hrule}%
        \noalign{\vskip\the\lineskip}%
        \mud{\copy\tempa}\cr}}%
                      \tempc=\wd\tempb
                      \advance\tempc by \wd\tempa
                      \divide\tempc by 2 }

\def\inrulean#1#2#3{{\inruleanhelp{#1}{#2}{#3}%
                     \hbox to \wd\tempa{\hss \box\tempb \hss}}}

\def\lowerhalf#1{\hbox{\raise -0.8\baselineskip\hbox{#1}}}

\def\inruleamhelp#1#2#3{\setbox\tempa=\hbox{$\displaystyle{\mathstrut #2}$}%
                        \setbox\tempd=\hbox{$\; #3$}%
                        \setbox\tempb=\vbox{\halign{##\cr
        \mud{#1}\cr
        \noalign{\vskip\the\lineskip}%
        \noalign{\hrule height 0pt}%
        \rig{\vbox to 0pt{\vss\hbox to 0pt{\copy\tempd \hss}\vss}}\cr
        \noalign{\hrule\vskip 1.5pt\hrule}%
        \noalign{\vskip\the\lineskip}%
        \mud{\copy\tempa}\cr}}%
                      \tempc=\wd\tempb
                      \advance\tempc by \wd\tempa
                      \divide\tempc by 2 }

\def\inruleam#1#2#3{{\inruleamhelp{#1}{#2}{#3}%
                     \hbox to \wd\tempa{\hss \box\tempb \hss}}}

\newcommand{\N}{{\setsofnumbersfont{N}}}

\newcommand{\iso}{\cong}

\DeclareMathOperator{\Prod}{\textstyle{\prod}}
\DeclareMathOperator{\Coprod}{\textstyle{\coprod}}

\newcommand{\loccit}{\emph{loc.\ cit.}}




\newcommand{\LinTypecat}{{\fixedcatfont{LinType}}}

\newcommand{\PFamAP}[1]%
  {\fixedcatfont{PFam(AP}(#1)\fixedcatfont{)}}
\newcommand{\PFamAPbot}[1]%
  {\fixedcatfont{PFam}(\APbot{{#1}})}
\newcommand{\APbot}[1]{\fixedcatfont{AP}(#1)_{\bot}}
\newcommand{\AdmRel}[2]{\text{AdmRel}(#1,#2)}

\newcommand{\rel}[2]{\co\mathsf{Rel}(#1, #2)}
\newcommand{\admrel}[2]{\co\mathsf{AdmRel}(#1, #2)}
\newcommand{\graph}[1]{\langle #1\rangle}

\newcommand{\inv}{^{-1}}

\newcommand{\congr}{\equiv}

\newcommand{\lapl}{\text{LAPL}}

\newcommand{\lpop}{\multimap}
\newcommand{\tensor}{\otimes}
\newcommand{\letexp}[3]{\text{let }#1\text{ be }#2\text{ in } #3}
\newcommand{\rec}{\text{rec }}
\newcommand{\comb}{\textit{pairing}}
\newcommand{\eq}{\textit{eq}}
\newcommand{\fold}{\textit{fold}}
\newcommand{\interm}{\textit{in}}
\newcommand{\pack}{\textit{pack}}
\newcommand{\unfold}{\textit{unfold}}
\newcommand{\out}{\textit{out}}
\newcommand{\apply}{r}

\newcommand{\linlambda}{\lambda^\circ}
\newcommand{\pill}{\text{PILL}}
\newcommand{\pilly}{\text{PILL}_{Y}}
\newcommand{\lily}{\text{Lily}}
\newcommand{\lists}{\textit{lists}}
\newcommand{\xypop}{\ar@{o}}
\newcommand{\copair}[2]{[#1,#2]}

\renewcommand{\perp}{\top}

\newcommand{\Wgraph}[2]{\xymatrix@C=10pt@R=4pt{#1\ar@<1ex>[dd]
    \ar@<-1ex>[dd] & & #1 \ar@<1ex>[dd] \ar@<-1ex>[dd] \\ 
& #1 \ar[dr] \ar[dl] \\
#2 \ar[uu] & & #2 \ar[uu]}
}

\theoremstyle{plain}\newtheorem{infrule}[thm]{Rule}
\theoremstyle{plain}\newtheorem{axiom}[thm]{Axiom}
\ifdraft
\fi
\def\eqalign#1{\null\,\vcenter{\openup\jot\mathsurround=0 pt
  \ialign{\strut\hfil$\displaystyle{##}$&$\displaystyle{{}##}$\hfil
      \crcr#1\crcr}}\,}

\def\doi{2 (5:2) 2006}
\lmcsheading%
{\doi}
{1--48}
{}
{}
{Apr.~20, 2006}
{Nov.~\phantom{0}3, 2006}
{}

\begin{document}
\title[Linear {A}badi \& {P}lotkin Logic]{Linear {A}badi \& {P}lotkin Logic}

\author[L.~Birkedal]{Lars Birkedal\rsuper a}
\address{{\lsuper{a,c}}IT University of Copenhagen}
\email{\{birkedal,rusmus\}@itu.dk}

\author[R.E.~M{\o}gelberg]{Rasmus Ejlers M{\o}gelberg\rsuper b} 
\address{{\lsuper b}LFCS, School of Informatics, University of Edinburgh}
\email{rasmus.mogelberg@ed.ac.uk}
\thanks{{\lsuper b}The majority of this work was conducted while
     this author was associated with the IT University of Copenhagen
     and with the University of Genova. Research partly supported by
     Danish Natural Science Research Council stipend no. 272-05-0031}

\author[R.L.~Petersen]{Rasmus Lerchedahl Petersen\rsuper c}

\keywords{parametric polymorphism, domain theory, recursive types}
\subjclass{F.4.1, D.3.3}
\titlecomment{}

\begin{abstract}
  \noindent We present a formalization of a version of Abadi and
  Plotkin's logic for parametricity for a polymorphic dual
  intuitionistic/linear type theory with fixed points, and show,
  following Plotkin's suggestions, that it can be used to define a
  wide collection of types, including existential types, inductive
  types, coinductive types and general recursive types. We show that
  the recursive types satisfy a universal property called
  dinaturality, and we develop reasoning principles for the
  constructed types. In the case of recursive types, the reasoning
  principle is a mixed induction/coinduction principle, with the
  curious property that coinduction holds for general relations, but
  induction only for a limited collection of ``admissible'' relations.
  A similar property was observed in Pitts' 1995 analysis of recursive
  types in domain theory.  In a future paper we will develop a
  category theoretic notion of models of the logic presented here, and
  show how the results developed in the logic can be transferred to
  the models.
\end{abstract}

\maketitle

\section*{Introduction}\label{sec:intro}


In 1983 Reynolds argued that parametric models of the second-order
lambda calculus are very useful for modeling data abstraction in
programming~\cite{Reynolds:83} (see also ~\cite{BPierce:tpl} for a
recent textbook description). For real programming, one is of course
not just interested in a strongly terminating calculus such as the
second-order lambda calculus, but also in a language with full
recursion.  Thus in \loccit\ Reynolds also asked for a parametric
\emph{domain-theoretic} model of polymorphism.  Informally, what is
meant \cite{Reynolds:2000} by this is a model of an extension of the
polymorphic lambda calculus~\cite{Reynolds:74a,Girard:72}, with a
polymorphic fixed-point operator
$Y\co\forall{\alpha}\ld(\alpha\to\alpha)\to\alpha$ such that
\begin{enumerate}
\item types are modeled as domains, the sublanguage without polymorphism
  is modeled in the standard way and $Y \sigma$ is the
  least fixed-point operator for the domain $\sigma$;
\item the logical relations theorem (also known as the abstraction
  theorem) is satisfied when the logical relations are admissible, 
  i.e., strict and closed under limits of chains;
\item every value in the domain representing some polymorphic type
  is parametric in the sense that it 
  satisfies the logical relations theorem (even if it is not the
  interpretation of any expression of that type).
\end{enumerate}

Of course, this informal description leaves room for different
formalizations of the problem. Even so, it has proved to be a
non-trivial problem. Unpublished work of
Plotkin~\cite{PlotkinGD:secotr} indicates one way to solve the problem
model-theoretically by using strict, admissible partial equivalence
relations over a domain model of the untyped lambda calculus but, as
far as we know, the details of this relationally parametric model have
not been worked out in the literature.

From a type theoretical perspective parametric polymorphism is
interesting because it allows for encodings of a large collection of
types from a small number of constructions. For example adding
parametric polymorphism as a reasoning principle to the second-order
lambda calculus gives encodings of products, coproducts, existential
types and general inductive and coinductive types from just $\to$  and
polymorphism \cite{Abadi:Plotkin:93,BirkedalL:catmap}. 

This strength of the typing system also complicates matters when
adding recursion. Simply adding a polymorphic fixed point combinator
to parametric second order lambda calculus would give a type theory
with coproducts, products, function spaces and fixed points, a
combination known to exist only in the trivial case of all types being
isomorphic \cite{Huwig:Poigne:90}. Inspired by domain theory Plotkin
suggested to consider a polymorphic dual intuitionistic/linear
lambda calculus and restrict the parametricity principle accordingly
to give encodings of coproducts and (co-)inductive types in the linear
part of the calculus but not the intuitionistic part. Moreover, the
existence of fixed points would provide solutions to general recursive
type equations using Freyd's theory of algebraically compact
categories~\cite{Freyd:90,Freyd:90a,Freyd:91}. This led Plotkin to
argue that such a calculus could serve as a very powerful metalanguage
for domain theory. 

Thus parametric domain-theoretic models of polymorphic intuitionistic
/ linear lambda calculus are of importance both from a programming
language perspective (for modeling data abstraction) and from a purely
domain-theoretic perspective.

Recently, Pitts and coworkers~\cite{BiermanGM:opeplp} 
have presented a syntactic approach to Reynolds' challenge, where the
notion of domain is essentially taken to be equivalence classes of
terms modulo a particular notion of contextual equivalence derived
from an operational semantics for a language called $\lily$, which is
essentially polymorphic intuitionistic/linear lambda calculus
endowed with an operational semantics.

In parallel with the work presented here, Rosolini and
Simpson~\cite{Rosolini:Simpson:04} have shown how to construct
parametric domain-theoretic models using synthetic domain-theory in
intuitionistic set-theory. Moreover, they have shown how to give a
computationally adequate denotational semantics of $\lily$.

This paper presents a formalization of Abadi \& Plotkin's logic adapted
to the case of Polymorphic Intuitionistic/Linear Lambda calculus
with a polymorphic fixed point combinator denoted $Y$ --- a language
which we shall call $\pilly$. $\pilly$ is a simple extension of
Barber and Plotkin's dual intuitionistic/linear lambda calculus
(DILL) with polymorphism and fixed points. By dual we mean that terms
have two contexts of term variables: an
intuitionistic and a linear one. 

Linear Abadi-Plotkin Logic (LAPL) presented in this paper is a logic
for reasoning about parametricity for $\pilly$. As mentioned above,
for the logic to be consistent, the parametricity principle has to be
restricted in some way, so that it can be used to prove universal
properties in the category of linear terms, but not in the category of
intuitionistic terms. To achieve this restriction, LAPL is equipped
with a
notion of admissible relation, and the parametricity principle is
formulated using these relations only. Admissible relations form a
subset of the set of definable relations between types, and the prime
example of an admissible relation in the logic is the graph of a
linear function, whereas the prime example of a relation that is not
admissible in general is the graph of an intuitionistic function.


Using the logic, we show how Plotkin's encodings of a large collection
of datatypes satisfy the usual universal properties with respect to
linear maps in the calculus, up to provability in the logic. In the
case of inductive types this means showing that the encodings give
initial algebras for certain functors induced by types, for
coinductive types we get final coalgebras, and for the general
recursive types, the encodings give initial dialgebras for the
bifunctors induced by type expressions. These results were sketched by
Plotkin in \cite{PlotkinGD:secotr}, but since the proofs are
non-trivial and have never appeared in the literature we include them
here. We treat recursive types in full generality, meaning that we
treat recursive types with parameters showing that nested recursive
types can be modeled.

We also present reasoning principles for the constructed types. Using
parametricity we get an induction principle for inductive types
holding only for admissible relations. For the coinductive types we
get a coinduction principle holding for all relations. These results
are extended to recursive types giving a mixed induction/coinduction
principle in which the induction part holds for admissible relations
only, but the coinduction part holds for all relations. Again these
principles are treated in full generality, i.e., also for recursive
types with parameters. A similar induction/coinduction principle
with the same restrictions was discovered by Pitts~\cite{Pitts:95a}
for recursive types in domain theory.

The present paper is the first in a series presenting an
axiomatization of domain theoretic models of parametricity. In
a forthcoming paper (based on~\cite{MogelbergR:lapl-tr}) we present a
sound and complete notion of parametric models of LAPL called
parametric LAPL-structures, and show how to transfer the results
proved in LAPL to these. In further papers we will show examples of
such parametric LAPL-structures, first 
treating Plotkin's idea of
using admissible pers over reflexive domains, and in further papers we
show how Rosolini and Simpson's
construction~\cite{Rosolini:Simpson:04} can be seen as constructing
parametric LAPL-structures and we construct LAPL-structures from
$\lily$ syntax in~\cite{lapl-lily-tr}. Finally
in~\cite{MogelbergR:lapl-completion-tr} we show how the parametric
completion process of Robinson \& Rosolini \cite{Robinson:Rosolini:94}
can be adapted to construct parametric LAPL-structures from internal
models of $\pilly$ in quasi toposes.

In each of these models the abstract notion of admissible relations in
LAPL is interpreted differently. For example, in the per model the
notion of admissible relations are certain subsets of the set of
equivalence classes of pers, and in the $\lily$ model admissible
relations are $\perp\perp$-closed sets of terms. The abstract notion
of admissible relations presented in this paper is general enough to
fit all these different cases.

We remark that one can see our notion of parametric LAPL-structure as
a suitable categorical axiomatization of a good category of domains.
In Axiomatic Domain Theory much of the earlier work has focused on
axiomatizing the adjunction between the category of predomains and
continuous functions and the category of predomains and partial
continuous functions~\cite[Page~7]{Fiore:96} -- here we axiomatize the
adjunction between the category of domains and strict functions and
the category of domains and all continuous functions and extend it
with parametric polymorphism, which then suffices to also model
recursive types.

\subsection*{Outline}
\label{sec:outline}

The remainder of this paper consists of two parts. The first part
(Section~\ref{sec:aplogic}) presents the calculus $\pilly$ and the
logic LAPL for reasoning about parametricity. The second part
(Section~\ref{sec:proofs-in-lapl}) gives detailed proofs of
correctness of encodings of a series of types including inductive,
coinductive and recursive types, and gives the reasoning principles
for these.

\section{Linear Abadi-Plotkin Logic}
\label{sec:aplogic}

In this section we define a logic for reasoning about parametricity for
Polymorphic Intuitionistic Linear Lambda calculus with fixed points
($\pilly$). The logic is based on Abadi and Plotkin's logic for
parametricity~\cite{Abadi:Plotkin:93} for the second-order
lambda calculus and thus we refer to the logic as Linear Abadi-Plotkin
Logic ($\lapl$).

The logic for parametricity is basically a higher-order logic over $\pilly$. 
Expressions of the logic are formulas in contexts of
variables of $\pilly$ and relations among types of
$\pilly$. Thus we start by defining $\pilly$. 

\subsection{$\pilly$}
$\pilly$ is essentially Barber and Plotkin's DILL \cite{Barber:97}
extended with polymorphism and a fixed point combinator.

Well-formed type expressions in $\pilly$ are expressions of the form: 
\[\alpha_1\co\Type,\ldots,\alpha_n\co\Type\ts \sigma\co \Type\]
where $\sigma$  is built using the syntax
\[\sigma::=\alpha\mid I\mid\sigma\tensor\sigma\mid \sigma\lpop\sigma\mid
\, \bang \sigma\mid \Prod\alpha\ld\sigma.\] and all the free variables
of $\sigma$ appear on the left hand side of the turnstile. The last
construction binds $\alpha$, so if we have a type
\[\alpha_1\co\Type,\ldots,\alpha_n\co\Type\ts \sigma\co \Type,\]
then we may form the type
\[\alpha_1\co\Type,\ldots,\alpha_{i-1}\co\Type, \alpha_{i+1}\co\Type\ldots
\alpha_n\co\Type\ts \Prod \alpha_i \ld\sigma\co \Type.\] We use
$\sigma$, $\tau$, $\omega$, $\sigma'$, $\tau'$\ldots to range over
types.  The list of $\alpha$'s is called the kind context, and is
often denoted simply by $\Xi$ or $\vec\alpha$. Since there is only one
kind the annotation $\co\Type$ is often omitted.

The terms of $\pilly$ are of the form:
\[\Xi\mid x_1\co\sigma_1,\ldots, x_n\co\sigma_n;x'_1\co\sigma'_1,\ldots,
x'_m\co\sigma'_m \ts t\co\tau\]
where the $\sigma_i$, $\sigma'_i$, and $\tau$ are well-formed types in the
kind context $\Xi$. The list of $x$'s is called the intuitionistic type context
and is often denoted $\Gamma$, and the list of $x'$'s is called the
linear type context, often denoted $\Delta$. No repetition of variable
names is allowed
in any of the contexts, but permutation akin to having an exchange
rule is. Note, that due to the nature of the axioms of the 
to-be-introduced formation rules, weakening and contraction can be
derived for all but the linear context.

The grammar for terms is:
\begin{displaymath}
  \begin{split}
    t \mathrel{::=}
  & x\mid \star\mid Y\mid \linlambda x\co\sigma.t\mid t\:t\mid t\tensor t
    \mid \bang t \mid \Lambda \alpha\co\Type\ld t\mid t(\sigma)\mid \\
  & \letexp{x\co\sigma\tensor y\co\tau}{t}{t}\mid \letexp{\bang
    x\co\sigma}{t}{t}\mid \letexp\star tt
  \end{split}
\end{displaymath}
We use $\linlambda$, which bear some graphical resemblance to $\lpop$,
to denote linear function abstraction. And we use $s$, $t$, $u$\ldots
to range over terms.

The formation rules are given in Figure~\ref{fig:termformation}.  A
term context $\Xi\mid\Gamma;\Delta$ is considered well-formed if for
all types $\sigma$ appearing in $\Gamma$ and $\Delta$, the type
construction $\Xi\ts\sigma\co\Type$ is well-formed. The linear
contexts $\Delta$ and $\Delta'$ are considered disjoint if the set of
variables appearing in $\Delta$ is disjoint from the set of variables
appearing in $\Delta'$.  We use $-$ to denote an empty context.  As
the types of variables in the let-constructions and function
abstractions are often apparent from the context, these will just as
often be omitted.

The fixed point combinator $Y$  appears as a term in the language, but
could equivalently have been given as an operator on terms as e.g. the
$\rec$  operator in $\lily$. By having it as a polymorphic term the
parametricity principle it satisfies becomes evident.
\begin{figure}[htbp]
  \centering
\[
\begin{prooftree}
\justifies 
\Xi\mid\Gamma;-\ts \star\co I
\end{prooftree}
\]
\[
\begin{prooftree}
\justifies 
\Xi\mid\Gamma;-\ts Y\co\Prod\alpha\ld \bang(\bang\alpha\lpop\alpha)\lpop\alpha
\end{prooftree}
\]
\[
\begin{prooftree}
\justifies 
\Xi\mid\Gamma,x\co\sigma;-\ts x\co\sigma
\end{prooftree}
\]
\[
\begin{prooftree}
\justifies 
\Xi\mid\Gamma;x\co\sigma\ts x\co\sigma
\end{prooftree}
\]
\[
\begin{prooftree}
\Xi\mid\Gamma;\Delta\ts t\co\sigma\lpop\tau
\quad
\Xi\mid\Gamma;\Delta'\ts u\co\sigma
\justifies 
\Xi\mid\Gamma;\Delta,\Delta'\ts t\:u\co\tau
\using
\Delta,\Delta'\text{ disjoint}
\end{prooftree}
\]
\[
\begin{prooftree}
\Xi\mid\Gamma;\Delta,x\co\sigma\ts u\co\tau
\justifies 
\Xi\mid\Gamma;\Delta\ts \linlambda x\co\sigma\ld u\co\sigma\lpop\tau
\end{prooftree}
\]
\[
\begin{prooftree}
\Xi\mid\Gamma;\Delta\ts t\co\sigma
\quad
\Xi\mid\Gamma;\Delta'\ts s\co\tau
\justifies 
\Xi\mid\Gamma;\Delta,\Delta'\ts t\tensor s\co\sigma\tensor\tau
\using
\Delta,\Delta'\text{ disjoint}
\end{prooftree}
\]
\[
\begin{prooftree}
\Xi\mid\Gamma;-\ts t\co\sigma
\justifies 
\Xi\mid\Gamma;-\ts \bang t\co\sigma
\end{prooftree}
\]
\[
\begin{prooftree}
\Xi,\alpha\co\Type\mid\Gamma;\Delta\ts t\co \sigma
\justifies
\Xi\mid\Gamma;\Delta \ts\Lambda\alpha\co\Type\ld t\co
\Prod\alpha\co\Type\ld\sigma 
\using
\Xi\mid\Gamma;\Delta\text{ is well-formed}
\end{prooftree}
\]
\[
\begin{prooftree}
\Xi\mid\Gamma;\Delta\ts t\co \Prod\alpha\co\Type\ld \sigma
\hypquad
\Xi\ts\tau\co\Type
\justifies
\Xi\mid\Gamma;\Delta\ts t(\tau)\co \sigma[\tau/\alpha]
\end{prooftree}
\]

\[
\begin{prooftree}
\Xi\mid\Gamma;\Delta\ts s\co\sigma\tensor\sigma'
\hypquad
\Xi\mid\Gamma;\Delta',x\co\sigma,y\co\sigma'\ts t\co\tau
\justifies
\Xi\mid\Gamma;\Delta,\Delta'\ts \letexp{x\co\sigma\tensor
  y\co\sigma'}{s}{t}\co\tau 
\using
\Delta,\Delta'\text{ disjoint}
\end{prooftree}
\]  
\[
\begin{prooftree}
\Xi\mid\Gamma;\Delta\ts s\co\bang\sigma
\hypquad
\Xi\mid\Gamma,x\co\sigma;\Delta'\ts t\co\tau
\justifies
\Xi\mid\Gamma;\Delta,\Delta'\ts \letexp{\bang x\co\bang\sigma}{s}{t}\co\tau
\using
\Delta,\Delta'\text{ disjoint}
\end{prooftree}
\]  
\[
\begin{prooftree}
\Xi\mid\Gamma;\Delta\ts t\co I
\quad
\Xi\mid\Gamma;\Delta'\ts s\co \sigma
\justifies
\Xi\mid\Gamma;\Delta,\Delta'\ts\letexp\star ts\co\sigma
\end{prooftree}
\]
  \caption{Formation rules for terms}
  \label{fig:termformation}
\end{figure}

\begin{lem}\label{lem:typing:uniqueness}
  Any term can in a given context be shown to have at most one type,
  i.e., if the typing judgements $\Xi\mid \Gamma; \Delta \ts t\co
  \tau$ and $\Xi\mid \Gamma; \Delta \ts t\co \tau'$ are derivable then
  $\tau = \tau'$.
\end{lem}

\begin{lem}\label{lem:typing:substitution}
  The following three substitution rules are derivable from the
  formation rules of $\pilly$. 
  \[
  \begin{prooftree}
    \Xi \mid \Gamma;\Delta, x\co \sigma \ts t\co \tau \qquad
    \Xi \mid \Gamma; \Delta' \ts u \co \sigma
    \justifies 
    \Xi \mid\Gamma;\Delta, \Delta' \ts t[u/x] \co \tau
  \end{prooftree}
  \]
  \[
  \begin{prooftree}
    \Xi \mid \Gamma, x\co \sigma; \Delta \ts t\co \tau \qquad
    \Xi \mid \Gamma; - \ts u \co \sigma
    \justifies 
    \Xi \mid\Gamma;\Delta \ts t[u/x] \co \tau
  \end{prooftree}
  \]
  \[
  \begin{prooftree}
    \Xi, \alpha \mid \Gamma;\Delta \ts t\co \tau \qquad \Xi \ts
    \sigma\co \Type \justifies \Xi \mid\Gamma[\sigma/
    \alpha];\Delta[\sigma/ \alpha] \ts t[\sigma/ \alpha] \co
    \tau[\sigma/ \alpha]
  \end{prooftree}
  \]
\end{lem}

What we have described above is called \emph{pure} $\pilly$. In
general we will consider $\pilly$ over polymorphic signatures
\cite[8.1.1]{Jacobs:99}. Informally, one may think of such a calculus
as pure $\pilly$ with added type-constants and term-constants. For
instance, one may have a constant type for integers or a constant type
for lists $\alpha\ts \lists (\alpha)\co\Type$. We will be particularly
interested in the internal languages of $\pilly$ models which in
general will be non-pure calculi.

We will also sometimes speak of the calculus $\pill$. This is $\pilly$
without the fixed point combinator $Y$.

\subsubsection{Equality}

The \emph{external equality} relation on $\pilly$ terms is the least
equivalence relation given by the rules in
Figure~\ref{fig:externaleq}. External equality is typed in the sense
that if in a given context two terms are externally equal, then they
have the same type. The definition makes use of the notion of a
\emph{context}, which, loosely speaking, is a term with exactly one
hole in it. Formally contexts are defined using the grammar:
\[
\begin{array}{rcl}
C[-] & ::= & - \mid \letexp\star{C[-]}t \mid \letexp\star t{C[-]} \mid
t\tensor C[-] \mid C[-]\tensor t \mid \\
&& \letexp{x\tensor y}{C[-]}t \mid \letexp{x\tensor y}t{C[-]} \mid
\linlambda x\co\sigma \ld C[-] \mid \\
&& C[-]\:t\mid t\:C[-] \mid \bang C[-]\mid \letexp{\bang x}{C[-]}t
\mid \letexp{\bang x}t{C[-]} \mid \\
&& \Lambda\alpha\co\Type\ld C[-] \mid C[-]\sigma
\end{array}
\]
A context $C[-]$ is called a $\Xi\mid\Gamma; \Delta\ts\sigma$ ---
$\Xi\mid\Gamma'; \Delta'\ts \tau$ context if for any well-formed term
$\Xi\mid\Gamma; \Delta\ts t\co \sigma$, the term $\Xi\mid\Gamma';
\Delta'\ts C[t]\co\tau$ is well-formed. A context is \emph{linear}, if
it does not contain a subcontext of the form $\bang C[-]$.



\begin{figure}
\[
\begin{prooftree}
\justifies
\Xi\mid\Gamma;\Delta \ts (\linlambda x\co\sigma\ld t) u = t[u/x]
\using 
\beta\text{-term}
\end{prooftree}
\]
\[
\begin{prooftree}
\justifies
\Xi\mid\Gamma;\Delta \ts (\Lambda \alpha\co\Type\ld t) \sigma =
t[\sigma/\alpha] 
\using 
\beta\text{-type}
\end{prooftree}
\]
\[
\begin{prooftree}
\justifies
\Xi\mid\Gamma;\Delta\ts \linlambda x\co\sigma\ld (t x) = t
\using
\eta\text{-term}
\end{prooftree}
\]
\[
\begin{prooftree}
\justifies
\Xi\mid\Gamma;\Delta\ts \Lambda \alpha\co\Type\ld (t \alpha) = t
\using
\eta\text{-type}
\end{prooftree}
\]
\[
\begin{prooftree}
\justifies
\Xi\mid\Gamma;\Delta\ts\letexp\star\star t=t
\using\beta-\star
\end{prooftree}
\]
\[
\begin{prooftree}
\justifies
\Xi\mid\Gamma;\Delta\ts\letexp\star t\star=t
\using\eta-\star
\end{prooftree}
\]
\[
\begin{prooftree}
\justifies
\Xi\mid\Gamma;\Delta\ts\letexp{x\tensor y}{s\tensor u}t=t[s,u/x,y]
\using\beta-\tensor
\end{prooftree}
\]
\[
\begin{prooftree}
\justifies
\Xi\mid\Gamma;\Delta\ts\letexp{x\tensor y}t{x\tensor y} =t
\using\eta-\tensor
\end{prooftree}
\]
\[
\begin{prooftree}
\justifies
\Xi\mid\Gamma;\Delta\ts \letexp{\bang x\co\sigma}{\bang u}t=t[u/x]
\using\beta-\bang
\end{prooftree}
\]
\[
\begin{prooftree}
\justifies
\Xi\mid\Gamma;\Delta\ts \letexp{\bang x\co\sigma}t{\bang x}=t
\using\eta-\bang
\end{prooftree}
\]

\[
\begin{prooftree}
\Xi\mid\Gamma;\Delta\ts t=s\co\sigma 
\quad
C[-]\text{ is a }\Xi\mid\Gamma;\Delta\ts\sigma -
\Xi\mid\Gamma';\Delta'\ts\tau \text{ context}
\justifies
\Xi\mid\Gamma';\Delta'\ts C[t]=C[s] 
\end{prooftree}
\]
\[
\begin{prooftree}
C[-]\text{ is a linear context}
\justifies 
\Xi\mid\Gamma;\Delta\ts \letexp\star t {C[u]}= C[\letexp\star tu]
\end{prooftree}
\]
\[
\begin{prooftree}
C[-]\text{ is a linear context and does not bind $x,y$  or contain
  them free}
\justifies 
\Xi\mid\Gamma;\Delta\ts \letexp{x\tensor y} t {C[u]}=
C[\letexp{x\tensor y}tu]
\end{prooftree}
\]
\[
\begin{prooftree}
C[-]\text{ is linear and does not bind $x$ or contain it free}
\justifies 
\Xi\mid\Gamma;\Delta\ts \letexp{\bang x}t{C[u]}= C[\letexp{\bang x}tu]
\end{prooftree}
\]
\[
\begin{prooftree}
\Xi\mid \Gamma; - \ts f\co\bang\sigma\lpop\sigma
\justifies 
\Xi\mid \Gamma;- \ts f\:\bang(Y\:\sigma\:(\bang f)) =
Y\:\sigma\:(\bang f)
\end{prooftree}
\]
\caption{Rules for external equality}\label{fig:externaleq}
\end{figure}
We prove a couple of useful lemmas about external equality.
\begin{lem}
Suppose $\Xi\mid \Gamma ;\Delta \ts f,g\co\bang \sigma\lpop\tau$  are
terms such that  
\[\Xi\mid \Gamma, x\co\sigma ;\Delta \ts f(\bang x)=g(\bang x).
\]
Then $f=g$. 
\end{lem}

\proof
Using the rules for external equality,  we conclude from the
assumption that
\[\Xi\mid \Gamma;\Delta, y\co\bang \sigma \ts \letexp{\bang
  x}{y}{f(\bang x)}= \letexp{\bang x}{y}{g(\bang x)}
\]
and further that
\[\Xi\mid \Gamma;\Delta, y\co\bang \sigma \ts f(\letexp{\bang
  x}{y}{\bang x})= g(\letexp{\bang x}{y}{\bang x}).
\]
Thus 
\[\Xi\mid \Gamma;\Delta, y\co\bang \sigma \ts f(y)= g(y),
\]
and hence $f=\linlambda y\co\bang \sigma\ld f(y) = \linlambda y\co\bang
\sigma\ld g(y) = g$. 
\qed

\subsubsection{Intuitionistic lambda abstraction}

We encode ordinary intuitionistic lambda abstraction using the Girard
encoding $\sigma\to\tau=\bang\sigma\lpop\tau$. The corresponding
lambda abstraction is defined as
\[\lambda x\co\sigma\ld t=\linlambda y\co\bang\sigma\ld\letexp{\bang x}yt\]
where $y$ is a fresh variable.
This gives us the rule
\[
\begin{prooftree}
\Xi\mid\Gamma,x\co\sigma;\Delta\ts t\co\tau
\justifies
\Xi\mid\Gamma;\Delta\ts \lambda x\co\sigma\ld t\co\sigma\to\tau
\end{prooftree}
\]
For evaluation we have the rule
\[
\begin{prooftree}
\Xi\mid\Gamma;-\ts t\co\sigma
\quad
\Xi\mid\Gamma;\Delta\ts f\co\sigma\to\tau
\justifies
\Xi\mid\Gamma;\Delta\ts f\:\bang t\co\tau
\end{prooftree}
\]
and the equality rules give
\[(\lambda x\co\sigma\ld t)\:\bang s = t[s/x].\]

Note that using this notation the constant $Y$ can obtain the more
familiar looking type
\[
  Y\co\Pi\alpha\ld(\alpha\to\alpha)\to\alpha.
\]
This notation also explains the occurrences of the $\bang$'s in the
last rule of Figure~\ref{fig:externaleq}.

\subsection{The logic}
\begin{figure}
\[\Xi :: = - \mid \Xi, \alpha \ts \Type
  \qquad \Gamma ::= - \mid \Gamma, x\co \sigma 
\]
\[
 \Theta ::= - \mid \Theta, R\rel{\sigma}\tau \mid \Theta,
  S\admrel{\sigma}\tau 
\]
\[
\Xi\co\Ctx \qquad 
\Xi\ts \sigma\co\Type \qquad
\Xi\mid\Gamma;\Delta\co\Ctx
\]
\[
\Xi\mid\Gamma\mid\Theta\co\Ctx \qquad 
\Xi\mid\Gamma;\Delta\ts t\co\sigma \qquad 
\Xi\mid\Gamma;\Delta\ts t=u
\] 
\[
\Xi\mid\Gamma\mid\Theta\ts \rho\rel{\sigma}{\tau} \qquad
\Xi\mid\Gamma\mid\Theta\ts \rho\admrel{\sigma}{\tau} \qquad
\] 
\[
\Xi\mid \Gamma\mid\Theta\ts \phi\co\Prop \qquad
\Xi\mid\Gamma\mid\Theta\mid\phi_1,\ldots,\phi_n\ts \psi
\] 
\caption{Types of judgments and grammar for LAPL
  contexts}\label{fig:judgements} 
\end{figure}
As mentioned, expressions of $\lapl$ live in contexts of variables of
$\pilly$ and relations among types of $\pilly$. The contexts look like
this:
\[\Xi\mid\Gamma\mid
R_1\rel{\tau_1}{\tau_1'},\ldots,R_n\rel{\tau_n}{\tau_n'},
S_1\admrel{\omega_1}{\omega_1'},\ldots,
S_m\admrel{\omega_m}{\omega_m'} \] where $\Xi\mid\Gamma;-$ is a
context of $\pilly$ and the $\tau_i,\tau_i',\omega_i, \omega_i'$ are
well-formed types in context $\Xi$, for all $i$. The list of $R$'s and
$S$'s is called the relational context and is often denoted $\Theta$.
As for the other contexts we do not allow repetition, but do allow
permutation of variables. 

The concept of admissible relations is taken from domain
theory. Intuitively admissible relations relate $\bot$  to $\bot$  and
are chain complete.

It is important to note that there is no linear component $\Delta$ in
the contexts --- the point is that the logic only allows for
\emph{intuitionistic} (no linearity) reasoning about terms of
$\pilly$, whereas $\pilly$ terms can behave linearly. This
simplification of the logic has been chosen since all parametricity
arguments in our knowledge involve purely intuitionistic reasoning.

Propositions in the logic are given by the syntax:
\[
\begin{array}{rcl}
\phi & ::= & (t=_\sigma u)\mid \rho(t,u)\mid \phi\imp\psi\mid \bot\mid
\top\mid\phi\meet\psi\mid \phi\vee\psi\mid \forall \alpha\co\Type\ld
\phi\mid\\  
& & \forall x\co\sigma\ld\phi \mid
\forall R\rel\sigma\tau\ld\phi \mid \forall
S\admrel\sigma\tau\ld\phi \mid \\ 
& & \exists \alpha\co\Type\ld \phi\mid \exists x\co \sigma\ld \phi\mid
\exists R\rel\sigma\tau\ld \phi\mid \exists
S\admrel\sigma\tau\ld \phi  
\end{array}
\]
where $\rho$ is a definable relation (to be defined below). 
The judgments of the logic are presented in Figure \ref{fig:judgements}.
In the following we give formation rules for the above. 

\begin{rem}\label{rem:otherarities}
Our Linear Abadi \& Plotkin logic is designed for reasoning about binary
relational parametricity. For reasoning about other arities of
parametricity, one can easily replace
binary relations in the logic by relations of other arities. In the case of
unary parametricity, for example, one would then have an interpretation
of types as predicates. See also \cite{takeuti:98,Wadler:04}
\end{rem}

We first have the formation rule for internal equality:
\[
\begin{prooftree}
\Xi\mid\Gamma;-\ts t\co \sigma
\qquad
\Xi\mid\Gamma;-\ts u\co \sigma
\justifies
\Xi\mid\Gamma\mid\Theta\ts t=_\sigma u\co\Prop
\end{prooftree}
\]
Notice here the notational difference between $t=u$ and $t=_\sigma u$.
The former denotes \emph{external} equality and the latter is a
proposition in the logic.  The rules for $\imp$, $\vee$ and $\meet$
are the usual ones, where $\imp$ denotes implication. $\top$, $\bot$
are propositions in any context. We use $\biimp$  for biimplication.

We have the following formation rules for universal quantification:
\[
\begin{prooftree}
\Xi\mid\Gamma,x\co\sigma\mid\Theta\ts \phi\co\Prop
\justifies
\Xi\mid\Gamma\mid\Theta\ts \forall x\co\sigma\ld\phi\co\Prop
\end{prooftree}
\]
\[
\begin{prooftree}
\Xi\mid\Gamma\mid\Theta,R\rel{\sigma}{\tau}\ts \phi\co\Prop
\justifies
\Xi\mid\Gamma\mid\Theta\ts \forall
R\rel{\sigma}{\tau}\ld\phi\co\Prop 
\end{prooftree}
\]
\[
\begin{prooftree}
\Xi\mid\Gamma\mid\Theta,S\admrel\sigma\tau\ts \phi\co\Prop
\justifies
\Xi\mid\Gamma\mid\Theta\ts \forall
S\admrel\sigma\tau\ld\phi\co\Prop 
\end{prooftree}
\]
\[
\begin{prooftree}
\Xi,\alpha\mid\Gamma\mid\Theta\ts \phi\co\Prop
\justifies
\Xi\mid\Gamma\mid\Theta\ts \forall \alpha\co\Type\ld\phi\co\Prop
\using
\Xi\mid\Gamma\mid\Theta\text{ is well-formed}
\end{prooftree}
\]
The side condition $\Xi\mid\Gamma\mid\Theta\text{ is well-formed}$ means
that all the types of variables in $\Gamma$ and of relation variables in
$\Theta$ are well-formed in $\Xi$ (i.e., all the free type variables of the
types occur in $\Xi$).

There are similar formation rules for the existential quantifier. 


Before we give the formation rule for $\rho(t,u)$, we discuss
definable relations.

\subsubsection{Definable relations}

Definable relations are given by the grammar:
\[\rho ::= R \mid (x\co\sigma,y\co\tau).\phi \mid \sigma[R] \]
Definable relations always have a domain and a codomain, just as terms
always have types.  The basic formation rules for definable relations are:
\[
\begin{prooftree}
\justifies
\Xi\mid\Gamma\mid \Theta, R\rel\sigma\tau\ts
 R\rel\sigma\tau
\end{prooftree}
\]
\[
\begin{prooftree}
\Xi\mid\Gamma,x\co\sigma,y\co\tau\mid \Theta\ts \phi\co\Prop
\justifies
\Xi\mid\Gamma\mid \Theta\ts
(x\co\sigma,y\co\tau)\ld\phi\rel\sigma\tau 
\end{prooftree}
\]
\[
\begin{prooftree}
\Xi\mid\Gamma\mid \Theta\ts \rho\admrel\sigma\tau
\justifies
\Xi\mid\Gamma\mid \Theta\ts \rho\rel\sigma\tau
\end{prooftree}
\]
Notice that in the second rule we can only abstract
\emph{intuitionistic} variables to obtain definable relations. In the
last rule, $\rho\admrel\sigma\tau$ is an admissible relation, a
concept to be discussed below. The rule says that the admissible
relations constitute a subset of the definable relations. The last
construction of the grammar refers to the relational interpretation of
types and will be discussed in Section~\ref{sec:adm:rel}.

An example of a definable relation is the graph relation of a
function:
\[\langle f\rangle=(x\co\sigma,y\co\tau)\ld fx=_\tau y,\]
for $f\co\sigma\lpop \tau$. The equality relation $\eq_\sigma$ is
defined as the graph of the identity map. 

If $\rho\rel{\sigma}{\tau}$ is a definable relation, and we are
given terms of the right types, then we may 
form the proposition stating that the two terms are related by the definable
relation: 
\begin{equation}
  \label{eq:relsubst}
\begin{prooftree}
\Xi\mid\Gamma\mid\Theta\ts \rho\rel{\sigma}{\tau}
\quad 
\Xi\mid\Gamma;-\ts t\co\sigma, s\co\tau
\justifies
\Xi\mid\Gamma\mid\Theta\ts \rho(t,s)\co \Prop
\end{prooftree}
\end{equation}
We shall also write $t\rho s$  for $\rho(t,s)$.

Relations can be reindexed along $\pilly$  maps as in the following
derivable rule
\[
\begin{prooftree}
  \Xi \mid \Gamma \mid \Theta \ts \rho\rel{\sigma}{\tau} \qquad
  \Xi \mid \Gamma ; - \ts f\co \sigma' \lpop \sigma, g\co \tau' \lpop \tau 
  \justifies 
  \Xi \mid \Gamma \mid \Theta \ts (x\co\sigma',y\co \tau')\ld
  \rho(f\:x,g\:y) \rel{\sigma'}{\tau'}
\end{prooftree}
\]
where $x,y$ are fresh variables.
We shall use the shorthand notation
$(f,g)^*\rho$ for 
\[(x\co\sigma',y\co \tau')\ld \rho(f\:x,g\:y).\]

\subsubsection{Constructions on definable relations}
\label{sec:conondefrel}

In this subsection we present some constructions on definable
relations - one for each type constructor of $\pilly$. These will be
used to give a relational interpretation of the types of $\pilly$.

If $\rho\rel{\sigma}{\tau}$ and $\rho'\rel{\sigma'}{\tau'}$ 
define
\[\rho\lpop\rho'=(f\co\sigma\lpop\sigma',g\co\tau\lpop\tau')\ld \forall
x\co\sigma\ld\forall y\co\tau \ld\rho( x, y)\imp
\rho'(fx,gy).
\]
for fresh variables $x,y,f,g$. Then the rule
\[
\begin{prooftree}
  \Xi \mid \Gamma \mid \Theta \ts \rho\rel{\sigma}{\tau},
  \rho'\rel{\sigma'}{\tau'} \justifies \Xi \mid \Gamma \mid \Theta
  \ts(\rho\lpop\rho')\rel{(\sigma\lpop\sigma')}{(\tau\lpop\tau')}
\end{prooftree}
\]
is derivable.

If 
\[\Xi,\alpha,\beta\mid \Gamma \mid \Theta, R\admrel\alpha\beta\ts
\rho\rel{\sigma}{\tau}\]  
is well-formed and $\Xi\mid\Gamma\mid \Theta$ is well-formed,
$\Xi,\alpha\ts\sigma\co\Type$, and $\Xi,\beta\ts\tau\co\Type$ we may
define the relation
\[
\Xi\mid\Gamma\mid \Theta \ts \forall (\alpha,\beta, R\admrel\alpha\beta)\ld
\rho\rel{(\Prod \alpha\co\Type\ld\sigma)}
{(\Prod \beta\co\Type\ld\tau)}\] 
as 
\[
\begin{array}{c}
\forall (\alpha,\beta, R\admrel\alpha\beta)\ld\rho =\\
(t:\Prod \alpha\co\Type\ld\sigma,u:\Prod
\beta\co\Type\ld\tau)\ld
\forall \alpha,\beta\co \Type\ld\forall R\admrel\alpha\beta\ld
\rho(t\alpha,u\beta).
\end{array}
\] 

In Section~\ref{sec:proofs-in-lapl} we will show how to encode the
type constructors $\tensor, \bang, I$ using $\lpop, \to$ and
polymorphism as in Figure~\ref{fig:typeencodings} below. At this point
we have not discussed parametricity and so can not use the encodings,
but we will still use these for the definitions of the constructions
on relations corresponding to $\tensor, I$ and $\bang$. The relational
interpretations of $\tensor, I, \bang$ are due to Alex Simpson, who
also uses this relational interpretation of $\bang$  in a more
general context in \cite{Simpson:06}.

First we define the tensor product of $\rho$ and $\rho'$
\[\rho\tensor\rho'\rel
{(\sigma\tensor\sigma')}{(\tau\tensor\tau')},\] for
$\rho\rel{\sigma}{\tau}$ and $\rho'\rel{\sigma'}{\tau'}$. We first
introduce the map
\[f_{\sigma,\sigma'}:\sigma\tensor\sigma'\lpop\Prod\alpha\ld
(\sigma\lpop\sigma'\lpop\alpha) \lpop\alpha
\] 
defined as
\[f_{\sigma,\sigma'}\:x=\letexp{x'\tensor x''\co\sigma\tensor\sigma' }{x}{
  \Lambda\alpha\ld\linlambda h\co\sigma\lpop\sigma' \lpop\alpha\ld
  h\:x'\:x''}.
\]
Then we define
\[\rho\tensor\rho'= (f_{\sigma,\sigma'},f_{\tau,\tau'})^* (\forall
(\alpha,\beta, R\admrel\alpha\beta)\ld (\rho\lpop\rho' \lpop R)\lpop
R),\] 
or, if we write it out,
\[
\begin{array}{rcl}
\rho\tensor\rho' & = & (x\co \sigma\tensor \sigma',y\co\tau \tensor
\tau')\ld \forall \alpha,\beta, R\admrel\alpha\beta \ld \\
&& \forall t\co \sigma\lpop\tau\lpop\alpha, t'\co
\sigma'\lpop\tau'\lpop\beta \ld (\rho\lpop\rho' \lpop R)(t,t')\imp \\
&& R(\letexp{x'\tensor x''}{x}{t\:x'\:x''},\letexp{y'\tensor
  y''}{y}{t'\:y'\:y''}).
\end{array}
\]
As a derivable rule we get 
\[
\begin{prooftree}
  \Xi \mid \Gamma \mid \Theta \ts \rho\rel{\sigma}{\tau},
  \rho'\rel{\sigma'}{\tau'} \justifies \Xi \mid \Gamma \mid \Theta
  \ts(\rho\tensor\rho')\rel{(\sigma\tensor\sigma')}{(\tau\tensor \tau')}
\end{prooftree}
\]


Following the same strategy, we define a relation
$I_{Rel}\rel II$ using the map 
\[f\co I\lpop \Prod\alpha\ld\alpha\lpop\alpha\]
defined as $\linlambda
x\co I\ld \letexp\star x{\id}$, where $\id= \Lambda\alpha\ld \linlambda
x\co\alpha\ld x$ and define
\[ I_{Rel} = (f,f)^* (\forall(\alpha,\beta, R\admrel \alpha\beta) \ld
R\lpop R), \]
which, if we write it out, is
\[
\begin{array}{c}
(x\co I,y\co I) \ld \forall(\alpha,\beta, R\admrel \alpha\beta) \ld
\forall z\co\alpha, w\co\beta \ld \\ 
z R w \imp (\letexp{\star}xz) R(\letexp \star yw).
\end{array}
\]
The relation $I_{Rel}$ types in any context, i.e., $\Xi \mid \Gamma
\mid \Theta \ts I_{Rel}\rel II$ is derivable for any well-formed
context $\Xi \mid \Gamma\mid \Theta$. 

The encoding of $\bang$ in Figure~\ref{fig:typeencodings}  uses $\to$,
which was defined above as $\sigma \to \tau = \bang \sigma\lpop
\tau$, but since $\to$  has a natural relational interpretation, we
will still use this to define the relational interpretation of
$\bang$. 

For  $\rho\rel{\sigma}{\tau}$ and $\rho'\rel{\sigma'}{\tau'}$ we
define 
\[\rho\to \rho' = (f\co \sigma \to \sigma', g\co \tau\to \tau') \ld
\forall x\co \sigma, y \co \tau \ld \rho(x,y) \imp \rho'(f(\bang x),
g(\bang y)) 
\]

Now, define for any type $\sigma$ the map $f_\sigma \co \bang \sigma
\lpop \Prod\alpha \ld (\sigma\to \alpha)\lpop \alpha$ as
\[\linlambda x\co \bang \sigma \ld \Lambda\alpha \ld \linlambda g\co
\sigma \to \alpha \ld g(x).
\]
The relation $\bang \rho \rel{\bang\sigma}{\bang\tau}$  is defined as
\[(f_\sigma, f_\tau)^*\forall(\alpha, \beta, R\admrel\alpha\beta )\ld
(\rho \to R)\lpop R.
\]
The derivable typing rule is 
\[
\begin{prooftree}
  \Xi \mid \Gamma \mid \Theta \ts \rho \rel{\sigma}{\tau}
  \justifies \Xi \mid \Gamma \mid \Theta \ts \bang\rho \rel{\bang
    \sigma}{\bang \tau}
\end{prooftree}
\]

\begin{rem}\label{rem:rel:interp}
  In~\cite{MogelbergR:lapl-tr} we show how the constructions on
  relations presented in this section gives rise to a $\pilly$-model
  of admissible relations. In other words $\tensor, \lpop$ defines a
  symmetric monoidal structure on relations, $\bang$ extends this to a
  linear structure, and $\forall (\alpha,\beta,
  R\admrel\alpha\beta)\ld$ defines a polymorphic product.
\end{rem}

\begin{rem}
  The definitions of $\rho \tensor \rho'$ and $\bang\rho$ involve an
  implicit admissible closure operator discussed in
  Section~\ref{sec:adm:closure} below. This operator helps secure that
  the collection of admissible relations is closed under the
  constructions above (see Proposition~\ref{prop:admrel} below).
\end{rem}

\subsubsection{Admissible relations} \label{sec:adm:rel}

As mentioned in the introduction, for the theory of parametricity to
be consistent in a type theory with recursion the parametricity
principle must be weakened. For this purpose we introduce a notion of
admissible relations axiomatized in Figure~\ref{fig:admrel}. In these
rules $\rho\congr\rho'$ is a shorthand for $\forall x,y\ld
\rho(x,y)\biimp \rho'(x,y)$.
\begin{figure}[htbp]
  \centering
\[
\begin{prooftree}
\justifies
\Xi\mid\Gamma\mid\Theta,R\admrel\sigma\tau\ts R \admrel\sigma\tau
\end{prooftree}
\]
\[
\begin{prooftree}
\justifies
\Xi\mid\Gamma\mid\Theta\ts \eq_\sigma \admrel\sigma\sigma
\end{prooftree}
\]
\[
\begin{prooftree}
\Xi\mid\Gamma\mid\Theta\ts \rho\admrel\sigma\tau
\qquad
\Xi\mid\Gamma; - \ts t\co\sigma'\lpop\sigma, u\co\tau'\lpop\tau
\quad
x,y\notin \Gamma
\justifies
\Xi\mid\Gamma\mid\Theta\ts (x\co\sigma',y\co\tau')\ld \rho(t\:x,u\:y)
\admrel{\sigma'}{\tau'} 
\end{prooftree}
\]
\[
\begin{prooftree}
\Xi\mid\Gamma\mid\Theta\ts \rho,\rho'\admrel\sigma\tau
\quad
x,y\notin \Gamma
\justifies
\Xi\mid\Gamma\mid\Theta\ts (x\co\sigma,y\co\tau)\ld \rho(x,y) \meet
\rho'(x,y) \admrel{\sigma}{\tau} 
\end{prooftree}
\]
\[
\begin{prooftree}
\Xi\mid\Gamma\mid\Theta\ts \rho\admrel\sigma\tau
\quad
x,y\notin \Gamma
\justifies
\Xi\mid\Gamma\mid\Theta\ts (x\co\tau,y\co\sigma)\ld \rho(y,x)
\admrel{\tau}{\sigma}
\end{prooftree}
\]
%
\[
\begin{prooftree}
x,y\notin \Gamma
\justifies
\Xi\mid\Gamma\mid\Theta\ts (x\co\sigma, y\co\tau)\ld \top \admrel\sigma\tau
\end{prooftree}
\]
\[
\begin{prooftree}
\Xi\mid\Gamma\mid\Theta\ts \rho\admrel\sigma\tau
\qquad
\Xi\mid\Gamma\mid\Theta\ts \phi\co\Prop
\quad
x,y\notin \Gamma
\justifies
\Xi\mid\Gamma\mid\Theta\ts (x\co\sigma,y\co\tau)\ld \phi\imp\rho(x,y)
\admrel{\sigma}{\tau}  
\end{prooftree}
\]  
\[
\begin{prooftree}
\Xi,\alpha\mid\Gamma\mid\Theta\ts \rho\admrel\sigma\tau
\qquad
\Xi\mid\Gamma\mid\Theta 
\qquad
\Xi\ts \sigma\co\Type
\qquad
\Xi\ts \tau\co\Type
\quad
x,y\notin \Gamma
\justifies
\Xi\mid\Gamma\mid\Theta\ts (x\co\sigma,y\co\tau)\ld \forall
\alpha\co\Type\ld \rho(x,y) \admrel{\sigma}{\tau}  
\end{prooftree}
\]  
\[
\begin{prooftree}
\Xi\mid\Gamma,z\co\omega\mid\Theta\ts \rho\admrel\sigma\tau
\quad
x,y\notin \Gamma
\justifies
\Xi\mid\Gamma\mid\Theta\ts (x\co\sigma,y\co\tau)\ld \forall
z\co\omega\ld \rho(x,y) \admrel{\sigma}{\tau}  
\end{prooftree}
\]  
\[
\begin{prooftree}
\Xi\mid\Gamma\mid\Theta,R\admrel{\omega}{\omega'}\ts \rho\admrel\sigma\tau
\quad
x,y\notin \Gamma
\justifies
\Xi\mid\Gamma\mid\Theta\ts (x\co\sigma,y\co\tau)\ld \forall
R\admrel{\omega}{\omega'}\ld \rho(x,y) \admrel{\sigma}{\tau}  
\end{prooftree}
\]  
\[
\begin{prooftree}
\Xi\mid\Gamma\mid\Theta,R\rel{\omega}{\omega'}\ts \rho\admrel\sigma\tau
\quad
x,y\notin \Gamma
\justifies
\Xi\mid\Gamma\mid\Theta\ts (x\co\sigma,y\co\tau)\ld \forall
R\rel{\omega}{\omega'}\ld \rho(x,y) \admrel{\sigma}{\tau}  
\end{prooftree}
\]  
\[
\begin{prooftree}
\Xi\mid\Gamma\mid\Theta \ts \rho\admrel\sigma\tau,
\rho'\rel{\sigma}{\tau} 
\qquad
\Xi\mid\Gamma\mid \Theta\mid \top \ts \rho\congr \rho'
\justifies
\Xi\mid\Gamma\mid \Theta\ts \rho'\admrel\sigma\tau
\end{prooftree}
\]
\[
\begin{prooftree} 
\begin{array}{c}
\alpha_1,\ldots,\alpha_n\ts\sigma(\vec\alpha)\co\Type \qquad 
\Xi\mid\Gamma\mid \Theta\ts \rho_1\admrel{\tau_1}{\tau_1'},
\ldots,\rho_n\admrel{\tau_n}{\tau_n'}
\end{array}
\justifies
\Xi\mid\Gamma \mid\Theta\ts \sigma[\vec\rho]\admrel
{\sigma(\vec\tau)}{\sigma(\vec\tau')}
\end{prooftree}
\]
  \caption{Rules for admissible relations}
  \label{fig:admrel}
\end{figure}

A few comments is needed for the last of the rules in
Figure~\ref{fig:admrel}. First observe that $\sigma[\vec\rho]$ is a
syntactic construction and is not obtained by substitution as
in~\cite{Abadi:Plotkin:93}. Still the notation
$\sigma[\rho_1/\alpha_1,\ldots,\rho_n/\alpha_n]$ might be more
complete, but this quickly becomes overly verbose.  In
\cite{Abadi:Plotkin:93} $\sigma [\vec\rho]$ is to some extent defined
inductively on the structure of $\sigma$, but in our case that is not
enough, since we will need to form $\sigma[\vec\rho]$ for type
constants (when using the internal language of a model of
$\lapl$). The inductive definition over the type structure is in stead
reflected in axioms~\ref{axiom:baserel} to \ref{axiom:bangrel}.

We call $\sigma[\vec\rho]$ the \emph{relational interpretation of the
  type $\sigma$}.
\begin{prop}\label{prop:admrel}
The class of admissible relations contains all graphs and is closed
under the constructions of Section~\ref{sec:conondefrel}, in fact the
following more general rules hold
\[
\begin{prooftree}
  \Xi \mid \Gamma ; - \ts f\co \sigma \lpop \tau 
  \justifies
  \Xi \mid \Gamma \mid \Theta \ts \graph{f} \admrel{\sigma}{\tau}
\end{prooftree}
\]
\[
\begin{prooftree}
  \Xi \mid \Gamma \mid \Theta \ts \rho\rel{\sigma}{\tau},
  \rho' \admrel{\sigma'}{\tau'} \justifies \Xi \mid \Gamma \mid \Theta
  \ts(\rho\lpop\rho')\admrel{(\sigma\lpop\sigma')}{(\tau\lpop\tau')}
\end{prooftree}
\]
\[
\begin{prooftree}
  \Xi \mid \Gamma \mid \Theta \ts \rho\rel{\sigma}{\tau},
  \rho' \admrel{\sigma'}{\tau'} \justifies \Xi \mid \Gamma \mid \Theta
  \ts(\rho\to\rho')\admrel{(\sigma\to\sigma')}{(\tau\to\tau')}
\end{prooftree}
\]
\[
\begin{prooftree}
  \Xi \mid \Gamma \mid \Theta \ts \rho\rel{\sigma}{\tau},
  \rho'\rel{\sigma'}{\tau'} \justifies \Xi \mid \Gamma \mid \Theta
  \ts(\rho\tensor\rho')\admrel{(\sigma\tensor\sigma')}{(\tau\tensor \tau')}
\end{prooftree}
\]
\[
\begin{prooftree}
  \Xi \mid \Gamma \mid \Theta \ts \rho \rel{\sigma}{\tau}
  \justifies \Xi \mid \Gamma \mid \Theta \ts \bang\rho \admrel{\bang
    \sigma}{\bang \tau}
\end{prooftree}
\]
\[
\begin{prooftree}
  \justifies
  \Xi \mid \Gamma \mid \Theta \ts I_{Rel} \admrel II
\end{prooftree}
\]
\[
\begin{prooftree}
\Xi,\alpha,\beta\mid \Gamma \mid \Theta, R\admrel\alpha\beta\ts
\rho\admrel{\sigma}{\tau} 
\qquad
\Xi,\alpha\ts\sigma\co\Type \qquad \Xi,\beta\ts\tau\co\Type
\justifies 
\Xi\mid\Gamma\mid \Theta \ts \forall (\alpha,\beta, R\admrel\alpha\beta)\ld
\rho \admrel{(\Prod \alpha\co\Type\ld\sigma)}
{(\Prod \beta\co\Type\ld\tau)}
\end{prooftree}
\]
where the last rule has the side condition that $\Xi\mid\Gamma\mid
\Theta$ must be well-formed.
\end{prop}
\proof
Graph relations are admissible since equality relations are and
admissible relations are closed under reindexing. For the
constructions of Section~\ref{sec:conondefrel},  we just give the
proof of $\lpop$.

We must prove that for $\rho
\rel{\sigma}{\tau},\rho'\admrel{\sigma'}{\tau'}$ relations in the
same context $\rho\lpop\rho'$ is admissible. Consider first the
relation 
\[(f\co \sigma \lpop \sigma', g\co \tau \lpop \tau')\ld \rho'(f\:x,g\:y)
\]
in the context where we have added fresh variables $x\co \sigma, y\co
\tau$ to the contexts of $\rho, \rho'$. This relation is a reindexing
of $\rho'$  along the evaluation maps, which are linear, and so the
relation is admissible. Since $f,g$  do not occur freely in $\rho$,
also
\[(f\co \sigma \lpop \sigma', g\co \tau \lpop \tau')\ld \rho(x,y) \imp
\rho'(f\:x,g\:y) 
\]
is admissible, and so since admissible relations are closed under
universal quantification, $\rho\lpop \rho'$  is admissible. 
\qed

\subsubsection{Axioms and Rules} \label{subsec:axioms}

The last judgment in Figure~\ref{fig:judgements} has not yet been
mentioned. It says that in the given context, the formulas
$\phi_1,\ldots,\phi_n$ collectively imply $\psi$. We will often write
$\Phi$ for $\phi_1,\ldots,\phi_n$.

Having specified the language of $\lapl$, it is time to
specify the axioms and inference rules. 
We have all the usual axioms and rules of predicate logic plus
the axioms and rules specified below. 

Rules for substitution:
\begin{infrule}\label{axiom:substterm}
\begin{center}
\begin{prooftree}
\Xi\mid\Gamma,x\co\sigma\mid\Theta\mid\top\ts \phi
\quad
\Xi\mid\Gamma\ts t\co\sigma
\justifies
\Xi\mid\Gamma\mid\Theta\mid\top\ts \phi[t/x]
\end{prooftree}
\end{center}
\end{infrule}

\begin{infrule}\label{axiom:substrel}
\begin{center}
\begin{prooftree}
\Xi\mid\Gamma\mid\Theta,R\rel{\sigma}{\tau}\mid\top\ts \phi
\quad
\Xi\mid\Gamma\mid\Theta\ts \rho\rel{\sigma}{\tau} 
\justifies
\Xi\mid\Gamma\mid\Theta\mid\top\ts \phi[\rho/R]
\end{prooftree}
\end{center}
\end{infrule}

\begin{infrule}\label{axiom:substadmrel}
\begin{center}
\begin{prooftree}
\Xi\mid\Gamma\mid\Theta,S\admrel\sigma\tau\mid\top\ts \phi
\quad
\Xi\mid\Gamma\mid\Theta\ts \rho\admrel\sigma\tau 
\justifies
\Xi\mid\Gamma\mid\Theta\mid\top\ts \phi[\rho/S]
\end{prooftree}
\end{center}
\end{infrule}

\begin{infrule}\label{axiom:substtype}
\begin{center}
\begin{prooftree}
\Xi,\alpha\mid\Gamma\mid\Theta\mid\top\ts \phi
\quad
\Xi\ts \sigma\co\Type
\justifies
\Xi\mid\Gamma[\sigma/\alpha]\mid\Theta[\sigma/\alpha]
\mid\top\ts \phi[\sigma/\alpha]
\end{prooftree}
\end{center}
\end{infrule}

The \emph{substitution} axiom:

\begin{axiom}\label{axiom:substitution}
  \begin{center}
    $\forall \alpha,\beta\co\Type.\forall x,x'\co\alpha.\forall
    y,y'\co\beta.\forall R\rel{\alpha}{\beta.} R(x,y)\meet$\\$
    x=_\alpha x'\meet y=_\beta y'\imp R(x',y')$
  \end{center}
\end{axiom}

Rules for $\forall$-quantification:
\begin{infrule}\label{axiom:foralltype}
\begin{center}
\begin{prooftree}
\Xi,\alpha\mid\Gamma\mid\Theta\mid \Phi\ts\psi
\Justifies
\Xi\mid\Gamma\mid\Theta\mid\Phi\ts\forall\alpha\co\Type . \psi
\using
\Xi\mid\Gamma\mid\Theta\ts \Phi
\end{prooftree}  
\end{center}
\end{infrule}

\begin{infrule}\label{axiom:forallvar}
\begin{center}
\begin{prooftree}
\Xi\mid\Gamma,x\co\sigma\mid\Theta\mid \Phi\ts\psi
\Justifies
\Xi\mid\Gamma\mid\Theta\mid \Phi\ts\forall x\co\sigma . \psi
\using
\Xi\mid\Gamma\mid\Theta\ts \Phi
\end{prooftree}
\end{center}
\end{infrule}

\begin{infrule}\label{axiom:forallrel}
\begin{center}
\begin{prooftree}
\Xi\mid\Gamma\mid\Theta,R\rel{\tau}{\tau'}\mid \Phi\ts\psi
\Justifies
\Xi\mid\Gamma\mid\Theta\mid \Phi\ts\forall
R\rel{\tau}{\tau'} . \psi 
\using
\Xi\mid\Gamma\mid\Theta\ts \Phi
\end{prooftree}
\end{center}
\end{infrule}

\begin{infrule}\label{axiom:foralladmrel}
\begin{center}
\begin{prooftree}
\Xi\mid\Gamma\mid\Theta,S\admrel\tau{\tau'}\mid \Phi\ts\psi
\Justifies
\Xi\mid\Gamma\mid\Theta\mid \Phi\ts\forall
S\admrel\tau{\tau'} . \psi 
\using
\Xi\mid\Gamma\mid\Theta\ts \Phi
\end{prooftree}
\end{center}
\end{infrule}

Rules for $\exists$-quantification:

\begin{infrule}\label{axiom:existstype}
\begin{center}
\begin{prooftree}
\Xi,\alpha\mid\Gamma\mid\Theta\mid \phi\ts\psi
\Justifies
\Xi\mid\Gamma\mid\Theta\mid \exists\alpha\co\Type . \phi\ts\psi
\using
\Xi\mid\Gamma\mid\Theta\ts \psi
\end{prooftree}
\end{center}
\end{infrule}

\begin{infrule}\label{axiom:existsvar}
\begin{center}
\begin{prooftree}
\Xi\mid\Gamma,x\co\sigma\mid\Theta\mid \phi\ts\psi
\Justifies
\Xi\mid\Gamma\mid\Theta\mid \exists x\co\sigma . \phi\ts \psi
\using
\Xi\mid\Gamma\mid\Theta\ts \psi
\end{prooftree}
\end{center}
\end{infrule}

\begin{infrule}\label{axiom:existsrel}
\begin{center}
\begin{prooftree}
\Xi\mid\Gamma\mid\Theta,R\rel{\tau}{\tau'}\mid \phi\ts\psi
\Justifies
\Xi\mid\Gamma\mid\Theta\mid \exists R\rel{\tau}{\tau'
.} \phi\ts\psi 
\using
\Xi\mid\Gamma\mid\Theta\ts \psi
\end{prooftree}
\end{center}
\end{infrule}

\begin{infrule}\label{axiom:existsadmrel}
\begin{center}
\begin{prooftree}
\Xi\mid\Gamma\mid\Theta,S\admrel\tau{\tau'}\mid \phi\ts\psi
\Justifies
\Xi\mid\Gamma\mid\Theta\mid \exists S\admrel\tau{\tau'}
. \phi\ts\psi 
\using
\Xi\mid\Gamma\mid\Theta\ts \psi
\end{prooftree}
\end{center}
\end{infrule}

External equality implies internal equality:

\begin{infrule}\label{axiom:exteq}
\begin{center}
\begin{prooftree}
\Xi \mid \Gamma; - \ts t,u \co \sigma
\qquad
\Xi\mid\Gamma;-\ts t=u 
\justifies
\Xi\mid\Gamma\mid\Theta\mid\top \ts t=_\sigma u
\end{prooftree}
\end{center}
\end{infrule}
There are also obvious rules expressing that
internal equality is an equivalence relation.

%
We have rules concerning the interpretation of types as relations:  
\begin{infrule}\label{axiom:baserel}
\begin{center}
\begin{prooftree}
\vec\alpha\ts\alpha_i\co\Type
\qquad
\Xi\mid\Gamma\mid\Theta\ts\vec\rho\admrel{\vec\tau}{\vec\tau'}
\justifies
\Xi\mid\Gamma \mid\Theta\mid\top\ts \alpha_i[\vec\rho]\congr\rho_i
\end{prooftree}
\end{center}
\end{infrule}

\begin{infrule}\label{axiom:lpoprel}
\begin{center}
\begin{prooftree}
\vec\alpha\ts\sigma\lpop\sigma'\co\Type
\qquad
\Xi\mid\Gamma\mid\Theta\ts \vec\rho\admrel{\vec\tau}{\vec\tau'}
\justifies
\Xi\mid\Gamma\mid \Theta\mid\top\ts (\sigma\lpop\sigma')[\vec\rho]\congr
(\sigma[\vec\rho]\lpop\sigma'[\vec\rho]) 
\end{prooftree}
\end{center}
\end{infrule}

\begin{infrule}\label{axiom:tensorrel}
\begin{center}
\begin{prooftree}
\vec\alpha\ts\sigma\tensor\sigma'\co\Type
\qquad
\Xi\mid\Gamma\mid\Theta\ts \vec\rho\admrel{\vec\tau}{\vec\tau'}
\justifies
\Xi\mid\Gamma\mid \Theta\mid\top\ts (\sigma\tensor\sigma')[\vec\rho]\congr
(\sigma[\vec\rho]\tensor\sigma'[\vec\rho]) 
\end{prooftree}
\end{center}
\end{infrule}

\begin{infrule}\label{axiom:Irel}
\begin{center}
\begin{prooftree}
\Xi\mid\Gamma\mid\Theta\ts \vec\rho\admrel{\vec\tau}{\vec\tau'}
\justifies
\Xi\mid\Gamma\mid \Theta\mid\top\ts I[\vec\rho]\congr I_{Rel}
\end{prooftree}
\end{center}
\end{infrule}

\begin{infrule}\label{axiom:polyrel}
\begin{center}
\begin{prooftree}
\vec\alpha\ts\Prod \beta\ld\sigma(\vec\alpha,\beta)\co\Type
\qquad
\Xi\mid\Gamma\mid\Theta\ts \vec\rho\admrel{\vec\tau}{\vec\tau'}
\justifies
\Xi\mid\Gamma\mid\top\ts (\Prod \beta\ld\sigma(\vec\alpha,\beta))[\vec\rho]
\congr \forall
(\beta,\beta',R\admrel{\beta}{\beta'})\ld\sigma[\vec\rho,R]) 
\end{prooftree}
\end{center}
\end{infrule}

\begin{infrule}\label{axiom:bangrel}
\begin{center}
\begin{prooftree}
\vec\alpha\ts\bang\sigma\co\Type
\qquad
\Xi\mid\Gamma\mid\Theta\ts\vec\rho\admrel{\vec\tau}{\vec\tau'}
\justifies
\Xi\mid\Gamma \mid\Theta\mid\top\ts
(\bang\sigma)[\vec\rho]\congr \bang(\sigma[\rho])
\end{prooftree}
\end{center}
\end{infrule}
%

If the definable relation $\rho$ is of the form
$(x\co\sigma,y\co\tau)\ld \phi(x,y)$, then $\rho(t,u)$ is equivalent
to $\phi$ with $x,y$ substituted by $t,u$:
\begin{infrule}\label{axiom:defrel}
\begin{center}
\begin{prooftree}
\Xi\mid\Gamma,x\co\sigma,y\co\tau\mid\Theta\ts\phi\co\Prop
\quad
\Xi\mid\Gamma;-\ts t\co\sigma, u\co\tau
\justifies
\Xi\mid\Gamma\mid\Theta\mid\top\ts((x\co\sigma,y\co\tau)\ld
\phi)(t,u) \biimp \phi[t,u/x,y]
\end{prooftree}
\end{center}
\end{infrule}

\begin{axiom}\label{axiom:Y}
  \begin{center}
    \begin{prooftree}
      \justifies \Xi\mid\Gamma \mid \Theta \mid \top \ts
      (\Prod\alpha\ld(\alpha\to\alpha)\to\alpha) (Y,Y)
    \end{prooftree}
  \end{center}
\end{axiom}
Given a definable relation $\rho$ we may construct a proposition
$\rho(x,y)$.
On the other hand, if $\phi$  is a proposition
containing two free variables $x$ and $y$, then we may construct the
definable relation $(x,y)\ld\phi$. The next lemma tells 
us that these constructions give a correspondence between definable
relations and propositions, which is bijective up to provable
equivalence in the logic. 
\begin{lem}\label{lem:proprel}
  Suppose $\Xi \mid \Gamma, x\co \sigma, y\co \tau \mid \Theta \ts
  \phi$ is a proposition. Then
  \[\Xi \mid \Gamma \mid \Theta \mid \top\ts ((x\co\sigma,y\co\tau)\ld
  \phi)( x, y)\biimp \phi\] 
  Suppose $\Xi \mid \Gamma \mid \Theta\ts \rho\rel{\sigma}{\tau}$ is a
  definable relation, then
  \[\Xi \mid \Gamma \mid \Theta\mid \top \ts \rho\congr
  (x\co\sigma,y\co\tau)\ld \rho( x, y).\] 
\end{lem}
%
%
The substitution axiom above implies the \emph{replacement} rule:
\begin{lem}\label{lem:replacement}
\[
\begin{prooftree}
\Xi\mid\Gamma\mid -\mid \top \ts t=_\sigma t'
\qquad
\Xi\mid\Gamma,x\co\sigma;-\ts u\co \tau
\justifies
\Xi\mid\Gamma\mid -\mid \top \ts u[t/x]=_\tau u[t'/x]
\end{prooftree}
\]
\end{lem}
\proof
Consider the definable relation
\[\rho=(y\co\sigma,z\co\sigma)\ld u[y/x]=_\tau u[z/x].\]
Clearly $\rho(t,t)$  holds, so by substitution $\rho(t,t')$  holds.
\qed
\begin{lem}\label{lem:tensorrel}
  Suppose $\Xi \mid \Gamma \mid \Theta \ts \rho \rel{\sigma}{\tau},
  \rho' \rel{\sigma'}{\tau'}$ and $x,x', y,y'$  are fresh variables.
  Then
  \[\Xi \mid \Gamma \mid \Theta \ts \forall x\co \sigma, x' \co \sigma',
  y\co \tau, y'\co \tau' \ld \rho(x,y)\meet \rho'(x', y')\imp \rho\tensor
  \rho'(x\tensor x',y\tensor y')\]
\end{lem}
\proof
  Suppose $\rho(x,y)\meet \rho'(x', y')$ and that
  $(\rho\lpop\rho' \lpop R)(t,t')$. Then clearly $R(t\:x\:x',t'\:y\:y')$
  and thus, since
  \[\letexp{x\tensor x'}{x\tensor x'}{t\:x\:x'} = t\:x\:x',\] 
  we conclude $\rho\tensor\rho' (x\tensor x',y\tensor y')$.
\qed
%
%
\begin{lem}\label{lem:bangrel}
  Suppose $\Xi \mid \Gamma \mid \Theta \ts \rho\rel{\sigma}{\tau}$
  then 
  \[\Xi \mid \Gamma \mid \Theta \mid \top \ts \forall x\co\sigma,y\co\tau \ld
  \rho (x,y)\imp (\bang\rho)(\bang x,\bang y)\] 
  If  $\Xi \mid \Gamma \mid \Theta \ts \rho\admrel{\sigma}{\tau}$  then 
  \[\Xi \mid \Gamma \mid \Theta \mid \top \ts \forall x\co\sigma,y\co\tau \ld
  \rho (x,y)\biimp (\bang\rho)(\bang x,\bang y)\] 
\end{lem}
\proof
  The first statement is clear from the definition of $\bang \rho$.
  For the right to left implication in the case of $\rho$ being
  admissible, observe that $(\lambda x\co \sigma \ld x, \lambda x\co
  \tau \ld x) \co \rho \to \rho$. Since $\bang x(\bang\rho)\bang y$
  this implies
  \[\rho ((\lambda x\co \sigma \ld x) (\bang x),
  (\lambda x\co \tau \ld \tau)(\bang y)),\]
  i.e., $\rho(x,y)$.
\qed
Recall that in Section~\ref{sec:conondefrel} the oconstruction $\to$
was defined directly on relations, whereas in $\pilly$ the type
constructor $\to$  is shorthand for $\bang (-) \lpop (=)$. The next
lemma shows that the relations $\rho \to \rho'$  and $\bang \rho \lpop
\rho'$  coincide in the case of $\rho'$  being admissible. 
\begin{lem}
  Suppose  $\Xi \mid \Gamma \mid \Theta \ts \rho\rel{\sigma}{\tau},
  \rho'\rel{\sigma'}{\tau'}$. Then 
  \[\Xi \mid \Gamma \mid \Theta \mid \top \ts \forall f\co \sigma
  \to \sigma', g\co \tau \to \tau' \ld (\rho \to \rho')(f,g) \imp
  (\bang \rho \lpop \rho')(f,g). 
  \]
  If $\Xi \mid \Gamma \mid \Theta \ts \rho\rel{\sigma}{\tau},
  \rho'\admrel{\sigma'}{\tau'}$ then 
  \[\Xi \mid \Gamma \mid \Theta \mid \top \ts (\rho \to \rho') \congr
  (\bang \rho \lpop \rho').
  \] 
\end{lem}
\proof
  For the first implication, suppose $(f,g)\co \bang \rho \lpop
  \rho'$. We must show that if $\rho(x,y)$ then $\rho'(f(\bang x),
  g(\bang y))$, which follows from the assumptions since $\bang
  \rho(\bang x, \bang y)$.
  
  For the second half, we must show that if $\rho'$ is admissible and
  $\rho\to \rho'(f,g)$ and $\bang \rho(x,y)$ then $\rho'(f(x),g(y))$.
  But this follows from the definition of $\bang \rho(x,y)$ when using
  that $\rho'$ is admissible.
\qed

\subsubsection{A closure operator for admissible
  relations}\label{sec:adm:closure} 

In this section we present a closure operator on relations giving the
least admissible relation containing a given relation. This closure
operator will be particularly useful for proving coinduction
principles later. Recall from Proposition~\ref{prop:admrel} that for
$\rho$ any relation and $\rho'$ admissible $\rho\lpop \rho'$ is
admissible. This means that for any relation $\rho \rel \sigma\tau$,
\[(\forall \alpha, \beta, S\admrel\alpha\beta) \ld (\rho \lpop S)
\lpop S
\]
is an admissible relation from $\Prod\alpha \ld(\sigma\lpop
\alpha) \lpop \alpha$  to $\Prod\alpha \ld(\tau\lpop
\alpha) \lpop \alpha$. We define $\Phi(\rho)$  to be the admissible
relation obtained by pulling back this relation along the canonical
maps $\sigma \lpop\Prod\alpha \ld(\sigma\lpop
\alpha) \lpop \alpha$  and $\tau \lpop \Prod\alpha \ld(\tau\lpop
\alpha) \lpop \alpha$, i.e.  $\Phi(\rho)$ is 
\[(x\co \sigma, y\co \tau)\ld(\forall \alpha, \beta,
S\admrel\alpha\beta) \ld \forall f, g\ld (\rho \lpop S)(f,g) \imp
S(f(x), g(y)). \]
\begin{lem}\label{lem:admissible:closure}
  The operator $\Phi$ preserves implication of relations and for any
  relation $\rho$, $\Phi(\rho)$ is the smallest admissible relation
  containing $\rho$, i.e., 
\[\eqalign{
  \hbox{if}\qquad
 &\Xi \mid \Gamma \mid \Theta \ts \rho
  \rel\sigma\tau, \rho' \admrel{\sigma}{\tau}\cr
  \hbox{then}\qquad 
 &\Xi\mid \Gamma \mid \Theta \mid \top \ts \forall x\co \sigma, y
  \co \tau \ld \rho(x,y) \imp \rho'(x,y)\cr
  \hbox{iff}\qquad
 &\Xi\mid \Gamma \mid \Theta \mid \top \ts \forall x\co \sigma, y
  \co \tau \ld \Phi(\rho)(x,y) \imp \rho'(x,y)\cr
}
\]
\end{lem}

In later a paper we will show how the programming language $\lily$
gives rise to a model of LAPL. In this concrete model the notion of
admissibility is modeled by the $\perp\perp$-closed relations, and so
the admissible closure operation presented here coincides with
$\perp\perp$-closure as defined in~\cite{BiermanGM:opeplp}.

\begin{rem}\label{rem:closure:operator}
  Lemma~\ref{lem:admissible:closure} provides an alternative way of
  viewing the constructions on relations presented in
  Section~\ref{sec:conondefrel}. In fact $\bang\rho$  is the smallest
  admissible relation containing all pairs of the form $(\bang x,\bang
  y)$ for $\rho(x,y)$. Likewise $\rho\tensor \rho'$  is the smallest
  admissible relation containing all pairs $(x\tensor x', y\tensor
  y')$  with $\rho(x,y)\meet \rho'(x',y')$, and $I_{Rel}$  is the
  smallest admissible relation containing $(\star, \star)$. 
\end{rem}

\subsubsection{Extensionality and Identity Extension Schemes}

Consider the two \emphdefn{extensionality schemes}:
\[
\begin{array}{c}
  (\forall x\co\sigma\ld t\:x=_\tau u\:x)\imp t=_{\sigma\lpop\tau}u\\
  (\forall \alpha\co\Type\ld t\:\alpha=_\tau u\:\alpha)\imp t=_{\prod
    \alpha\co\Type\ld\tau}u. 
\end{array}
\]
These are taken as axioms in \cite{Abadi:Plotkin:93}, but we shall not
take these as axioms as we would like to be able to talk about models
that are not necessarily extensional. 
\begin{lem}\label{lem:intext}
  It is provable in the logic that 
  \[\forall f,g\co\sigma\to\tau\ld(\forall x\co\sigma\ld f(\bang
  x)=_\tau g(\bang x))\imp \forall x\co\bang\sigma\ld f(x)=_\tau g(x).\]
  In particular, extensionality implies
  \[\forall f,g\co\sigma\to\tau\ld(\forall x\co\sigma\ld f(\bang
  x)=_\tau g(\bang x))\imp f=_{\sigma\to\tau}g\]
\end{lem}
\proof
  The first formula of the theorem is just the statement that $(f,g)\co
  \eq_\sigma\to \eq_\tau$  implies $(f,g)\co \bang\eq_\sigma \lpop
  \eq_\tau$. The second formula follows from the first.
\qed
The schema
\[-\mid -\mid -\ts \forall\vec\alpha\co\Type\ld \sigma[\eq_{\vec\alpha}]\congr
\eq_{\sigma(\vec\alpha)}\] 
is called the \emphdefn{identity extension schema}. Here $\sigma$
ranges over all types, and $\eq_{\vec\alpha}$ is short notation for
$\eq_{\alpha_1}, \ldots,\eq_{\alpha_n}$.

For any type $\beta,\alpha_1,\ldots,\alpha_n\ts \sigma(\beta,\vec\alpha)$
we can form the \emphdefn{parametricity schema}:
\[-\mid -\mid -\ts \forall\vec\alpha\forall u\co(\Prod\beta\ld\sigma) \ld
\forall\beta,\beta'\ld\forall R\admrel{\beta}{\beta'}\ld
(u\:\beta)\sigma[R,\eq_{\vec\alpha}](u\:\beta'),
\] 
where, for readability, we have omitted $\co\Type$  after
$\beta,\beta'$.

We remark that the reason that parametricity and identity extension
are formulated as schemas is that we cannot quantify over type
constructors. 
\begin{prop}
  The identity extension schema implies the parametricity schema. 
\end{prop}
\proof
  The identity extension schema tells us that
  \[\forall\vec\alpha\forall u\co(\Prod\beta\ld\sigma) \ld
  u(\Prod\beta\ld\sigma)[\eq_{\vec\alpha}] u.
  \] 
  Writing out this expression using Rule~\ref{axiom:polyrel}
  for the relational interpretation
  of polymorphic types, one obtains the parametricity schema.
\qed

In the case of second-order lambda-calculus, the parametricity schema
implied identity extension for the pure calculus, since it provided
the case of polymorphic types in a proof by induction. It is
interesting to notice that this does not seem to be the case for
$\pilly$, since it seems that we need identity extension to prove for
example $\eq_\sigma\tensor\eq_\tau\congr \eq_{\sigma\tensor \tau}$.


\subsection{Logical Relations Lemma}

We end our presentation of Linear Abadi \& Plotkin Logic with the
logical relations lemma.

\begin{lem}[Logical Relations Lemma]\label{lem:lrl}
  In pure $\lapl$, for any closed term $-\mid-;-\ts t\co\tau$,
  \[t\tau t.\]
  In words, any closed term of closed type, is related to itself in
  the relational interpretation of the type. More generally, for any
  open term
  \[\vec\alpha\mid \vec x'\co\vec\sigma'(\alpha); \vec
  x\co\vec\sigma(\alpha) \ts t(\vec\alpha,\vec x',\vec x)\co\tau\] in the
  pure calculus; the proposition
  \[
  \begin{array}{c}
  \forall\vec\alpha,\vec\beta\ld\forall \vec R\admrel
  {\vec\alpha}{\vec\beta} \ld\forall \vec x\co\vec\sigma(\vec\alpha),
  \vec y\co\vec\sigma(\vec\beta)\ld \forall \vec
  x'\co\vec\sigma'(\vec \alpha), \vec y'\co\vec \sigma'(\vec \beta) \ld \\
  \vec x\vec\sigma[\vec R] \vec y\meet \vec x'\vec\sigma'[\vec R] \vec
  y' \imp t(\vec\alpha,\vec x',\vec x)\tau[\vec R]t(\vec\beta,\vec
  y',\vec y)
  \end{array}
  \]
  holds in the logic.
\end{lem}
%
A detailed proof of the Logical Relations Lemma can be found
in~\cite{Mogelberg:PhDThesis}.

\section{Encoding datatypes using parametricity}
\label{sec:proofs-in-lapl}

In this section we show how to use the logic to prove correctness of
encodings of a large class of data types in $\pilly$ using
parametricity. These encoding are due to Plotkin, and many of them are
listed Figure~\ref{fig:typeencodings}. In
Figure~\ref{fig:typeencodings} there are two sorts of equations. The
first four equations are isomorphisms between types already present in
$\pilly$. In these cases we shall show that the isomorphisms hold in a
category of \emph{linear} maps, where maps are considered equal up to
provability in the logic. We shall give a precise definition of this
category shortly. 

The other type of equation in Figure~\ref{fig:typeencodings} defines
encodings of types not already present in $\pilly$. We shall show
correctness of these encodings, by which we mean that they satisfy the
usual universal properties with respect to the above mentioned
category of linear maps. In the last two encodings, $\sigma$ is
assumed to be a type expression of $\pilly$ in which $\alpha$ occurs
only positively (see Section~\ref{sec:functors}) in which case
$\mu\alpha \ld \sigma$ defines an initial algebra for the functor
induced by $\sigma$ and $\nu\alpha \ld \sigma$ defines a final
coalgebra. We will also discuss reasoning principles for the encoded
types.

We will prove that the fixed point combinator $Y$ causes the initial
algebras and final coalgebras to coincide --- a phenomenon called
algebraic compactness. As a special case we have the coincidence of
the initial and final object (0=1) as can be seen in
Figure~\ref{fig:typeencodings}. Following
Freyd~\cite{Freyd:90,Freyd:90a,Freyd:91} we show how algebraic
compactness implies the existence of general recursive types in
Section~\ref{sec:rectypes}.

In the following we shall write ``using extensionality'' and ``using
identity extension'' to mean that we assume the extensionality schemes
and the identity extension schema, respectively.

%
\subsection{A category of linear functions}
\begin{figure}[tbp]
  \centering
  \[
  \begin{array}{rcl}
    \sigma & \iso & \Prod\alpha\ld (\sigma\lpop\alpha)\lpop\alpha\\
    \sigma\tensor \tau & \iso & \Prod\alpha\ld (\sigma\lpop \tau\lpop
    \alpha)\lpop \alpha \\
    \bang \sigma & \iso & \Prod\alpha \ld (\sigma \to \alpha) \lpop
    \alpha \\
    I & \iso & \Prod\alpha\ld\alpha\lpop\alpha\\
    0 & = & \Prod\alpha\ld\alpha\\
    1 & = & \Prod\alpha\ld\alpha\\
    \sigma + \tau & = & \Prod\alpha\ld (\sigma\lpop\alpha) \to
    (\tau\lpop \alpha) \to \alpha \\
    \sigma\times \tau & = & \Prod\alpha\ld (\sigma\lpop\alpha) + (\tau
    \lpop \alpha)\lpop \alpha \\
    \N & = & \Prod\alpha\ld (\alpha\lpop\alpha)\to \alpha\lpop
    \alpha\\
    \Coprod\alpha \ld \sigma & = & \Prod\beta\ld (\Prod\alpha\ld
    \sigma\lpop \beta)\lpop \beta\\
    \mu\alpha\ld \sigma & = & \Prod\alpha\ld (\sigma\lpop
    \alpha)\to\alpha \\
    \nu \alpha \ld \sigma & = & \Coprod\alpha \ld \bang
    (\alpha\lpop\sigma)\tensor \alpha
  \end{array}
  \]
  \caption{Types definable using parametricity}
  \label{fig:typeencodings}
\end{figure}
%
The precise formulation of correctness of encodings of the datatypes
presented in this section will be that they satisfy the usual
universal properties. To state this precisely, we introduce for each
kind context $\Xi$ the category $\LinTypecat_\Xi$ as follows:
\begin{description}
\item[Objects] are types $\Xi\mid -;-\ts\sigma\co\Type$.
\item[Morphisms] $[\Xi\mid -;-\ts f\co\sigma\lpop\tau]$ are equivalence
classes of terms of type $\sigma\lpop\tau$; the equivalence
relation on these terms being provable equality in LAPL using
extensionality and identity extension.
\end{description}
Composition in this category is given by lambda abstraction,
i.e. $f\co\sigma\lpop\tau$ composed with $g\co\omega\lpop\sigma$
yields $\linlambda x\co\omega\ld f(g x)$.

We start by proving that under the assumption of identity extension
and extensionality, for all types $\Xi\ts \sigma\co\Type$  we have an
isomorphism of objects of $\LinTypecat_\Xi$:
\[\sigma\iso\Prod\alpha\ld(\sigma\lpop\alpha)\lpop\alpha\]
for $\alpha$ not free in $\sigma$. We can define terms
\[f\co\sigma\lpop\Prod\alpha\ld
((\sigma\lpop\alpha) \lpop\alpha)\]
and 
\[g\co\Prod\alpha\ld
((\sigma\lpop\alpha) \lpop\alpha)\lpop\sigma\]
by 
\[f=\linlambda x\co\sigma\ld \Lambda\alpha\ld\linlambda
h\co\sigma\lpop \alpha\ld h\:x\]
and
\[g=\linlambda x\co\Prod\alpha\ld
((\sigma\lpop\alpha) \lpop\alpha)\ld x\:\sigma\:\id_{\sigma}\]

Clearly
\[g\:(f\:x)=(f\:x)\:\sigma\:\id_{\sigma}=x\]
so $g\circ f=\id_{\sigma}$. Notice that this only involve external equality
and thus we did not need extensionality here.
\begin{prop}
  Using identity extension and extensionality, one may prove
  that $f\circ g$ is internally equal to the identity.
\end{prop}
\proof
  For a term $a\co\Prod\alpha\ld(\sigma\lpop\alpha)\lpop\alpha$  we have
  \[f\circ g\:a=\Lambda\alpha\ld\linlambda h\co\sigma\lpop\alpha\ld
  h(a\:\sigma\: \id_{\sigma}).\] Using extensionality, it suffices to prove
  that
  \[\Xi,\alpha\mid h\co\sigma\lpop\alpha\mid-\ts
  h(a\:\sigma\:\id_{\sigma})=_\alpha a\:\alpha\:h\] holds in the internal
  logic.
  
  By the parametricity schema we know that for any admissible relation
  $\rho\admrel{\tau}{\tau'}$
  \[(a\:\tau)((\eq_{\sigma}\lpop\rho)\lpop\rho)(a\:\tau')\]
  If we instantiate this with the admissible relation $\langle h\rangle$,
  we get
  \[(a\:\sigma)((\eq_{\sigma}\lpop\langle h\rangle)\lpop\langle
  h\rangle)(a\:\alpha)\] 
  Since $ \id_{\sigma}(\eq_{\sigma}\lpop\langle h\rangle) h$  we
  know that $ (a\:\sigma\:\id_{\sigma})\langle h\rangle
  (a\:\alpha\:h)$, i.e.,
  \[h(a\:\sigma\:\id_{\sigma})=_{\alpha}a\:\alpha\:h,\]
  as desired.
\qed
This proof may essentially be found in \cite{BirkedalL:catmap}.

Intuitively, what happens here is that $\sigma$  is a subtype of
$\Prod\alpha \ld (\sigma\lpop\alpha)\lpop \alpha$,  where the
inclusion $f$  maps $x$  to application at $x$. We use parametricity
to show that $\Prod\alpha \ld (\sigma\lpop\alpha)\lpop \alpha$ does
not contain anything that is not in $\sigma$. 

\subsection{Tensor types}

The goal of this section is to prove
\[\sigma\tensor\tau\iso
\Prod\alpha\ld(\sigma\lpop\tau\lpop\alpha)\lpop\alpha\] 
using identity extension and extensionality, for
$\Xi\ts\sigma\co\Type$ and $\Xi\ts\tau\co\Type$ types in the same
context. The isomorphism is in the category $\LinTypecat_\Xi$.

This isomorphism leads to the question of whether tensor types are
actually superfluous in the language. The answer is yes in the
following sense: Call the language without tensor types (and $I$) $t$
and the language as is $T$. Then there are transformations $p:T\to t$
and $i:t\to T$, $i$ being the inclusion, such that $p\circ i = id_T$
and $i\circ p\iso id_t$. This is all being stated more precisely, not
to mention proved, in~\cite{MogelbergR:lapl-sdt-tr}. In this paper we
settle for the isomorphism above.

We can construct terms 
\[f:\sigma\tensor\tau\lpop\Prod\alpha\ld
(\sigma\lpop\tau\lpop\alpha) \lpop\alpha\]
and 
\[g:(\Prod\alpha\ld
(\sigma\lpop\tau\lpop\alpha) \lpop\alpha)\lpop\sigma\tensor\tau\]
by
\[f\:y=\letexp{x\tensor x'\co\sigma\tensor\tau }{y}{
  \Lambda\alpha\ld\linlambda h\co\sigma\lpop\tau\lpop\alpha\ld h\:x\:x'}\]
and 
\[g\:y=y\:(\sigma\tensor\tau)\: \comb,\]
where the map $\comb:\sigma\lpop\tau\lpop\sigma\tensor\tau$  is
\[\comb=\linlambda x\co\sigma\ld\linlambda x'\co\tau\ld x\tensor x'.\]

Let us show that the composition $g\circ f$ is the identity.
\[
\begin{array}{rcl}
g\circ f\:y & = & g(\letexp{x\tensor x'\co\sigma\tensor\tau }{y}{
  \Lambda\alpha\ld\linlambda h\co\sigma\lpop\tau\lpop\alpha\ld
  h\:x\:x'})\\
& = & (\letexp{x\tensor x'\co\sigma\tensor\tau }{y}{
  \Lambda\alpha\ld\linlambda h\co\sigma\lpop\tau\lpop\alpha\ld
  h\:x\:x'})\:(\sigma\tensor\tau)\: \comb \\
& = & (\Lambda\alpha\ld\linlambda h\co\sigma\lpop\tau\lpop\alpha\ld
\letexp{x\tensor x'\co\sigma\tensor\tau }{y}{  
  h\:x\:x'})\:(\sigma\tensor\tau)\: \comb \\
& = & \letexp{x\tensor x'\co\sigma\tensor\tau }{y}{x\tensor x'} \\
& = & y.
\end{array}
\]
\begin{prop}
Using extensionality and identity extension one may prove that the
composition 
\[f \circ g\co (\Prod\alpha\ld(\sigma\lpop\tau\lpop\alpha) \lpop\alpha)
\lpop (\Prod\alpha\ld(\sigma\lpop\tau\lpop\alpha) \lpop\alpha)\]
is internally equal to the identity.
\end{prop}
\proof
We compute
\[
\begin{array}{c}
f\circ g\:(y)=f(y\:(\sigma\tensor\tau)\:\comb)=\\
\letexp{x\tensor
  x'\co\sigma\tensor\tau}{(y\:\sigma\tensor\tau\:\comb)}
  {\Lambda\alpha\ld\linlambda h\co\sigma\lpop\tau\lpop\alpha\ld h\:x\:x'}
\end{array}
\]
Suppose we are given a type $\alpha$  and a map
$h\co\sigma\lpop\tau\lpop\alpha$.  We can define
$\phi_h\co\sigma\tensor\tau\lpop\alpha$  as 
\[\phi_h=\linlambda y\co \sigma\tensor \tau\ld \letexp{x\tensor
  x'\co\sigma\tensor\tau}{y}{h\:x\:x'}.\] 
Then $\phi_h(\comb\:x\:x')=h\:x\:x'$, which means that
$\comb(\eq_\sigma \lpop\eq_\tau \lpop\langle\phi_h\rangle) h$.  By the
parametricity schema 
\[
\begin{array}{c}
\Xi,\alpha\mid h\co\sigma\lpop\tau\lpop\alpha,y\co \Prod\alpha\ld
(\sigma\lpop\tau\lpop\alpha) \lpop\alpha\mid-\mid\top\ts\\
(y\:\sigma\tensor\tau)((\eq_\sigma \lpop\eq_\tau 
\lpop\langle\phi_h\rangle) \lpop \langle\phi_h\rangle) (y\:\alpha)
\end{array}
\]
so
\[ (y\:\sigma\tensor\tau\:\comb)\langle\phi_h\rangle
(y\:\alpha\:h),\] 
i.e,
\[\phi_h(y\:\sigma\tensor\tau\:\comb)=_{\alpha}y\:\alpha\:h.\]
Writing this out we get 
\[
\begin{array}{c}
\Xi,\alpha\mid h\co\sigma\lpop\tau\lpop\alpha,y\co \Prod\alpha\ld
(\sigma\lpop\tau\lpop\alpha) \lpop\alpha\mid-\mid\top\ts\\
\letexp{x\tensor
  x'\co\sigma\tensor\tau}{(y\:\sigma\tensor\tau\:\comb)}{h\:x\:x'}=_\alpha
  y\:\alpha\:h.
\end{array}
\] 
Using extensionality we get 
\[\Lambda\alpha\ld\linlambda h\co\sigma\lpop\tau\lpop\alpha\ld \letexp{x\tensor
  x'\co\sigma\tensor\tau}{(y\:\sigma\tensor\tau\:\comb)}
  {(h\:x\:x')}=_\alpha y.\]
This is enough, since by the rules for external equality the left hand side is
\[\letexp{x\tensor
  x'\co\sigma\tensor\tau}{(y\:\sigma\tensor\tau\:\comb)}
  {(\Lambda\alpha\ld\linlambda 
  h\co\sigma\lpop\tau\lpop\alpha\ld h\:x\:x')}.
\]
\qed

\subsection{Unit object}

The goal of this section is to prove that identity extension together with
extensionality implies 
\[I\iso \Prod\alpha\ld\alpha\lpop\alpha.\]
The isomorphism holds in $\LinTypecat_\Xi$  for all $\Xi$. 

We first define maps $f\co I\lpop \Prod\alpha\ld\alpha\lpop\alpha$
and $g\co (\Prod\alpha\ld\alpha\lpop\alpha) \lpop I$ as
\[
\begin{array}{c}
f= \linlambda x\co I\ld \letexp\star x{\id}, \\
g = \linlambda t\co \Prod\alpha\ld \alpha\lpop\alpha \ld t\:I\:\star,
\end{array}
\]
where 
\[\id = \Lambda\alpha\ld \linlambda y\co\alpha \ld y.\]
We first notice that 
\[
\begin{array}{rcl}
g(f(x)) & = & (\letexp\star x{\id})\:I\:\star  \\ 
& = & \letexp\star x{(\id\:I\:\star)} \\
& = & \letexp\star x{\star} \\
& = & x.
\end{array}
\]
\begin{prop}
Using identity extension and extensionality, we have that
$f\circ g$  is internally equal to the identity on
$\Prod\alpha\ld\alpha\lpop\alpha$. 
\end{prop}
\proof
First we write out the definition
\[f\circ g= \linlambda t\co(\Prod\alpha\ld\alpha\lpop\alpha) \ld
\letexp{\star}{(t\:I\:\star)}{\id}.
\]
We show that for any $t\co \Prod\alpha\ld\alpha\lpop \alpha$,
for any type $\sigma$,  and any $x\co\sigma$ we have
$f\circ g(t)\: \sigma\: x=_\sigma t\:\sigma\:x$. 

Given $\sigma, x$  as above, we can define $h\co I\lpop \sigma$  as
$h=\linlambda z\co I\ld \letexp{\star}zx$. Then $\graph h$  is
admissible, so by identity extension 
\[(t\:I)(\graph h\lpop\graph h)(t\:\sigma).\]
Since $h(\star)=x$  we have $h(t\:I\:\star)=_\sigma t\:\sigma \:x$,
and by definition 
\[
\begin{array}{rcl}
h(t\:I\:\star)&=&\letexp{\star}{(t\:I\:\star)}x 
\\ &=& \letexp{\star}{(t\:I\:\star)}(\id \:\sigma\:x) \\
&=& (\letexp{\star}{(t\:I\:\star)}\id) \:\sigma\:x \\ 
&=& f\circ g(t)\:\sigma\: x.
\end{array}
\]
\qed

\subsection{Initial objects and coproducts}

We define 
\[0 = \Prod\alpha\ld \alpha
\]
For each $\Xi$ this defines a weak initial object in $\LinTypecat_\Xi$,
since for any type $\Xi\ts \sigma$, there exists a term $0_\sigma\co
0\lpop \sigma$, defined as
\[\linlambda x\co 0\ld x\:\sigma
\]

\begin{prop}
  Suppose $f\co 0\lpop\sigma$  for some type $\Xi\ts\sigma$. Using
  identity extension and extensionality it is provable that
  $f=_{0\lpop\sigma} 0_\sigma$. Thus, $0$  is an
  initial object in $\LinTypecat_\Xi$ for each $\Xi$. 
\end{prop}

\proof
  First notice that for any map $h\co \sigma\lpop \tau$, by identity
  extension $(x\:\sigma)\graph h (x\:\tau)$ for any $x\co 0$. Thus, by
  extensionality, $h\circ 0_\sigma=_{0\lpop\tau} 0_\tau$ for any $h\co
  \sigma\lpop\tau$. In particular, for any type $\sigma$, the case
  $h=0_\sigma$ gives us $x\:0\:\sigma=_\sigma x\:\sigma$, i.e.,
  $0_0=_{0\lpop 0}\id_0$. If $f\co 0\lpop\sigma$, by the above we have
  $0_\sigma=_{0\lpop \sigma} f\circ 0_0=_{0\lpop \sigma} f$
\qed

Next, suppose $\Xi\ts \sigma,\tau$  are types in the same context. We
define 
\[\sigma + \tau = \Prod\alpha\ld (\sigma\lpop\alpha)\to
(\tau\lpop\alpha)\to \alpha
\]
and show under the assumption of identity extension and extensionality
that this defines a coproduct of $\sigma$ and $\tau$ in
$\LinTypecat_\Xi$.

First define terms $\interm_\sigma\co \sigma\lpop \sigma + \tau$,
$\interm_\tau\co \tau\lpop \sigma + \tau$ as
\[
\begin{array}{rcl}
\interm_\sigma & = & \linlambda x\co \sigma\ld \Lambda \alpha\ld \lambda
f\co \sigma\lpop \alpha \ld \lambda g\co \tau\lpop \alpha\ld f(x) \\
\interm_\tau & = & \linlambda y\co \tau\ld \Lambda \alpha\ld \lambda
f\co \sigma\lpop \alpha \ld \lambda g\co \tau\lpop \alpha\ld g(y) 
\end{array}
\]
For any pair of maps $f\co \sigma \lpop \omega$, $g\co \tau \lpop
\omega$  define the copairing $\copair fg\co \sigma+\tau\lpop \omega$ as
\[\copair fg = \linlambda x\co \sigma + \tau \ld x\:\omega \:\bang
f\:\bang g,
\]
then clearly $\copair fg (\interm_\sigma (x))= f(x)$ and $\copair fg
(\interm_\tau (y))= g(y)$,  and so $\sigma+\tau$  is a weak coproduct
of $\sigma$  and $\tau$  in $\LinTypecat_\Xi$. We remark that the
copairing constructor can also be defined as a polymorphic term
\[\copair -- \co \Lambda\alpha \ld (\sigma\lpop \alpha) \to (\tau\lpop
\alpha) \to \sigma+\tau \lpop \alpha
\]
of \emph{intuitionistic} function type. Of course we can define an
even more general copairing by abstracting $\sigma,\tau$ as well.

\begin{lem}
  If $h\co \omega\lpop \omega'$, $f\co \sigma\lpop \omega$  and $g\co
  \tau\lpop \omega$, then using extensionality and identity extension,
  it is provable that $\copair{h\circ f}{h\circ g}=_{\sigma+\tau \lpop
  \omega'} h\circ \copair fg$. 
\end{lem}
\proof
  Since 
  \[
  \begin{array}{c}
    f(\eq_\sigma \lpop \graph h)h\circ f\\
    g(\eq_\tau \lpop \graph h)h\circ g
  \end{array}
  \]
  for any $x\co \sigma+\tau$, 
  \[(x\:\omega\:\bang f\:\bang g)\graph h (x\:\omega'\:\bang (h\circ
  f)\:\bang (h\circ g)) 
  \]
  by identity extension, i.e., $h(\copair fg (x))= \copair {h\circ
  f}{h\circ g}(x)$. 
\qed

\begin{lem}
  Using extensionality and identity extension, 
  $\copair {\interm_\sigma}{\interm_\tau}=_{\sigma+\tau \lpop
  \sigma+\tau} \id_{\sigma+\tau}$ is provable. 
\end{lem}

\proof
  Given any $\omega, a\co \sigma\lpop \omega, b\co \tau\lpop\omega$,
  we have 
  \[\copair ab (\copair {\interm_\sigma}{\interm_\tau} (x))=_\omega  \copair
  {\copair ab \circ \interm_\sigma}{\copair ab \circ \interm_\tau} (x)
  =_\omega \copair ab (x)
  \]
  for any $x\co \sigma+\tau$. By unfolding the definition of $\copair
  ab$ in the above equality we get
  \[\copair {\interm_\sigma}{\interm_\tau} (x)\:\omega \:\bang
  a\:\bang b =_\omega x\:\omega \:\bang a\:\bang b.
  \]
  Since $\omega,a,b$ were arbitrary, extensionality (and
  Lemma~\ref{lem:intext}) implies $\copair
  {\interm_\sigma}{\interm_\tau} (x)=_{\sigma + \tau} x$ for all $x$.
\qed
\begin{prop}
  For any $f\co \sigma\lpop \omega$, $g\co \tau\lpop \omega$ and $h\co
  \sigma+\tau\lpop \omega$, if $h\circ \interm_\sigma =_{\sigma\lpop
    \omega} f$ and $h\circ \interm_\tau =_{\tau\lpop \omega} g$, then
  it is provable using identity extension and extensionality that
  $h=_{\sigma+\tau\lpop \omega}\copair fg$. Thus $\sigma+\tau$ is a
  coproduct of $\sigma$ and $\tau$ in $\LinTypecat_\Xi$.
\end{prop}
\proof
  \[\copair fg =_{\sigma+\tau\lpop \omega} \copair {h\circ
  \interm_\sigma}{h\circ \interm_\tau} =_{\sigma+\tau\lpop \omega}
  h\circ \copair{\interm_\sigma}{\interm_\tau} =_{\sigma+\tau\lpop
  \omega} h
  \]
\qed

\subsection{Terminal objects and products}

The initial object $0$  is also weakly terminal, since for any type
$\sigma$, 
\[\Omega_{\sigma\lpop 0} = Y\:\sigma\: \bang\id_{\sigma\lpop 0}
\]
is a term of type $\sigma\lpop 0$. In fact, using parametricity, $0$
can be proved to be terminal.

\begin{prop}
  Suppose $f,g\co \sigma\lpop 0$. Using identity extension and
  extensionality it is provable that $f=_{\sigma\lpop 0} g$. Thus $0$
  is a terminal object in $\LinTypecat_\Xi$ for any $\Xi$.
\end{prop}

\proof
  We will prove 
  \[\forall x,y\co 0\ld x=_0 y
  \]
  which, by extensionality, implies the proposition. Suppose we are
  given $x,y\co 0$. The term \[\linlambda z\co 0\ld z\:0\lpop
  \sigma \: y\]  has type $0\lpop \sigma$,  and thus is equal to
  $0_\sigma$. This means that $x\:\sigma =_\sigma x\:0\lpop\sigma
  y$. Likewise $x\:0\lpop\sigma y=_\sigma y\:\sigma$, so
  $x\:\sigma=_\sigma y\:\sigma$. Since this holds for all $\sigma$, by
  extensionality $x=_0 y$. 
\qed

Suppose $\sigma,\tau$  are types in the same context $\Xi$. Define
\[\sigma\times \tau=\Prod\alpha\ld (\sigma\lpop\alpha)+(\tau\lpop
\alpha) \lpop \alpha.
\]
This defines a weak product in $\LinTypecat_\Xi$  with projections
$\pi_\sigma \co\sigma\times \tau\lpop\sigma$  and $\pi_\tau
\co\sigma\times \tau\lpop\tau$ defined as
\[
\begin{array}{rcl}
\pi_\sigma & = & \linlambda x\co \sigma\times \tau \ld x\:\sigma\:
(\interm_{\sigma\lpop\sigma} \id_\sigma)\\
\pi_\tau & = & \linlambda x\co \sigma\times \tau \ld x\:\tau\:
(\interm_{\tau\lpop\tau} \id_\tau)
\end{array}
\]
The pairing of terms $f\co \omega\lpop\sigma$  and $g\co
\omega\lpop\tau$  is $\pair fg\co \omega\lpop\sigma\times \tau$  defined
as
\[\pair fg = \linlambda x\co \omega \ld \Lambda \alpha\ld \linlambda
h\co (\sigma\lpop\alpha) + (\tau\lpop\alpha)\ld \copair {\linlambda
  z\co \sigma\lpop \alpha\ld z\circ f} {\linlambda z\co \tau\lpop
  \alpha\ld z\circ g} \:h\:x
\]
Then
\[\pi_\sigma(\pair fg (x)) = \pair fg (x)\:\sigma\:
(\interm_{\sigma\lpop \sigma} \id_\sigma) = (\linlambda
z\co\sigma\lpop\sigma\ld z\circ f)\: \id_\sigma\:x = f(x)
\]
and so $\pi_\sigma\circ \pair fg=f$  and likewise $\pi_\tau\circ \pair
fg=g$  proving that $\sigma\times \tau$  defines a weak product.
\begin{lem}
  Using identity extension and extensionality it is provable that for
  any $f\co\omega\lpop\sigma$, $g\co\omega\lpop\tau$, $k\co \omega'\lpop
  \omega$, 
  \[\pair fg\circ k =_{\omega'\lpop\sigma\times\tau} \pair{f\circ k}{g\circ k}
  \]
\end{lem}

\proof
  The lemma is easily proved by the following direct computation using
  properties of coproducts established above. The notation $(-\circ k)$
  below denotes the term $\linlambda y\co \omega\lpop \alpha\ld y\circ
  k$  of type $(\omega\lpop\alpha)\lpop \omega'\lpop\alpha$.
  \[
  \begin{array}{rcl}
    & & \pair{f\circ k}{g\circ k}(x) \\
    & =_{\sigma\times\tau} & \Lambda
    \alpha\ld \linlambda 
    h\co (\sigma\lpop\alpha) + (\tau\lpop\alpha)\ld \copair {\linlambda
      z\co \sigma\lpop \alpha\ld z\circ f\circ k} {\linlambda z\co \tau\lpop
      \alpha\ld z\circ g\circ k} \:h\:x  \\
    & =_{\sigma\times\tau} & \Lambda \alpha\ld \linlambda 
    h\ld \copair {(-\circ
    k)\circ (\linlambda
      z\co \sigma\lpop \alpha\ld z\circ f)} {(-\circ k)\circ
    (\linlambda z\co \tau\lpop \alpha\ld z\circ g)} \:h\:x 
    \\ 
    & =_{\sigma\times\tau} &
    \Lambda \alpha\ld \linlambda 
    h\ld (-\circ k)\circ \copair {(\linlambda
      z\co \sigma\lpop \alpha\ld z\circ f)} {(\linlambda z\co
    \tau\lpop \alpha\ld z\circ g)} \:h\:x \\ 
    & =_{\sigma\times\tau} & \Lambda \alpha\ld \linlambda 
    h\ld \copair {(\linlambda 
      z\co \sigma\lpop \alpha\ld z\circ f)} {(\linlambda z\co
    \tau\lpop \alpha\ld z\circ g)} \:h\:(k(x)) \\ 
    & =_{\sigma\times\tau} & \pair{f}{g}\circ k(x)
  \end{array}
  \]
\qed
\begin{lem}
  Identity extension and extensionality implies that
  $\pair{\pi_\sigma}{\pi_\tau}=_{\sigma\times\tau \lpop
  \sigma\times\tau}\id_{\sigma\times\tau}$. 
\end{lem}
\proof
  We must show that for any $x\co \sigma\times\tau$,  any $\alpha$  and
  any $h\co (\sigma\lpop\alpha)+(\tau\lpop\alpha)$ 
\[\copair{\linlambda z\co \sigma\lpop \alpha\ld z\circ
  \pi_\sigma}{\linlambda z\co \sigma\lpop \alpha\ld z\circ \pi_\tau}\:
  h\: x=_\alpha x\:\alpha\:h
\]
In fact, since we are dealing with coproducts, it suffices to show
that for any $l\co \sigma\lpop\alpha$  and $k\co \tau\lpop\alpha$ 
\[
\begin{array}{rcl}
  l(\pi_\sigma (x)) & =_\alpha & x\:\alpha \:(\interm_{\sigma\lpop\alpha}\:l)\\
  k(\pi_\tau (x)) & =_\alpha & x\:\alpha \:(\interm_{\tau\lpop\alpha}\:k)
\end{array}
\]
We just prove the first of these equations. Since 
\[\id_\sigma(\eq_\sigma \lpop \graph l) l
\]
by parametricity of a polymorphic version of $\interm$,
\[\interm_{\sigma\lpop\sigma}(\id_\sigma) ((\eq_\sigma\lpop \graph l) +
(\eq_\tau \lpop \graph l) \interm_{\sigma\lpop \alpha}(l) 
\]
and so by parametricity of $x\co \sigma\times\tau$
\[x\:\sigma \:(\interm_{\sigma\lpop\sigma}\:\id_\sigma)\graph l
x\:\alpha \:(\interm_{\sigma\lpop\alpha}\: l)
\]
i.e.
\[\pi_\sigma(x)\graph l x\:\alpha\:(\interm_{\sigma\lpop\alpha}\: l)
\]
as desired.
\qed

\begin{prop}
  Suppose $h\co \omega\lpop\sigma\times\tau$  is such that $\pi_\sigma
  \circ h=_{\omega\lpop\sigma} f$  and $\pi_\tau
  \circ h=_{\omega\lpop\tau} g$  then it is provable using identity
  extension and extensionality that $h=_{\omega\lpop \sigma\times\tau}
  \pair fg$. Thus $\sigma\times\tau$  is a product of $\sigma$  and
  $\tau$  in $\LinTypecat_\Xi$. 
\end{prop}

\proof
  \[h =_{\omega\lpop\sigma\times\tau} \pair{\pi_\sigma}{\pi_\tau}
  \circ h =_{\omega\lpop\sigma\times\tau} \pair{\pi_\sigma\circ
  h}{\pi_\tau\circ h} =_{\omega\lpop\sigma\times\tau} \pair fg.
  \]
\qed

\subsection{Natural Numbers}

We define the type of natural numbers as
\[\N= \Prod\alpha\ld (\alpha\lpop\alpha)\to\alpha\lpop\alpha.\]
We further define terms $0\co \N$,  $s\co\N\lpop\N$  as 
\[0=\Lambda\alpha\ld\lambda f\co\alpha\lpop\alpha\ld \linlambda
x\co\alpha\ld x, \quad
s= \linlambda y\co\N\ld \Lambda \alpha\ld \lambda f\co\alpha\lpop\alpha\ld
\linlambda x\co \alpha\ld f(y\:\alpha\: \bang f\:x)\]
and prove that $(\N,0,s)$ is a weak natural numbers object in each 
$\LinTypecat_\Xi$, and, using parametricity and extensionality,
an honest natural numbers object.

Suppose we are given a type $\sigma$,  a term $a\co\sigma$    and a
morphism $b\co\sigma\lpop\sigma$. We can then define $h\co\N\lpop
\sigma$  as $h(y)=y\:\sigma\: \bang b\: a$. Then clearly
$h(0)=a$,  and $h(s\:x)= b(x\:\sigma\:\bang b\: a)= b(h(x))$,  so
$(\N, 0,s)$  is a weak natural numbers object.

We can express the weak natural numbers
object property as: for all $a,b$,  there exists an $h$ such that
\[
\xymatrix{
I \xypop[r]^0 \xypop[dr]_a & \N \xypop[d]^h \xypop[r]^s & \N \xypop[d]^h \\
& \sigma \xypop[r]^b & \sigma 
}
\]
commutes.

\begin{lem}\label{lem:natnumhelp}
  Identity Extension and extensionality implies
\[\forall x\co \N \ld x\:\N\:\bang s\:0=_\N x
\]
\end{lem}
\proof
  Suppose we are given $\sigma, a,b$ and define $h$ as above. Since
  $b\circ h= h\circ s$ and $h\:0=a$, we have $s(\graph h\lpop \graph
  h)b$ and $0\graph h a$, by parametricity of $x$, $(x\:\N\: \bang
  s\:0)\graph h (x\:\sigma\: \bang b\:a)$, i.e.,
\[(x\:\N\: \bang s\:0)\:\sigma\: \bang b\: a=_\sigma
x\:\sigma\: \bang b\:a.\] 
Letting $\sigma$  range over all types and $a,b$  over all terms,
using extensionality and Lemma~\ref{lem:intext},  we have 
\[x\:\N\: \bang s\:0=_\N x,\]
as desired.  
\qed
We can now prove that $\N$  is a natural numbers object in each
$\LinTypecat_\Xi$. 
\begin{lem}
Assuming identity extension and extensionality,  given $\sigma, a,b$, the
map $h$    defined as above is up to internal equality the unique $h'$
such that $h'(0)=a$,  $h'(s\:x)=b(h'\:x)$. 
\end{lem}
\proof
Suppose $h'$  satisfies the requirements of the lemma. Then
$s(\graph {h'}\lpop \graph {h'})b$ and  $0\graph {h'} a$ (this is just a
reformulation of the requirements),  so for arbitrary $x\co\N$, by
parametricity of $x$,
\[x\:\sigma\:\bang b\: a=_\sigma h'(x\:\N\: \bang s\: 0)
=_\sigma h'(x). \]
Thus, by extensionality, $h'=_{\N\lpop\sigma} h$. 
\qed

\subsubsection{Induction principle}

The parametricity principle for the natural numbers implies, that if
$R\admrel\N\N$,  and $x\co \N$,  then 
\[(x\:\N)((R\lpop R)\to R\lpop R) (x\:\N).\]
So if $s(R\lpop R)s$ and  $R(0,0)$,  then 
\[(x\:\N\:\bang s\:0)R(x\:\N\:\bang s\:0).\]
By Lemma~\ref{lem:natnumhelp}, $x\:\N\:\bang s\: 0=_\N x$, so we can
conclude that $R(x,x)$.  If $\phi$ is a proposition on $\N$ such that
$(x\co \N,y\co\N)\ld \phi(x)$ is admissible, then from parametricity
we obtain the usual induction principle
\[(\phi(0)\meet \forall x\co \N \ld\phi(x)\imp \phi(s(x)))\imp \forall x\co
\N\ld \phi(x).\]

\subsection{Types as functors}\label{sec:functors}

\begin{defi}\label{defn:inductive}
We say that $\vec\alpha\ts\sigma\co\Type$  is an inductively
constructed type, if it can be constructed from free variables
$\vec\alpha$ and closed types using the type constructors of $\pilly$,
i.e., $\lpop,\tensor,I,\bang$ and $\Prod\alpha\ld$.
\end{defi}

For example, all types of pure $\pilly$ are inductively defined, and
if $\sigma$ is a closed type then $\Prod\alpha\ld\sigma\times\alpha$
is an inductively constructed type. However, some models may contain
types that are not inductively constructed! For example, in
syntactical models, any basic open type, such as the type
$\alpha\ts\lists(\alpha)$ is not inductively constructed.

We define positive and negative occurrences of free type variables in
inductively defined types as usual. The type variable $\alpha$ occurs
positive in the type $\alpha$ and the positive occurrences of a type
variable $\alpha$ in $\sigma\lpop\tau$ are the positive occurrences of
$\alpha$ in $\tau$ and the negative in $\sigma$. The negative
occurrences of $\alpha$ in $\sigma\lpop\tau$ are the positive in
$\sigma$ and the negative in $\tau$. The positive and negative
occurrences of $\alpha$ in $\Prod\beta\ld \sigma$ are the positive and
negative occurrences in $\sigma$ for $\alpha\neq \beta$. The rest of
the type constructors preserve positive and negative occurrences of
type variables.

If $\sigma(\alpha,\beta)$ is an inductively defined type in which the
free type variable $\alpha$ appears only negatively and the free type
variable $\beta$ appears only positively, then we can consider
$\sigma$ as a functor
$\LinTypecat_{\Xi}^{op}\times\LinTypecat_{\Xi}\to\LinTypecat_{\Xi}$
for each $\Xi$ by defining the term
\[M_{\sigma(\alpha,\beta)}\co\Prod\alpha,\beta,\alpha',\beta'\ld
(\alpha'\lpop\alpha) \to(\beta\lpop\beta')\to
\sigma(\alpha,\beta)\lpop \sigma(\alpha',\beta'),\] which behaves as
the morphism part of a functor, i.e., it respects composition and
preserves identities. We define $M_{\sigma(\alpha,\beta)}$ by
structural induction on $\sigma$.  This construction immediately
generalizes to types with fewer or more than two free
type variables, each of which appear only positively or only
negatively. This idea of the functorial interpretation of types being
representable by polymorphic terms has also been used in second order
lambda calculus (see e.g.~\cite{Reynolds:Plotkin:90}).


For the base case of the induction, if $\sigma(\alpha,\beta)=\beta$,
define
\[M_{\beta}=
\Lambda \alpha, \beta, \alpha', \beta'\ld \lambda f,g\ld g.
\]
In the case $\sigma(\beta,\alpha)\lpop\tau(\alpha,\beta)$ we
define the term
\[
\begin{array}{c}
M_{\sigma(\beta,\alpha)\lpop\tau(\alpha,\beta)} \co \\
\Prod\alpha,\beta, \alpha',\beta' \ld
(\alpha'\lpop\alpha)\to(\beta\lpop\beta')\to
(\sigma(\beta,\alpha)\lpop \tau(\alpha,\beta))\lpop
\sigma(\beta',\alpha')\lpop \tau(\alpha',\beta')
\end{array}
\]
by
\[
\begin{array}{c}
M_{\sigma(\beta,\alpha)\lpop\tau(\alpha,\beta)}=
\Lambda \alpha,\beta,\alpha',\beta'\ld\lambda f,g\ld\\
\quad\linlambda h\co \sigma(\beta,\alpha)\lpop \tau(\alpha,\beta) \ld
(M_\tau\:\alpha \:\beta\: \alpha'\: \beta'\:f\:g)\circ h\circ
(M_\sigma \:\beta'\: \alpha'\: \beta\:\alpha \:g\:f).
\end{array}
\]
For bang types, we define:
\[
\begin{array}{c}
M_{\bang\sigma(\alpha,\beta)}=\Lambda\alpha,\beta,\alpha',\beta'\ld
  \lambda f\co\alpha'\lpop\alpha\ld \lambda
  g\co\beta\lpop\beta'\ld\linlambda x\co\bang\sigma(\alpha,\beta)\ld\\
\letexp
  {\bang y}x{\bang(M_{\sigma(\alpha,\beta)}\:\alpha\:\beta\:\alpha'\:\beta'\:
  f\: g\: y)}.
\end{array}
\]
For tensor types, we define:
\[
\begin{array}{c}
M_{\sigma(\alpha,\beta)\tensor\tau(\alpha,\beta)}=
\Lambda\alpha,\beta,\alpha',\beta'\ld\lambda f,g\ld
\linlambda z\co\sigma(\alpha,\beta)\tensor\tau(\alpha,\beta)\ld\\
\letexp{x\tensor y\co\sigma(\alpha,\beta)\tensor\tau(\alpha,\beta)}z
{(M_\sigma \alpha\:\beta\:\alpha'\:\beta'\:f\:g\:x)\tensor 
(M_\tau \alpha\:\beta\:\alpha'\:\beta'\:f\:g\:y)}.
\end{array}
\]
The last case is the case of polymorphic types:
\[
\begin{array}{c}
M_{\prod\omega\ld\sigma(\alpha,\beta)}=
\Lambda\alpha,\beta,\alpha',\beta'\ld\lambda f,g\ld
\linlambda z\co\Prod\omega\ld\sigma(\alpha,\beta)\ld\\
\Lambda\omega\co\Type \ld M_{\sigma(\alpha,\beta)}\:\alpha\:
\beta\: \alpha'\: \beta'\: f\: g\: (z\:\omega).
\end{array}
\]
\begin{lem}\label{lem:typefunctor}
The term $M_{\sigma}$  respects composition and preserves identities,
i.e., for $f\co\alpha'\lpop\alpha$, $f'\co\alpha''\lpop\alpha'$, 
$g\co\beta\lpop\beta'$, and $g'\co\beta'\lpop\beta''$,  
\begin{itemize}
\item $M_{\sigma(\alpha,\beta)}\:\alpha \:\beta \:\alpha'' \:\beta''
  \bang(f\circ f')\:\bang(g'\circ g)=$
  $(M_{\sigma(\alpha,\beta)}\:\alpha' \:\beta' 
  \:\alpha'' \:\beta'' \:\bang f'\:\bang g')\circ
  (M_{\sigma(\alpha,\beta)}\:\alpha \:\beta \:\alpha' \:\beta'
  \:\bang f\:\bang g),$
\item
  $M_{\sigma(\alpha,\beta)}\alpha\:\beta\:\alpha\:\beta\:\bang \id_\alpha
  \bang\id_\beta = \id_{\sigma(\alpha,\beta)}.$
\end{itemize}
\end{lem}
\proof
  The proof proceeds by induction over the structure of $\sigma$, and most
  of it is the same as in \cite{Abadi:Plotkin:93}, except the case of
  tensor-types and $\bang$. These cases are essentially proved in
  \cite{Barber:97}.
\qed
Notice that in the proof of Lemma~\ref{lem:typefunctor} we do not need
parametricity. 
Suppose \[\Xi\mid -;-\ts
f\co\alpha'\lpop\alpha,g\co\beta\lpop\beta'.\] We shall write 
$\sigma(f,g)$ for 
\[M_{\sigma(\alpha,\beta)}\alpha\:\beta\:\alpha'\:\beta'\:\bang
f\:\bang g.\]
The type of $\sigma(f,g)$ is
$\sigma(\alpha,\beta)\lpop\sigma(\alpha',\beta')$. Notice that we 
apply $M$  to $\bang f,\bang g$,  since $M$  is of intuitionistic
function type ($\to$  instead of $\lpop$). By the previous lemma,
$\sigma$  defines a bifunctor
$\LinTypecat_{\Xi}^{op}\times\LinTypecat_{\Xi} \to\LinTypecat_{\Xi}$
for each $\Xi$. 

First we consider this in the case of only one argument:
\begin{lem}[Graph lemma]\label{lem:graph}
  Assuming identity extension, for any type $\alpha\ts\sigma$ with $\alpha$
  occurring only positively and any map $f\co\tau\lpop\tau'$
  \[\sigma[\langle f\rangle]\congr\langle \sigma(f)\rangle.\]
  Likewise, suppose $\alpha\ts \sigma'$ is a type with $\alpha$ only
  occurring negatively. Then identity extension implies
 \[\sigma[\langle f\rangle]\congr\langle \opcat{\sigma(f)\rangle},\]
where $\opcat{\langle\sigma(f)\rangle}$  is $(x\co \sigma(\tau), y\co
\sigma(\tau'))\ld \langle\sigma(f)\rangle(y,x)$.
\end{lem}
\proof
  We will only prove the first half of the lemma; the other half is proved
  the same way. Since $\alpha$ occurs only positively in $\sigma$, we will
  assume for readability that $M_\sigma$ has type $\Prod \alpha,\beta \ld
  (\alpha\lpop \beta) \to \sigma (\alpha)\lpop \sigma(\beta)$.
  
  By parametricity of $M_\sigma$, we have, for any pair of admissible relations
  $\rho\admrel\alpha{\alpha'}$ and $\rho'\admrel\beta{\beta'}$,
  \begin{equation}
    \label{eq:graphlem}
    (M_\sigma\:\alpha\:\beta)((\rho\lpop\rho')\to(\sigma[\rho]\lpop
    \sigma[\rho']))
    (M_\sigma\:\alpha'\:\beta').
  \end{equation}
  Let $f:\tau\lpop\tau'$ be arbitrary. If we instantiate
  (\ref{eq:graphlem}) with $\rho=eq_{\tau}$ and 
  $\rho'=\langle f\rangle$, we get
  \[(M_\sigma\:\tau\:\tau)((eq_\tau\lpop\langle
  f\rangle)\to(eq_{\sigma(\tau)}\lpop\sigma[\langle f\rangle]))
  (M_\sigma\:\tau\:\tau'),\]
  using the identity extension schema. Since
  $\id_\tau(eq_\tau\lpop\langle f\rangle) f$, 
  \[\bang\id_\tau\bang(eq_\tau\lpop\langle f\rangle)\bang f,\]
  and using $M_\sigma\:\tau\:\tau'\:\bang f=\sigma(f)$  we get 
  \[\id_{\sigma(\tau)}(eq_{\sigma(\tau)}\lpop\sigma[\langle
  f\rangle])\sigma(f),\]
  i.e.,
  \[\forall x\co\sigma(\tau)\ld  x(\sigma[\langle f\rangle])
  (\sigma(f) x).\]
  We have thus proved $\langle\sigma(f)\rangle$ 
  implies $\sigma[\langle f\rangle]$.
  
  To prove the other direction, instantiate (\ref{eq:graphlem}) with the
  admissible relations $\rho=\langle f\rangle$, $\rho'=eq_{\tau'}$ for
  $f\co \tau\lpop\tau'$. Since $ f(\langle f\rangle \lpop
  eq_{\tau'})\id_{\tau'}$,
  \[\sigma(f)(\sigma[\langle
  f\rangle]\to eq_{\sigma(\tau')})\id_{\sigma(\tau')}.\] 
  So for any $x\co\sigma(\tau)$ and $y\co\sigma(\tau')$  we have
  $x(\sigma[\langle f\rangle]) y$ implies
  $\sigma(f)x=_{\sigma(\tau')}y$. This just means that $\sigma[\langle
  f\rangle]$ implies  $\langle \sigma(f)\rangle$. 
\qed

\subsection{Existential types}

In this section we consider existential or sum types. If
$\Xi,\alpha\ts \sigma$  is a type, we define the type $\Xi\ts
\Coprod\alpha\ld \sigma$  as
\[\Coprod\alpha\ld \sigma = \Prod\beta\ld (\Prod\alpha\ld
\sigma\lpop\beta)\lpop \beta
\]
In fact, this defines a functor 
\[\LinTypecat_{\Xi,\alpha} \to \LinTypecat_\Xi
\]
with functorial action as defined in Section~\ref{sec:functors}. In
this section we show that this functor is left adjoint to the
weakening functor
\[\LinTypecat_\Xi \to \LinTypecat_{\Xi,\alpha}
\]
mapping a type $\Xi\ts\sigma$ to $\Xi,\alpha\ts \sigma$. In other
words, we show that for any type $\Xi\ts \tau$, there is a one-to-one
correspondence between terms $\Xi\ts t\co
(\Coprod\alpha\ld\sigma)\lpop \tau$ and terms $\Xi,\alpha\ts \sigma\lpop
\tau$ if we consider terms up to internal equality provable using
identity extension and extensionality.

First define the term 
\[\pack\co \Prod\alpha\ld (\sigma\lpop \Coprod\alpha\ld \sigma)
\]
as $\Lambda\alpha\ld \linlambda x\co \sigma\ld \Lambda\beta\ld
\linlambda f\co \Prod\alpha\ld (\sigma\lpop\beta)\ld f\:\alpha\:x$. 
The correspondence is as follows. Suppose first $\Xi,\alpha\ts t\co
\sigma\lpop \tau$. Then $\Xi\ts \hat t\co (\Coprod\alpha\ld \sigma) \lpop
\tau$  is $\linlambda x\co (\Coprod\alpha\ld \sigma)\ld  x\: \tau \:
(\Lambda\alpha\ld t)$. If $\Xi\ts s\co (\Coprod\alpha\ld \sigma) \lpop
\tau$  then $\Xi,\alpha \ts\tilde s\co \sigma\lpop \tau$  is defined to
be $\lambda x\co \sigma\ld s (\pack\:\alpha\:x)$.

Now, suppose we start with a term $\Xi,\alpha \ts t\co \sigma\lpop
\tau$ then 
\[
\begin{array}{rcl}
\tilde{\hat t} & = & \linlambda x\co \sigma \ld (\linlambda y\co
\Coprod\alpha\ld \sigma\ld y\:\tau\:(\Lambda\alpha\ld t))\:(\pack
\:\alpha\: x) \\
& = & \linlambda x\co \sigma\ld \pack\:\alpha\: x\: \tau\:
(\Lambda\alpha\ld t) \\
& = & \linlambda x\co \sigma\ld (\Lambda \alpha\ld t) \: \alpha\: x \\
& = & t.
\end{array}
\]
It remains to prove that $\hat{\tilde s}$  is equal to $s$ for any
$\Xi\ts s\co (\Coprod\alpha\ld \sigma) \lpop \tau$. For this we need
to use identity extension. 
\begin{lem}\label{lem:existenstypelem1}
  Suppose $x\co\Coprod\alpha\ld \sigma$, $\tau,\tau'$  are types and
  $f\co \tau\lpop\tau', g\co \Prod\alpha\ld \sigma\lpop\tau$. Then
  using identity extension and extensionality,
  \[x\:\tau'\:(\Lambda\alpha\ld f\circ (g\:\alpha)) =_{\tau'}
  f\:(x\:\tau\:g)
  \]
\end{lem}
\proof
  Using identity extension on $g$ it is easy to see that
  $g(\Prod\alpha\ld \sigma\lpop \graph f)\Lambda \alpha\ld f\circ
  (g\:\alpha)$. If $x\co\Coprod\alpha\ld \sigma$ then by identity
  extension
  \[x\:\tau \:g\graph f x\:\tau'\:(\Lambda \alpha\ld f\circ(g\:\alpha))
  \]
  which is what we needed to prove.
\qed
\begin{lem}\label{lem:existenstypelem2}
  It is provable using identity extension and extensionality that 
  \[\forall x\co (\Coprod\alpha\ld \sigma) \ld x\:\Coprod\alpha\ld\sigma
  \: \pack =_{\coprod\alpha\ld\sigma} x
  \]
\end{lem}
\proof
  Suppose we are given $\beta$  and $f\co \Prod\alpha\ld \sigma\lpop
  \beta$.  We show that 
\[x\:\beta\:f =_\beta x\:(\Coprod\alpha\ld \sigma)\: \pack \: \beta \: f
\]
Define $f'=\linlambda x\co (\Coprod\alpha\ld \sigma)\: x\: \beta \: f$
of type $(\Coprod\alpha\ld\sigma) \lpop \beta$. By
Lemma~\ref{lem:existenstypelem1} 
\[x\:\beta\: (\Lambda\alpha\ld f'\circ (\pack\:\alpha)) =_\beta f'(x\:
\Coprod\alpha \ld \sigma\: \pack) =_\beta x\: \Coprod\alpha\ld
\sigma\: \pack\: \beta \: f
\]
so we just need to show that $\Lambda\alpha\ld f'\circ (\pack\:\alpha)$
is internally equal to $f$. But
\[\Lambda\alpha\ld f'\circ (\pack\:\alpha)\:\alpha\:y =_\beta
f'\:(\pack \:\alpha\: y) =_\beta \pack\:\alpha \:y\:\beta\: f=_\beta
f\:\alpha \: y.
\]
\qed
\begin{prop}
  Suppose $\Xi\ts s\co (\Coprod\alpha\ld \sigma)\lpop \tau$. It is
  provable using identity extension and extensionality that $\hat{\tilde
  s}$  is internally equal to $s$. 
\end{prop}
\proof
  \[\hat{\tilde s}(x)=_\tau x\:\tau \: (\Lambda\alpha\ld \linlambda x'\co
  \sigma\ld s\:(\pack\: \alpha\: x')) =_\tau s \:(x\:
  \Coprod\alpha \ld \sigma\:\pack)=_\tau s\:x
  \]
  where for the second equality we have used
  Lemma~\ref{lem:existenstypelem1}. 
\qed

Parametricity induces the following reasoning principle for
existential types.

\begin{prop} \label{prop:existential}
  For $x,y \co \Coprod\alpha \ld \sigma(\alpha)$ the following is
  equivalent to internal equality of $x$  and $y$.
  \[\exists \alpha, \beta, R\co \AdmRel\alpha\beta,  x' \co
  \sigma(\alpha), y'\co \sigma(\beta) \ld x=\pack \:\alpha\: x' \meet
  y=\pack \:\beta\: y'  \meet \sigma[R](x',y').
  \]
  As a special case we get the following principle:
  \[\forall x\co \Coprod\alpha\ld \sigma(\alpha)\ld \exists \alpha,
  x'\co \sigma(\alpha) \ld x=_{\coprod\alpha\ld \sigma(\alpha)}  \pack
  \:\alpha\: x' 
  \]
\end{prop}

\proof
  Let us for simplicity write $\chi$  for 
  \[(x,y)\ld \exists \alpha, \beta, R\co \AdmRel\alpha\beta,  x' \co
  \sigma(\alpha), y\co \sigma(\beta) \ld x=\pack \:\alpha\: x' \meet
  y=\pack \:\beta\: y' \meet \sigma[R](x',y').
  \]

  We now prove that, for any pair of types $\tau, \tau'$,
  any admissible  relation $S\co \AdmRel{\tau}{\tau'}$, and any
  pair of maps $t,t'$  we have 
  \[(t,t')\co \eq_{\coprod \alpha\ld \sigma} \lpop S
  \]
  iff
  \[(t,t') \co \chi \lpop S
  \]
  and the two implications of the first part of the proposition
  follows from the case of $t,t'$ both being identity and taking $S$
  to be respectively $\chi$  and $\eq_{\coprod \alpha\ld \sigma}$. 
  
  First notice that 
  \[
  \begin{prooftree}
    \[
    \Xi\mid x,y \ts \chi(x,y) \imp S(t(x),t'(y))
    \Justifies 
    \Xi, \alpha,\beta\mid R\co \AdmRel\alpha\beta \mid x,y, x', y' \ts
    \sigma[R](x',y') \imp S(t(\pack \:\alpha \:x'), t'(\pack \:\beta \: y'))
    \]
    \Justifies 
    \Xi, \alpha,\beta\mid R\co \AdmRel\alpha\beta \mid x', y' \ts
      \sigma[R](x',y') \imp S(\tilde t(x'), \tilde t'(y'))
  \end{prooftree}
  \]
  so it suffices to show that 
  \[
  \begin{prooftree}
    \Xi\mid x,y \ts x=_{\coprod \alpha\ld \sigma(\alpha)}y \imp S(t(x),t'(y))
    \Justifies 
    \Xi, \alpha,\beta\mid R\co \AdmRel\alpha\beta \mid x', y' \ts
    \sigma[R](x',y') \imp S(\tilde t(x'), \tilde t'(y'))
  \end{prooftree}
  \]
  i.e., that $(t,t')$  preserve relations iff $(\tilde t, \tilde t')$
  do. 

  First assume $(t,t')$  preserve relations. By parametricity of
  $\pack$, 
  \[(\pack \:\alpha, \pack \: \beta)\co \sigma[R] \lpop \eq,\]
  and so since $\tilde t= t\circ (\pack\:\alpha)$  and $\tilde t'= t'\circ
  (\pack\:\beta)$  the pair $(\tilde t,\tilde t')$  preserve relations. On
  the other hand, if $(\tilde t, \tilde t')$  preserve relations then
  \[(\Lambda\alpha \ld \tilde t, \Lambda\beta \ld \tilde t') \co \forall
  \alpha, \beta, R\co \AdmRel\alpha\beta \ld \sigma[R] \lpop S,\] 
  and so by parametricity, if $\eq_{\coprod\alpha\ld
    \sigma(\alpha)}(x,y)$ then
  \[(t(x), t'(y)) = (x\: \Coprod\alpha\ld \sigma(\alpha) \:(\Lambda\alpha
  \ld \tilde t),
  y\:\Coprod\alpha\ld \sigma(\alpha) \:(\Lambda\beta\ld \tilde t')) \in S
  \]
\qed


\subsection{Initial algebras}\label{sec:initalg}

Suppose $\alpha\ts\sigma\co\Type$ is an inductively constructed type
in which $\alpha$ occurs only positively. As we have seen earlier,
such a type induces a functor
\[\LinTypecat_\Xi \to \LinTypecat_\Xi
\]
for each $\Xi$. We aim to define an initial algebra for this type. 

Define the closed type
\[\mu\alpha\ld\sigma(\alpha)= \Prod\alpha\ld
(\sigma(\alpha)\lpop\alpha) \to\alpha,\] 
and define
\[\fold\co\Prod \alpha\ld(\sigma(\alpha)\lpop\alpha)\to
(\mu\alpha\ld\sigma(\alpha)\lpop \alpha)\]
as
\[\fold=\Lambda\alpha\ld\lambda f\co
\sigma(\alpha)\lpop\alpha\ld\linlambda u\co\mu\alpha\ld\sigma(\alpha)\ld
u\:\alpha\:\bang f,\] 
and
\[\interm\co\sigma(\mu\alpha\ld\sigma(\alpha))\lpop\mu\alpha\ld\sigma(\alpha)\]
as
\[\interm\:z=\Lambda\alpha\ld\lambda f\co\sigma(\alpha)\lpop\alpha\ld
f(\sigma(\fold\:\alpha\:\bang f)\:z).\]
\begin{lem}
  For any algebra $f\co\sigma(\tau)\lpop\tau$, $\fold\:\tau\:\bang f$ is a
  map of algebras from $(\mu\alpha\ld\sigma(\alpha),\interm)$ to
  $(\tau,f)$, i.e., the diagram 
\[
\xymatrix{
\sigma(\mu\alpha\ld \sigma(\alpha)) \xypop[r]^\interm
\xypop[d]_{\sigma(\fold\:\tau \:\bang f)} & \mu\alpha\ld\sigma(\alpha)
\xypop[d]^{\fold \: \tau \:\bang f} \\
\sigma(\tau) \xypop[r]^f & \tau
}
\]
commutes.
\end{lem}
\proof
  For $x\co\sigma(\mu\alpha\ld\sigma(\alpha))$
\[
\begin{array}{c}
(\fold\:\tau\:\bang f)\circ \interm\:x= \interm \:x\:\tau\:\bang f=
f(\sigma(\fold\:\tau\:\bang f)\:x),
\end{array}
\]
as desired.
\qed
In words we have shown that $\interm$  defines a weakly initial
algebra for the functor defined by $\sigma$  in $\LinTypecat_\Xi$  for
each $\Xi$. Notice that parametricity was not needed in this proof.
\begin{lem}\label{lem:fold}
Suppose $\Xi\mid\Gamma;-\ts f\co\sigma(\tau)\lpop\tau$ and
$\Xi\mid\Gamma;-\ts g\co\sigma(\omega)\lpop\omega$ are algebras
for $\sigma$, and $\Xi\mid\Gamma;-\ts h\co\tau\lpop\omega$ is a map of
algebras, i.e., $h\: f = g\:\sigma(h)$. Then, assuming
identity extension and extensionality,
\[h\circ(\fold\:\tau\:\bang f)=_{\mu\alpha\ld\sigma(\alpha)\lpop\omega}
\fold\:\omega\:\bang g.\]
\end{lem}
%
%
\proof
Since $h$ is a map of algebras 
\[ f (\langle\sigma(h)\rangle \lpop \langle h\rangle) g ,\]
so by the Graph Lemma (\ref{lem:graph})
\[ f(\sigma[\langle h\rangle]\lpop\langle h\rangle) g\]
and by Lemma \ref{lem:bangrel} 
\[ \bang f(\bang(\sigma[\langle h\rangle]\lpop\langle
h\rangle)) \bang g.\]
Clearly
$(\fold,\fold)\in \eq_{\prod \alpha\ld(\sigma(\alpha)\lpop\alpha)\to
(\mu\alpha\ld\sigma(\alpha)\lpop \alpha)}$, and thus, by identity
extension,
\[(\fold,\fold)\in\Prod\alpha\ld (\sigma(\alpha)\lpop\alpha)\to
(\beta\lpop \alpha)[eq_{\mu\alpha\ld\sigma(\alpha)}/\beta],\]
so for any $x\co\mu\alpha\ld\sigma(\alpha)$,
\[(\fold\:\tau\: \bang f\:x)\langle h\rangle 
(\fold\:\omega\:\bang g\:x),\] 
i.e.,
\[h\circ(\fold\:\tau\:\bang f)=_{\mu\alpha\ld\sigma(\alpha)\lpop\omega}
\fold\:\omega\:\bang g,\]
as desired.
\qed
\begin{lem}\label{lem:foldtoid}
Using identity extension and extensionality,
\[\fold\:\mu\alpha\ld\sigma(\alpha)\:\bang
\interm=_{\mu\alpha\ld\sigma(\alpha)\lpop 
  \mu\alpha\ld\sigma(\alpha)} \id_{\mu\alpha\ld\sigma(\alpha)}.\]
\end{lem}
\proof
By Lemma \ref{lem:fold} we know that for any type $\tau$,
$f\co\sigma(\tau)\lpop\tau$ and 
$u\co\mu\alpha\ld\sigma(\alpha)$ 
\[(\fold\:\tau\:\bang f)\circ(\fold\:\mu\alpha\ld\sigma(\alpha)\:\bang
\interm)\:u =_\tau \fold\:\tau\: \bang f\:u.\]
The left hand side of this equation becomes
\[\fold\:\tau\: \bang f\:(u\:\mu\alpha\ld\sigma(\alpha)\:\bang \interm) =
(u\:\mu\alpha\ld\sigma(\alpha)\:\bang \interm) \tau\: \bang f\]
and, since the right hand side is simply 
\[u\:\tau\: \bang f,\] 
the lemma follows from Lemma~\ref{lem:intext}.
\qed
\begin{thm}\label{thm:initalg}
Consider an algebra $\Xi\mid-;-\ts f\co\sigma(\tau)\lpop\tau$ and a
map of algebras $\Xi\mid-;-\ts h\co\mu\alpha\ld\sigma(\alpha)\lpop\tau$
from $\interm$ to $f$.  Then if we assume identity extension and
extensionality, 
$h=_{\mu\alpha\ld\sigma(\alpha)\lpop\tau} \fold\:\tau\:\bang f.$
\end{thm}
\proof
By Lemma~\ref{lem:fold} we have
\[h\circ(\fold\:\mu\alpha\ld\sigma(\alpha)\:\bang \interm)
=_{\mu\alpha\ld\sigma(\alpha) \lpop \tau}\fold\:\tau\:\bang f.\]
Lemma \ref{lem:foldtoid} finishes the job.
\qed
We have shown that $\interm$  defines an initial algebra. 

In the logic, the initial algebras also satisfy an induction
principle. We now show the following (relational) induction principle.

\begin{thm}[Induction]\label{thm:induction}
  Suppose $R \co \AdmRel{\mu\alpha\ld \sigma(\alpha)}{\mu\alpha\ld
  \sigma(\alpha)}$  satisfies \[(\interm,\interm) \co \sigma[R] \lpop
  R.\] Then 
  \[\forall x\co \mu\alpha\ld \sigma(\alpha)\ld R(x,x)
  \]
\end{thm}

\begin{rem}
  The induction principle speaks about relations since it is obtained
  as a consequence of binary parametricity. In case one also has unary
  parametricity available (for some notion of admissible
  propositions), applying the proof of Theorem~\ref{thm:induction} to
  unary parametricity will yield the well-known propositional
  induction principle: If $\phi$  is an \emph{admissible} proposition
  on $\mu\alpha\ld \sigma(\alpha)$, then 
  \[(\forall x\co \sigma(\mu\alpha\ld \sigma(\alpha)) \ld
  \sigma[\phi](x)\imp \phi (\interm\: x) ) \imp \forall x\co
  \mu\alpha\ld \sigma(\alpha) \ld \phi(x)
  \]
\end{rem}

\proof[Proof of Theorem~\ref{thm:induction}]
  By parametricity, for any $x\co \mu\alpha\ld \sigma(\alpha)$, 
  \[x(\forall \alpha,\beta, R\co \AdmRel\alpha\beta \ld (\sigma[R]
  \lpop R)\to R)x
  \]
  The assumption states that $(\interm, \interm)\co \sigma[R]\lpop R$
  and so by Lemma~\ref{lem:bangrel}
  \[(\bang \interm,
  \bang \interm)\co \bang(\sigma[R]\lpop R).\]
  Thus 
  \[R(x\: \mu\alpha\ld \sigma(\alpha) \: \bang\interm, x\: \mu\alpha\ld
  \sigma(\alpha) \: \bang\interm).
  \]
  Finally, Lemma~\ref{lem:foldtoid} tells us that $x\: \mu\alpha\ld
  \sigma(\alpha) \: \bang\interm= x$, which proves the theorem.
\qed

\subsection{Final Coalgebras}\label{sec:finalcoalg}

As in section \ref{sec:initalg} we will assume that $\alpha\ts
\sigma(\alpha)\co\Type$ is a type in which $\alpha$ occurs only
positively, and this time we construct final coalgebras for the
induced functor. 

Define
\[\nu\alpha\ld\sigma(\alpha) = \Coprod \alpha\ld \bang(\alpha\lpop
\sigma(\alpha)) \tensor \alpha= \Prod\beta\ld\left(\Prod\alpha\ld
                                (\bang(\alpha\lpop 
\sigma(\alpha))\tensor \alpha\lpop\beta)\right)\lpop\beta\]
with combinators 
\[
\begin{array}{c}
\unfold\co \Prod\alpha\ld (\alpha\lpop\sigma(\alpha))\to
\alpha\lpop\nu\alpha\ld \sigma(\alpha), \\
\out\co \nu\alpha\ld \sigma(\alpha) \lpop\sigma(\nu\alpha\ld\sigma(\alpha))
\end{array}
\]
defined by
\[
\begin{array}{rcl}
\unfold &=& \Lambda\alpha\ld\linlambda f\co
\bang(\alpha\lpop\sigma(\alpha))\ld 
\linlambda x\co\alpha\ld \pack\:\alpha\:(f\tensor x)\\
\out &=& \linlambda x\co\nu\alpha\ld\sigma(\alpha) \ld x\:
\sigma(\nu\alpha\ld\sigma(\alpha)) \: \apply,
\end{array}
\]
where
\[
\begin{array}{c}
\apply:\Prod\alpha\ld\bang (\alpha\lpop\sigma(\alpha))\tensor\alpha
\lpop \sigma(\nu\alpha\ld\sigma(\alpha)) \\
\apply = \Lambda\alpha \ld \linlambda
y\co\bang (\alpha\lpop\sigma(\alpha))\tensor \alpha
\ld \letexp {w\tensor z}y{\sigma(\unfold\:\alpha\:w)(\letexp{\bang f}w{f\:z})}.
\end{array}
\]
\begin{lem}\label{lem:weakfinal}
For any coalgebra $f\co\tau \lpop\sigma(\tau)$, the map $\unfold \:\tau\:
\bang f$  is a map of coalgebras from $f$  to $\out$.
\end{lem}
\proof
We need to prove that the following diagram commutes
\[\xymatrix{\tau \ar@{-o}[r]^f \ar@{-o}[d]_{\unfold\: \tau \: \bang f}
  &  \sigma(\tau) \ar@{-o}[d]^{\sigma(\unfold\: \tau \: \bang f)} \\ 
  \nu\alpha\ld\sigma(\alpha) \ar@{-o}[r]^\out &
  \sigma(\nu\alpha\ld\sigma(\alpha)).} 
\]
But this is done by a simple computation
\[
\begin{array}{rcl}
\out (\unfold \:\tau \:\bang f\: x) & = & \out(\pack\: \tau (\bang f)\tensor
x) \\
& = & \pack \:\tau (\bang f)\tensor x\: \sigma(\nu\alpha\ld \sigma(\alpha))
\: \apply \\
& = &  \apply \: \tau\:((\bang f)\tensor x) \\
& = & \sigma(\unfold\:\tau\:(\bang f))\: (f\:x).
\end{array}
\]
\qed
Lemma~\ref{lem:weakfinal}  shows that $\out$  is a weakly final
coalgebra for the functor induced by $\sigma$  on $\LinTypecat_\Xi$
for each $\Xi$. Notice that parametricity was not needed here. 
\begin{lem}\label{lem:coalglem}
  Suppose $h\co\tau \lpop \tau'$ is a map of coalgebras from
  $f\co\tau\lpop\sigma(\tau)$ to $f'\co\tau'\lpop\sigma(\tau')$. If we
  assume identity extension, then the diagram
\[
\xymatrix{\tau\ar@{o}[d]_{h} \ar@{o}[rr]^-{\unfold\: \tau\: \bang f}  &&
  \nu\alpha\ld\sigma(\alpha) \\ 
  \tau' \ar@{o}[urr]_{\unfold\: \tau'\: \bang f'}}
\]
commutes internally.
\end{lem}
\proof
Using the Graph Lemma, the notion of $h$ being a map of coalgebras can
be expressed as
\[f(\graph h \lpop\sigma[\graph h])f'.\]
Now, by parametricity of $\unfold$,
\[\unfold\:\tau \: \bang f(\graph h\lpop\eq_{\nu\alpha\ld\sigma(\alpha)})
\unfold\:\tau' \:\bang f',\]
which is exactly what we wanted to prove.
\qed

\begin{lem}\label{lem:tensorrel:forall}
Given linear contexts $C$ and $C'$, suppose
\[
\forall x\co\sigma\ld\forall y\co\tau\ld C[x\tensor y] =_\omega
C'[x\tensor y] .
\]
then
\[
\forall z\co\sigma\tensor\tau\ld
 \letexp{x\tensor y}z{C[x\tensor y]} =_\omega
 \letexp{x\tensor y}z{C'[x\tensor y]}
\]
\end{lem}
\proof
Consider
\[
f = \linlambda x\co\sigma\ld\linlambda y\co\tau\ld C[x\tensor y]
\qquad
f' = \linlambda x\co\sigma\ld\linlambda y\co\tau\ld C'[x\tensor y]
\]
then
\[
 f\:(eq_\sigma\lpop eq_\tau\lpop eq_\omega)\: f' .
\]

If $z\co\sigma\tensor\tau$  then by identity extension
$\eq_{\sigma}\tensor\eq_\tau (z,z)$. By definition of
$\eq_{\sigma}\tensor\eq_\tau$  we have 
\[
\letexp{x\tensor x'}z{f x x'}=_{\omega}
\letexp{x\tensor x'}z{f' x x'}
\]
which proves the lemma.
\qed
\begin{lem}\label{lem:finalcoalg}
Using extensionality and identity extension,
\[\unfold\:\nu\alpha\ld\sigma(\alpha)\:\bang\out\]
is internally equal to the identity on $\nu\alpha\ld\sigma(\alpha)$.
\end{lem}
\proof
Set $h=\unfold\:\nu\alpha\ld\sigma(\alpha)\:\bang\out$ in the following.

By Lemma~\ref{lem:weakfinal} $h$  is a map of coalgebras from $\out$
to $\out$,  so by Lemma~\ref{lem:coalglem}, $h=h^2$. Intuitively, all
we need to prove now is that $h$ is ``surjective''.

Consider any
$g:\Prod\alpha\ld(\bang(\alpha\lpop\sigma(\alpha))\tensor\alpha\lpop\beta)$.
For any coalgebra map $k: \alpha \lpop \alpha'$ from
$f:\alpha\lpop\sigma(\alpha)$ to $f':\alpha'\lpop\sigma(\alpha')$, we
must have, by Lemmas~\ref{lem:graph}, \ref{lem:bangrel},
and~\ref{lem:tensorrel},
\[(\bang f\tensor x)(\bang(\graph k \lpop \sigma[\graph k])\tensor
\graph k)(\bang
f'\tensor kx),\]
so by identity extension and parametricity of $g$,
\[\forall x\co\alpha\ld g\:\alpha\:(\bang f)\tensor x=_\beta
g\:\alpha'\:(\bang f')\tensor k(x).\] 
Using this on the coalgebra map $\unfold\:\alpha\:\bang f$ from $f$ to
$\out$ we obtain 
\[\forall x\co\alpha\ld g\: \alpha\: (\bang f)\tensor x=_\beta
g\:\nu\alpha\ld\sigma(\alpha)\: (\bang\out)\tensor
\unfold\:\alpha\:\bang f\:x.\] 
By Lemma~\ref{lem:intext} this implies that
\[\forall f\co\bang(\alpha\lpop\sigma(\alpha)), x\co\alpha\ld g\:
\alpha\: f\tensor x=_\beta 
g\:\nu\alpha\ld\sigma(\alpha)\: (\bang\out)\tensor
\unfold\:\alpha\: f\:x,\] 
which implies
\[\forall z\co \bang(\alpha\lpop\sigma(\alpha)) \tensor\alpha\ld g\:
\alpha\: z=_\beta 
g\:\nu\alpha\ld\sigma(\alpha)\: (\letexp{f\tensor x}z{(\bang\out)\tensor
\unfold\:\alpha\: f\:x})\]
using Lemma~\ref{lem:tensorrel:forall}.

In other words, if we define
\[k\co\Prod\alpha\ld(\bang(\alpha\lpop\sigma(\alpha))\tensor\alpha\lpop\tau),\]
where
$\tau=\bang (\nu\alpha\ld\sigma(\alpha)\lpop\sigma(\nu\alpha\ld\sigma(\alpha)))
\tensor\nu\alpha\ld\sigma(\alpha)$, to be 
\[k=\Lambda\alpha\ld\linlambda y:\bang
(\alpha\lpop\sigma(\alpha))\tensor\alpha\ld 
\letexp{f\tensor x}{y}{(\bang \out)\tensor \unfold\:\alpha\: f\:x},\] 
then
\begin{equation}
  \label{eq:prooffinalcoalg}
\forall \alpha\ld
g\:\alpha=_{\bang(\alpha\lpop\sigma(\alpha))\tensor\alpha\lpop\beta}
(g\:\nu\alpha\ld\sigma(\alpha))\circ(k\:\alpha).  
\end{equation}
Now, suppose we are given $\alpha,\alpha', R\rel{\alpha}{\alpha'}$
and terms $f,f'$  such that \[f(\bang (R\lpop\sigma[R])\tensor R)f'.\] 
Then, by (\ref{eq:prooffinalcoalg}) and parametricity of $g$
\[g\:\alpha\:f=_\beta g\:\alpha'\:f'=_\beta
(g\:\nu\alpha\ld\sigma(\alpha)) (k\:\alpha'\:f'),\]
from which we conclude
\[g(\forall (\alpha,\beta,R\rel{\alpha}{\beta)}\ld(\bang(R\lpop
\sigma[R])\tensor R\lpop\langle 
g\:\nu\alpha\ld\sigma(\alpha)\rangle^{op}) )k.\] 
(Here we use $S^{op}$ for the inverse relation of $S$.)
Using parametricity, this implies that, 
for any $x\co\nu\alpha\ld\sigma(\alpha)$,
we have
\[x\:\beta\:g=_\beta g\:\nu\alpha\ld\sigma(\alpha)\:(x\:\tau\: k).\]

Thus, since $g$  was arbitrary, we may apply the above to $g=k$  and get
\[x\:\tau\: k=_{\tau}
k\:\nu\alpha\ld\sigma(\alpha)\:(x\:\tau\:k)=
\letexp{f\tensor z}{(x\:\tau\:k)}{(\bang \out)\tensor \unfold\:\alpha\:
  f\:z}.\] 
If we write 
\[l=\lambda x\co\nu\alpha\ld\sigma(\alpha)\ld 
\letexp{f\tensor z}{(x\:\tau\:k)}{\unfold\:\alpha\:  f\:z},\] 
then, since $k$ is a closed term, so is $l$,  and from the above
calculations we conclude that we have
\[\forall \beta\ld\forall
g:\Prod\alpha\ld\bang(\alpha\lpop\sigma(\alpha))\tensor\alpha\lpop\beta\ld
x\:\beta\:g=_\beta
g\:\nu\alpha\ld\sigma(\alpha)\:(\bang\out)\tensor (l\:x).\]

Now, finally,
\[
\begin{array}{rcl}
h(l\:x) & = & \unfold\:\nu\alpha\ld\sigma(\alpha)\:\bang\out\:(l\:x) \\
& = & \pack\:\nu\alpha\ld\sigma(\alpha)\:\bang\out\tensor(l\:x)\\
& = & \Lambda\beta\ld\lambda
g:\Prod\alpha\ld(\bang(\alpha\lpop\sigma(\alpha))\tensor\alpha\lpop\beta)\ld
g\:\nu\alpha\ld\sigma(\alpha)\:\bang\out\tensor(l\:x) \\
& =_{\nu\alpha\ld\sigma(\alpha)} &
\Lambda\beta\ld\lambda
g:\Prod\alpha\ld(\bang(\alpha\lpop\sigma(\alpha))\tensor\alpha\lpop\beta)\ld 
x\:\beta\:g \\
& = & x,
\end{array}
\] 
where we have used extensionality. Thus $l$  is a right inverse to
$h$, and we conclude
\[h\:x=_{\nu\alpha\ld\sigma(\alpha)}h^2(l\:x)=_{\nu\alpha\ld\sigma(\alpha)}
h(l\:x)=_{\nu\alpha\ld\sigma(\alpha)} x.\] 
\qed
\begin{thm}\label{thm:finalcoalg}
Suppose $\Xi\mid-;-\ts f\co\tau\lpop\sigma(\tau)$ is a coalgebra and
$\Xi\mid-;-\ts h\co\tau\lpop\mu\alpha\ld\sigma(\alpha)$  is a map
of algebras from $f$ to $\out$. Then if we assume identity extension and
extensionality 
$h=_{\tau\lpop\mu\alpha\ld\sigma(\alpha)} \unfold\:\alpha\:\bang f$.
\end{thm}
\proof
Consider a map of coalgebras into $\out$:
\[\xymatrix{\tau\ar@{o}[r]^f \ar@{o}[d]^h &
  \sigma(\tau)\ar@{o}[d]^{\sigma(h)}\\ 
\nu\alpha\ld\sigma(\alpha) \ar@{o}[r]^{\out} &
\sigma(\nu\alpha\ld\sigma(\alpha)).}
\]
By Lemmas~\ref{lem:coalglem}  and \ref{lem:finalcoalg},
\[\unfold\:\tau\:\bang f=_{\tau\lpop\nu\alpha\ld\sigma(\alpha)}
(\unfold\:\nu\alpha\ld\sigma(\alpha)\:\bang\out)\circ 
g =_{\tau\lpop\nu\alpha\ld\sigma(\alpha)} g.\] 
\qed
Theorem~\ref{thm:finalcoalg} shows that $\out$ is a final coalgebra
for the endofunctor on $\LinTypecat_\Xi$ induced by $\sigma$ for each
$\Xi$.

We now show how the final coalgebras satisfy a coinduction
principle. 

\begin{thm}[Coinduction]\label{thm:coinduction}
  Suppose that $R\co \AdmRel{\nu\alpha \ld \sigma(\alpha)} {\nu\alpha \ld
  \sigma(\alpha)}$ is such that 
  \[(\out,\out)\co R\lpop \sigma[R].\]
  We then have that
  \[\forall x,y\co \nu\alpha \ld \sigma(\alpha) \ld R(x,y) \imp
  x=_{\nu\alpha \ld \sigma(\alpha)} y.
  \]
\end{thm}
\proof
  Suppose $R\co \AdmRel{\nu\alpha \ld \sigma(\alpha)} {\nu\alpha \ld
  \sigma(\alpha)}$ satisfies $(\out,\out)\co R\lpop \sigma[R]$ and
  $R(x,y)$. By parametricity of 
  \[\pack\co \Prod\alpha\ld \bang(\alpha\lpop \sigma(\alpha))\tensor
  \alpha \lpop \nu\alpha\ld \sigma(\alpha)
  \]
  we have
  \[\pack \: \nu\alpha\ld \sigma(\alpha) \:\bang\out \tensor x
  =_{\nu\alpha\ld \sigma(\alpha)} \pack \: \nu\alpha\ld \sigma(\alpha)
  \:\bang\out \tensor y
  \]
  and by \ref{lem:finalcoalg},
  \[
  \begin{array}{c}
    \pack \: \nu\alpha\ld \sigma(\alpha) \:\bang\out \tensor x
    =_{\nu\alpha\ld \sigma(\alpha)} x \\
    \pack \: \nu\alpha\ld \sigma(\alpha) \:\bang\out \tensor y
    =_{\nu\alpha\ld \sigma(\alpha)} y
  \end{array}
  \]
  which proves the theorem.
\qed

The next theorem is an interesting generalization of
Theorem~\ref{thm:coinduction}, stating that the assumption of
admissibility in the coinduction principle is unnecessary. A similar
result was proved by Pitts in the setting of coinductive types in the
category of domains \cite{Pitts:95a}. To state this theorem we need
again to use the general hypothesis of this section that $\sigma$ is
an inductively defined type, since in this case we can define
$\sigma[R]$ for general (not just admissible) relations inductively
over the structure of $\sigma$  using the constructions of
Section~\ref{sec:conondefrel}. Recall that for more general types
$\sigma$  the construction $\sigma[R]$  is defined as in
Figure~\ref{fig:admrel}  for admissible relations $R$  only. 

\begin{thm}[General coinduction
  principle]\label{thm:coinduction:general} Suppose $R \rel{\nu\alpha
    \ld \sigma(\alpha)}{\nu\alpha \ld \sigma(\alpha)}$ is a relation
  such that $(\out,\out)\co R\lpop \sigma[R]$, then
  \[\forall x,y\co \nu\alpha \ld \sigma(\alpha) \ld R(x,y) \imp
  x=_{\nu\alpha \ld \sigma(\alpha)} y
  \]
\end{thm}

\proof
  Suppose $R\rel{\nu\alpha \ld \sigma(\alpha)}{ \nu\alpha \ld
    \sigma(\alpha)}$ is any relation satisfying $(\out,\out)\co R\lpop
  \sigma[R]$. The idea of the proof is to use
  Theorem~\ref{thm:coinduction} on the admissible relation $\Phi(R)$.
  Since by Lemma~\ref{lem:admissible:closure} $\Phi$ is a functor,
  \[(\out,\out)\co \Phi(R) \lpop \Phi(\sigma[R]),\]
  and since $\sigma[\Phi(R)]$  is an admissible relation containing
  $\sigma[R]$ , and $\Phi(\sigma[R])$ is the smallest such,
  we have $\Phi(\sigma[R]) \subset \sigma[\Phi(R)]$  and so
  \[(\out,\out)\co \Phi(R) \lpop \sigma[\Phi(R)].\]
  Now, the coinduction principle for admissible relations gives us 
  \[\forall x,y\co \nu\alpha\ld\sigma(\alpha) \ld \Phi(R)(x,y) \imp
  x =_{\nu\alpha \ld \sigma(\alpha)} y 
  \]
  and so the theorem follows from $R\subset \Phi(R)$. 
\qed

\subsection{Recursive type equations}\label{sec:rectypes}

In this section we consider inductively constructed types $\alpha\ts
\sigma(\alpha)$ and construct closed types $\rec\alpha \ld
\sigma(\alpha)$ such that $\sigma(\rec\alpha \ld \sigma(\alpha))\iso
\rec\alpha \ld \sigma(\alpha)$. In Sections~\ref{sec:initalg}
and~\ref{sec:finalcoalg} we solved the problem in the special case of
$\alpha$ occurring only positively in $\sigma$, by finding initial
algebras and final coalgebras for the functor induced by $\sigma$.

This section details the sketch of \cite{PlotkinGD:secotr}, but the
theory is due to Freyd~\cite{Freyd:90,Freyd:90a,Freyd:91}. In short,
the main observation is that because of the presence of fixed points,
the initial algebras and final coalgebras of
Sections~\ref{sec:initalg}, \ref{sec:finalcoalg} coincide
(Theorem~\ref{thm:compactness} below). This phenomenon is called
algebraic compactness, and was studied by Freyd in $\loccit$. Using
Freyd's techniques we find solutions to recursive type equations as
advertised, and show that they satisfy a universal property called the
initial dialgebra property. Moreover, we generalize the induction and
coinduction properties of Theorems~\ref{thm:induction},
\ref{thm:coinduction:general} to a combined induction/coinduction
property for recursive types. In
Section~\ref{sec:recursive:parameters} we treat the case of recursive
type equations with parameters.


Before we start, observe that we may split the occurrences of $\alpha$
in $\sigma$ into positive and negative occurrences. So our standard
assumption in this section is that we are given a type
$\alpha,\beta\ts \sigma(\alpha,\beta)$, in which $\alpha$ occurs only
negatively and $\beta$ only positively, and we look for a type
$\rec\alpha \ld \sigma(\alpha, \alpha)$ isomorphic to
$\sigma(\rec\alpha
\ld \sigma(\alpha, \alpha),\rec\alpha \ld \sigma(\alpha, \alpha))$. 
In this notation, $\rec \alpha\ld \sigma(\alpha, \alpha)$  binds
$\alpha$ in $\sigma$. 

\subsubsection{Parametrized initial algebras}

Set $\omega(\alpha)=\mu\beta\ld\sigma(\alpha,\beta)= \Prod\beta\ld
(\sigma(\alpha,\beta)\lpop\beta)\lpop \beta$. Now,  $\omega$
induces a contravariant functor from types to types.
\begin{lem}\label{lem:paraminitalg}
Assuming identity extension and extensionality, for
$f\co\alpha'\lpop\alpha$, up to internal equality
$\omega(f)\co\omega(\alpha)\lpop\omega(\alpha')$ is the unique $h$
such that
\[
\xymatrix{\sigma(\alpha,\omega(\alpha)) \ar@{o}[d]_{\sigma(\id,h)}
  \ar@{o}[r]^-{\interm} & \omega(\alpha) \ar@{o}[dd]^h \\
\sigma(\alpha,\omega(\alpha')) \ar@{o}[d]_{\sigma(f,\id)} \\
\sigma(\alpha',\omega(\alpha')) \ar@{o}[r]^-\interm & \omega(\alpha')
}
\]
commutes internally.
\end{lem}
\proof
One may define $\interm$  as a polymorphic term
\[\interm\co\Prod\alpha\ld \sigma(\alpha,\omega(\alpha))
\lpop \omega(\alpha)\]
by
\[
\interm = \Lambda\alpha\ld
\linlambda z\co\sigma(\alpha,\omega(\alpha))\ld
\Lambda\beta\ld \lambda f\co\sigma(\alpha,\beta)\lpop\beta\ld
f(\sigma(\lambda x\co\alpha\ld x, \fold\:\beta\:\bang f)\:z).
\]
By parametricity we have 
\[\interm\:\alpha'(\sigma(\graph f,\omega(\graph f)) \lpop
\omega(\graph f)) \interm\:\alpha,\]
which, by the Graph Lemma (Lemma~\ref{lem:graph}),  means that
\[\interm\:\alpha'(\opcat{\graph{\sigma(f,\omega(f))}} \lpop
\opcat{\graph{\omega(f)}}) \interm\:\alpha,\]
which in turn amounts to 
internal commutativity of the diagram of the lemma.

Uniqueness is by initiality of $\interm$ (in $\LinTypecat_\alpha$,
proved as before) used on the diagram
\[\xymatrix{
\sigma(\alpha,\omega(\alpha)) \xypop[d]_{\sigma(\id,h)}
\xypop[rr]^-\interm && 
\omega(\alpha) \xypop[d]^h \\
\sigma(\alpha,\omega(\alpha')) \xypop[r]^-{\sigma(f,\id)} &
\sigma(\alpha', \omega(\alpha')) 
\xypop[r]^-\interm & \omega(\alpha').
}\]
\qed

\subsubsection{Dialgebras}

\begin{defi}\label{defn:dialg}
  A dialgebra for $\sigma$ is a quadruple $(\tau, \tau', f, f')$ such
  that $\tau$ and $\tau'$ are types, and
  $f\co\sigma(\tau',\tau)\lpop\tau$ and $f'\co \tau'\lpop
  \sigma(\tau,\tau')$ are morphisms.  A morphism of dialgebras from
  dialgebra $(\tau_0,\tau_0', f_0,f_0')$ to $(\tau_1,\tau_1',f_1,f_1')$ is a
  pair of morphisms $h\co\tau_0\lpop\tau_1$,
  $h'\co\tau_1'\lpop\tau_0'$, such that
\[
\xymatrix{\sigma(\tau_0',\tau_0) \ar@{o}[r]^-{f_0} \ar@{o}[d]_{\sigma(h',h)}  &
   \ar@{o}[d]^h  \tau_0 \\ 
\sigma(\tau_1',\tau_1) \ar@{o}[r]_-{f_1} & \tau_1
}
\qquad
\xymatrix{\tau_1' \ar@{o}[r]^-{f_1'} \ar@{o}[d]_{h'} &
   \sigma(\tau_1,\tau_1')\ar@{o}[d]^{\sigma(h,h')} \\  
\tau_0' \ar@{o}[r]_-{f_0'} &  \sigma(\tau_0,\tau_0').
}
\]
\end{defi}
\begin{lem}
If $(h,h')$  is a map of dialgebras and $h,h'$  are isomorphisms, then
$(h,h')$  is an isomorphism of dialgebras.
\end{lem}
\proof
  The only thing to prove here is that $(h\inv,(h')\inv)$  is in fact a
  map of dialgebras, which is trivial.
\qed
\begin{rem}
  If we for the type $\alpha,\beta\ts \sigma\co\Type$ consider for
  each $\Xi$ the endofunctor
\[\pair{\opcat\sigma}{\sigma}\co \opcat\LinTypecat_{\Xi}
\times\LinTypecat_{\Xi} \to \opcat\LinTypecat_{\Xi}
\times\LinTypecat_{\Xi}\]
defined by $(\alpha,\beta) \mapsto
(\sigma(\beta,\alpha),\sigma(\alpha,\beta))$, then dialgebras for
$\sigma$ are exactly the algebras for $\pair{\opcat\sigma}{\sigma}$,
maps of dialgebras are maps of algebras for
$\pair{\opcat\sigma}{\sigma}$ and initial dialgebras correspond to
initial algebras. Dialgebras as considered here are a special case of
what Hagino  calls $F$, $G$-dialgebras in his thesis~\cite{Hagino:87},
for $F$  being $\pair{\opcat\sigma}{\sigma}$ and $G$  being the
identity functor. 
\end{rem}
\begin{thm}\label{thm:initdialg}
Assuming identity extension and extensionality, initial dialgebras exist
for all functors induced by types $\sigma(\alpha,\beta)$,  up to
internal equality.
\end{thm}
\proof
In this proof,  commutativity of diagrams will mean commutativity up
to internal equality.

Set $\omega(\alpha)=\mu\beta\ld\sigma(\alpha,\beta)$. Then, $\omega$
defines a contravariant functor. Define
\[\tau' = \nu\alpha\ld\sigma(\omega(\alpha),\alpha), \qquad \tau =
\omega(\tau') = \mu\beta \ld\sigma(\tau',\beta).\]
Since $\tau'$  is defined as the final coalgebra for a functor,  we
have a morphism 
\[\out\co\tau'\lpop \sigma(\omega(\tau'),\tau')=\sigma(\tau,\tau'), \]
and since $\tau$  is defined to be an initial algebra, we get a
morphism
\[\interm \co \sigma(\tau',\tau)\lpop \tau.\] 
We will show that $(\tau,\tau',\interm,\out)$  is an initial
dialgebra. 

Suppose we are given a dialgebra $(\tau_0,\tau_0', g,g')$.  Since
$\interm$  is an initial algebra, there exists a unique map
$a$,  such that
\[\xymatrix{\sigma(\tau_0',\omega(\tau_0')) \ar@{o}[r]^-\interm
  \ar@{o}[d]_{\sigma(\id,a)} & \omega(\tau_0') \ar@{o}[d]^a \\
\sigma(\tau_0',\tau_0) \ar@{o}[r]^-g & \tau_0,
}
\]
and thus, since $\out$  is a final coalgebra, we find a map $h'$ making
the diagram 
\begin{equation}
  \label{eq:h'}
\xymatrix{\tau_0' \ar@{o}[r]^-{g'} \ar@{o}[d]_{h'} &
  \sigma(\tau_0,\tau_0') \ar@{o}[r]^-{\sigma(a,\id)} &
  \sigma(\omega(\tau_0'),\tau_0') \ar@{o}[d]^{\sigma(\omega(h'),h')}\\
\tau' \ar@{o}[rr]^\out & & \sigma(\omega(\tau'),\tau')
}
\end{equation}
commute. Set $h=a\circ\omega(h')$.  We claim that $(h,h')$  defines a
map of dialgebras. The second diagram of Definition~\ref{defn:dialg}
is simply~(\ref{eq:h'}). The first diagram of~\ref{defn:dialg} follows
from the commutativity of the composite diagram
\begin{equation}
  \label{eq:initdialg}
\xymatrix{\sigma(\tau',\omega(\tau')) \ar@{o}[r]^-\interm
  \ar@{o}[d]_{\sigma(h',\omega(h'))} & \omega(\tau') \ar@{o}[d]^{\omega(h')}\\
\sigma(\tau_0',\omega(\tau_0')) \ar@{o}[r]^-\interm
  \ar@{o}[d]_{\sigma(\id,a)} & \omega(\tau_0') \ar@{o}[d]^a \\
\sigma(\tau_0',\tau_0) \ar@{o}[r]^-g & \tau_0,
}
\end{equation}
where the top diagram commutes by Lemma~\ref{lem:paraminitalg}. 

Finally, we will prove that $(h,h')$  is the unique dialgebra
morphism. Suppose we are given a map of dialgebras $(k,k')$ from
$(\tau,\tau',in,out)$ to $(\tau_0,\tau_0',g,g')$. By the
first diagram of Definition~\ref{defn:dialg},  we have a commutative
diagram
\[
\xymatrix{\sigma(\tau',\tau) \ar@{o}[rr]^-\interm
  \ar@{o}[d]_{\sigma(\id,k)} && \tau \ar@{o}[d]^{k} \\
\sigma(\tau',\tau_0) \ar@{o}[r]^-{\sigma(k',\id)} &
  \sigma(\tau_0',\tau_0) \ar@{o}[r]^-g & \tau_0.
}
\]
Since clearly~(\ref{eq:initdialg})  also commutes when $k'$ is
substituted for $h'$,  by (strong) initiality of $\interm$,  we conclude
that $k=_{\tau\lpop\tau'}a\circ\omega(k')$. Finally, by the second diagram of
Definition~\ref{defn:dialg} we have commutativity of
\[\xymatrix{\tau_0' \ar@{o}[d]_{k'} \ar@{o}[r]^-{g'} &
  \sigma(\tau_0,\tau_0') \ar@{o}[r]^{\sigma(a,\id)} &
  \sigma(\omega(\tau_0'),\tau_0') \ar@{o}[d]^{\sigma(\omega(k'),k')}
  \\
\tau'\ar@{o}[rr]^\out && \sigma(\omega(\tau'),\tau').
}
\]
So since $\out$ is a final coalgebra  we conclude
$k'=_{\tau_0'\lpop\tau'} h'$. 
\qed

\subsubsection{Algebraic compactness}


As advertised in the introduction to this section, the presence of
fixed points makes initial algebras and final coalgebras coincide.

\begin{thm}[Algebraic compactness]\label{thm:compactness}
Assuming identity extension and extensionality, for all types
$\alpha\ts\sigma(\alpha)$  in which $\alpha$  occurs only positively,
$\interm\inv$  is internally a final coalgebra and $\out\inv$  is
internally an initial algebra. Furthermore $\interm\inv$ and
$\out\inv$ can be written as terms of $\pilly$.
\end{thm}
\proof
By Theorems~\ref{thm:initalg} and~\ref{thm:finalcoalg}
$\interm$  is an initial algebra,  and $\out$  is a final coalgebra
for $\sigma$. Consider 
\[h=Y\:(
\nu\alpha\ld\sigma(\alpha))
\lpop 
\mu\alpha\ld \sigma(\alpha) 
\:(\lambda h\co
\nu\alpha \ld\sigma(\alpha)
\lpop
\mu\alpha\ld\sigma(\alpha)
\ld \interm\circ \sigma(h)\circ \out).\]
Since $Y$ is a fixed-point operator,  we know that 
\[\xymatrix{\sigma(\nu\alpha\ld\sigma(\alpha)) \ar@{o}[d]_{\sigma(h)}
  & \nu\alpha\ld\sigma(\alpha) \ar@{o}[l]_\out \ar@{o}[d]^h \\
\sigma(\mu\alpha\ld\sigma(\alpha)) \ar@{o}[r]^\interm &
\mu\alpha\ld\sigma(\alpha) 
}
\]
commutes. Since $\interm\inv$  is a coalgebra,  we also have a map $k$
going the other way, and since $\out$ is a final coalgebra,
$k \circ h=_{\nu\alpha\ld\sigma(\alpha) \lpop
  \nu\alpha\ld\sigma(\alpha)}\id_{\nu\alpha\ld\sigma(\alpha)}$.  Since
$\interm$  is an initial algebra,  we know that
$h\circ k=_{\mu\alpha\ld\sigma(\alpha)\lpop
  \mu\alpha\ld\sigma(\alpha)}\id_{\mu\alpha\ld\sigma(\alpha)}$. So
$\interm\inv\iso \out$  as coalgebras and $\out\inv\iso \interm$  as
algebras, internally.
\qed
\begin{lem}\label{lem:compactness}
  Assume identity extension and extensionality. Let $(\tau,\tau',\interm,\out)$
  be the initial dialgebra from the proof of Theorem~\ref{thm:initdialg}.
  Then $(\tau',\tau,\out\inv,\interm\inv)$ is also an initial dialgebra
  internally.
\end{lem}
\proof
In this proof, commutativity of diagrams is up to internal equality.

Suppose we are given a dialgebra $(\tau_0, \tau_0', g,g')$.  We will
show that there exists a unique morphism of dialgebras  from
$(\tau',\tau,\out\inv,\interm\inv)$  to $(\tau_0, \tau_0', g,g')$.

By Theorem~\ref{thm:compactness},  for all types $\alpha$,
$\interm\inv\co \omega(\alpha)\lpop \sigma(\alpha,\omega(\alpha))$  is a
final coalgebra for the functor $\beta\mapsto \sigma(\alpha,\beta)$,
and $\out\inv\co \sigma(\tau,\tau') \lpop \tau'$  is an initial
algebra for the functor $\alpha\mapsto
\sigma(\omega(\alpha),\alpha)$. 

Let $a$  be the unique map making the diagram
\[\xymatrix{\tau_0' \ar@{o}[r]^{g'} \ar@{o}[d]_a &
  \sigma(\tau_0,\tau_0') \ar@{o}[d]^{\sigma(\id,a)} \\
\omega(\tau_0) \ar@{o}[r]^{\interm\inv} & \sigma(\tau_0,\omega(\tau_0))
}
\]
commute. Define $h$ to be the unique map making
\begin{equation}
  \label{eq:hdiag}
\xymatrix{\sigma(\tau,\tau') \ar@{o}[rr]^-{\out\inv}
  \ar@{o}[d]_{\sigma(\omega(h),h)} & & \tau' \ar@{o}[d]^h \\
\sigma(\omega(\tau_0),\tau_0) \ar@{o}[r]^-{\sigma(a,\id)} &
\sigma(\tau_0',\tau_0) \ar@{o}[r]^-g & \tau_0
}
\end{equation}
commute. We define $h'$ to be $\omega(h)\circ a$ and prove that $(h,h')$
is a map of dialgebras. The first diagram of
Definition~\ref{defn:dialg}  is simply~(\ref{eq:hdiag}).
Commutativity of the second diagram follows from commutativity of
\begin{equation}
  \label{eq:h'diag}
\xymatrix{\tau_0' \ar@{o}[r]^{g'} \ar@{o}[d]_a &
  \sigma(\tau_0,\tau_0') \ar@{o}[d]^{\sigma(\id,a)} \\
\omega(\tau_0) \ar@{o}[d]_{\omega(h)} \ar@{o}[r]^{\interm\inv} &
\sigma(\tau_0,\omega(\tau_0)) \ar@{o}[d]^{\sigma(h,\omega(h))} \\
\omega(\tau') \ar@{o}[r]^{\interm\inv} & \sigma(\tau',\omega(\tau')),
}  
\end{equation}
where commutativity of the last diagram follows from
Lemma~\ref{lem:paraminitalg}. 

Finally, we will show that if $(k,k')$ is another map of dialgebras
from the dialgebra $(\tau',\tau,\out\inv,\interm\inv)$ to $(\tau_0,
\tau_0', g,g')$ then $h=_{\tau'\lpop\tau_0}k$ and $h'=_{\tau_0'\lpop
\tau}k'$. By the second diagram of Definition~\ref{defn:dialg} we know
that
\begin{equation}
  \label{eq:neweq}
\xymatrix{\tau_0' \ar@{o}[r]^-{g'} \ar@{o}[d]_{k'}  &
  \sigma(\tau_0,\tau_0') \ar@{o}[r]^{\sigma(k,\id)} &
  \sigma(\tau',\tau_0') \ar@{o}[d]^{\sigma(\id,k')} \\
\tau \ar@{o}[rr]^{\interm\inv} && \sigma(\tau',\tau)
}  
\end{equation}
commutes. Clearly,  if we substitute $k$  for $h$  in
(\ref{eq:h'diag}),  we obtain a diagram that commutes by
Lemma~\ref{lem:paraminitalg}. So, using the fact that $\interm\inv$  is
a final coalgebra  on (\ref{eq:neweq}), we get $k'=_{\tau_0'\lpop
  \tau}\omega(k)\circ a$. 

The first diagram of Definition~\ref{defn:dialg}  implies that
\[\xymatrix{\sigma(\tau,\tau')\ar@{o}[rr]^{\out\inv}
  \ar@{o}[d]_{\sigma(\omega(k), k)} && \tau' \ar@{o}[d]^k \\
\sigma(\omega(\tau_0),\tau_0) \ar@{o}[r]^-{\sigma(a,\id)} &
\sigma(\tau_0',\tau_0) \ar@{o}[r]^-g & \tau_0
}
\]
commutes.
Comparing this to (\ref{eq:hdiag}) we obtain $h=_{\tau'\lpop\tau_0}k$,
by initiality of $\out\inv$. 
\qed
%
%
\begin{thm}\label{lem:sym:dialg}
  Assuming identity extension and extensionality, for all types
  $\sigma(\alpha,\beta)$ where $\alpha$ occurs only negatively and
  $\beta$ only positively, there exists a type  $\rec\alpha\ld
  \sigma(\alpha, \alpha)$ and an isomorphism \[i\co\sigma(\rec\alpha\ld
  \sigma(\alpha, \alpha),\rec\alpha\ld \sigma(\alpha,
  \alpha))\lpop\rec\alpha\ld \sigma(\alpha, \alpha),\] such that
  $(\rec\alpha\ld \sigma(\alpha, \alpha),\rec\alpha\ld \sigma(\alpha,
  \alpha),i,i\inv)$ is an initial dialgebra up to internal equality.
\end{thm}

\proof
As usual commutativity of diagrams will be up to internal equality.

We have a unique map of dialgebras 
\[(h,h')\co (\tau,\tau',\interm,\out)\to
(\tau',\tau,\out\inv,\interm\inv)\] We claim that $(h',h)$ is also a
map of dialgebras from $(\tau,\tau',\interm,\out)$ to
$(\tau',\tau,\out\inv,\interm\inv)$.  To prove this we need to prove
commutativity of the diagrams
\[
\xymatrix{\sigma(\tau',\tau)\ar@{o}[r]^-\interm
  \ar@{o}[d]_{\sigma(h,h')} & \tau \ar@{o}[d]^{h'} \\
\sigma(\tau,\tau') \ar@{o}[r]^-{\out\inv} & \tau'}
\qquad
\xymatrix{\tau \ar@{o}[r]^-{\interm\inv} \ar@{o}[d]_{h}
   & \sigma(\tau',\tau)  \ar@{o}[d]^{\sigma(h',h)}  \\
\tau' \ar@{o}[r]^-{\out} & \sigma(\tau,\tau')},
\]
but the fact that $(h,h')$  is a map of dialgebras tells us exactly
that 
\[
\xymatrix{\sigma(\tau',\tau)\ar@{o}[r]^-\interm
  \ar@{o}[d]_{\sigma(h',h)} & \tau \ar@{o}[d]^{h} \\
\sigma(\tau,\tau') \ar@{o}[r]^-{\out\inv} & \tau'}
\qquad
\xymatrix{\tau \ar@{o}[r]^-{\interm\inv} \ar@{o}[d]_{h'}
   & \sigma(\tau',\tau)  \ar@{o}[d]^{\sigma(h,h')}  \\
\tau' \ar@{o}[r]^-{\out} & \sigma(\tau,\tau'),}
\]
and these two diagram are the same as the above but in opposite order.
Thus,  by uniqueness of maps of dialgebras out of
$(\tau,\tau',\interm,\out)$, we get 
$h=_{\tau\lpop\tau'}h'$. Since $(h,h)$
is a map between initial dialgebras, $h$  is an isomorphism.

Now define $f\co \sigma(\tau,\tau)\lpop\tau$  to be $\interm\circ
\sigma(h\inv,\id_\tau)$.  Then clearly $(\id_\tau,h\inv)$  is a morphism
of dialgebras  from $(\tau,\tau,f,f\inv)$  to
$(\tau,\tau',\interm,\out)$, since the diagrams proving $(\id_\tau,h\inv)$
to be a map of dialgebras are
\[
\xymatrix{\sigma(\tau,\tau) \ar@{o}[d]_{\sigma(h\inv,\id)}
  \ar@{o}[r]^{\sigma(h\inv,\id)} \ar@{o}@/_/[rr]_{f} &
  \sigma(\tau',\tau) \ar@{o}[r]^\interm & \tau \ar@{o}[d]^{\id} \\
\sigma(\tau',\tau)  \ar@{o}[rr]^\interm && \tau}
\quad
\xymatrix{\tau' \ar@{o}[rr]^\out \ar@{o}[d]_{h\inv} && \sigma(\tau,\tau')
  \ar@{o}[d]^{\sigma(\id,h\inv)} \\
\tau \ar@{o}[r]^{\interm\inv} \ar@{o}@/_/[rr]_{f\inv} & \sigma(\tau',\tau)
  \ar@{o}[r]^{\sigma(h, \id)} & \sigma(\tau,\tau).}
\]
Clearly the first diagram commutes,  and the second diagram is just
part of the definition of $(h,h)$  being a map of dialgebras.
Thus $(\id_\tau, h\inv)$ defines an isomorphism of dialgebras from
$(\tau,\tau,f,f\inv)$  to $(\tau,\tau', \interm, \out)$, as desired.
\qed
%
Notice that the closed terms $\rec\alpha\ld \sigma(\alpha,
\alpha)\lpop \sigma(\rec\alpha\ld \sigma(\alpha, \alpha),\rec\alpha\ld
\sigma(\alpha, \alpha))$ and 
\[\sigma(\rec\alpha\ld \sigma(\alpha,
\alpha),\rec\alpha\ld \sigma(\alpha, \alpha))\lpop\rec\alpha\ld
\sigma(\alpha, \alpha)\] always exist, independent of the assumption of
parametricity. Parametricity implies that they are each others
inverses.

\subsection{A mixed induction/coinduction principle}
\label{sec:induction:coinduction}

Here we prove the following reasoning principle for the recursive type
$\rec \alpha\ld \sigma(\alpha,\alpha)$. This principle is the same as
the one obtained by Pitts for recursive types in the category domains
\cite[Cor~4.10]{Pitts:95a}. Again, as noted before
Theorem~\ref{thm:coinduction:general} we must assume that $\sigma$ is
an inductively defined type to make sense of the relational
interpretation of  $\sigma$ at general non-admissible relations.

\begin{thm} \label{thm:induction:recursive:types}
  Suppose $\alpha, \beta \ts \sigma(\alpha, \beta)$  is an inductively
  defined type in which $\alpha$  occurs only positively and $\beta$
  only negatively. Suppose further 
  \[
  \begin{array}{c}
    R^-\rel{\rec \alpha\ld \sigma(\alpha,\alpha)}{ \rec
    \alpha\ld \sigma(\alpha,\alpha)} \text{ and} \\
    R^+\co \admrel{\rec \alpha\ld
      \sigma(\alpha,\alpha)}{\rec \alpha\ld \sigma(\alpha,\alpha)}
  \end{array}
  \]
  are relations. Then the following principle holds
  \[
  \begin{prooftree}
    (i\inv ,i \inv) \co R^-\lpop \sigma(R^+,R^-) \qquad (i,i)\co
    \sigma(R^-, R^+) \lpop R^+
    \justifies 
    R^- \subset \eq_{\rec \alpha\ld \sigma(\alpha,\alpha)} \subset R^+
  \end{prooftree}
  \]
  where $i$  denotes the isomorphism 
  \[\sigma(\rec \alpha\ld \sigma(\alpha,\alpha), \rec \alpha\ld
  \sigma(\alpha,\alpha)) \lpop \rec \alpha\ld \sigma(\alpha,\alpha). 
  \]
\end{thm}

\proof
  We first prove the rule in the case of both relations being
  admissible. The proof in this case is a surprisingly simple
  consequence of parametricity. 
  
  The proof of Theorem~\ref{lem:sym:dialg} is constructive in the
  sense that there is a construction of the maps $h,h'$ constituting
  the unique dialgebra map out of the initial dialgebra from the given
  types $\omega, \omega'$ and terms $t,t'$. In fact, from the proof we
  can derive terms
  \[
  \begin{array}{c}
    k\co \Prod\omega,\omega' \ld (\sigma(\omega', \omega)\lpop
    \omega)\lpop (\omega' \lpop \sigma(\omega,\omega')) 
    \lpop \rec \alpha\ld \sigma(\alpha,\alpha) \lpop \omega \\
    k'\co \Prod\omega,\omega' \ld (\sigma(\omega', \omega)\lpop
    \omega)\lpop (\omega' \lpop \sigma(\omega,\omega')) \lpop \omega'
    \lpop \rec \alpha\ld \sigma(\alpha,\alpha) 
  \end{array}
  \]
  such that the maps $h, h'$  can be obtained as 
  \[
  \begin{array}{rcl}
    h &=& k\:\omega\:\omega' \: t\:t' \\
    h' &=& k'\:\omega\:\omega' \: t\:t'
  \end{array}
  \]
  The exact constructions of $k,k'$  are not of interest us right now
  --- what matters to us is that we can use the assumption of
  parametricity on them. We consider the case $\omega= \omega' = \rec \alpha\ld
  \sigma(\alpha,\alpha)$  and $t=i$   and $t'=i\inv$. In this case of
  course $h=h'=\id$. If we use parametricity of $k'$ by substituting
  the relation $R^-$ for the type $\omega'$  and  $R^+$  for
  $\omega$  then we get  since 
  \[\id = k\:\rec \alpha\ld \sigma(\alpha,\alpha) \: \rec \alpha\ld
  \sigma(\alpha,\alpha)\: i \: i\inv \] 
  $(\id,\id) \co R^- \lpop
  \eq_{\rec \alpha\ld \sigma(\alpha,\alpha)}$.  Likewise, using
  parametricity of $k$ we get
  \[(\id,\id) \co \eq_{\rec \alpha\ld \sigma(\alpha,\alpha)} \lpop R^+
  \]
  which proves the theorem in the case of $R^-$  being admissible. 
  
  For the general case, we just need a simple application of the
  closure operator of Lemma~\ref{lem:admissible:closure}. So assume
  again 
  \[
  \begin{array}{l}
    (i\inv, i\inv) \co R^-\lpop \sigma(R^+, R^-), \\
    (i,i)\co \sigma(R^-, R^+)\lpop R^+,
  \end{array}
  \]
  and $R^+$ is admissible, but $R^-$ may
  not be. The idea is to use the case above on $\Phi(R^-)$ and $R^+$
  which are both admissible, but we need to check that the hypothesis
  still holds for this case. First, by $\Phi$ being a functor
  \[(i\inv, i\inv) \co \Phi(R^-) \lpop \Phi (\sigma(R^+, R^-)) .
  \]
  But, since $\sigma(R^+, \Phi(R^-))$  is an admissible relation
  containing $\sigma(R^+, R^-)$, 
  \[\Phi (\sigma(R^+, R^-)) \subset \sigma(R^+, \Phi(R^-))\]
  and so 
  \begin{equation}
    \label{eq:hyp1}
    (i\inv, i\inv) \co \Phi(R^-) \lpop \sigma(R^+,\Phi(R^-)).
  \end{equation}
  Since $\sigma(\Phi(R^-), R^+) \subset \sigma(R^-, R^+)$  we also
  have
  \begin{equation}
    \label{eq:hyp2}
    (i,i) \co \sigma(\Phi(R^-), R^+) \lpop R^+.
  \end{equation}
  Using the case of admissible relation proved above on
  (\ref{eq:hyp1})  and (\ref{eq:hyp2}),  we get 
  \[\Phi(R^-) \subset \eq_{\rec \alpha\ld \sigma(\alpha, \alpha)}
  \subset R^+
  \]
  which together with $R^-\subset \Phi(R^-)$  proves the theorem in
  the general case. 
\qed

\subsection{Recursive type equations with parameters}
\label{sec:recursive:parameters}

We now consider recursive type equations with parameters, i.e., we
consider types $\vec\alpha,\alpha\ts \sigma(\vec\alpha,\alpha)$ and
look for types $\vec\alpha\ts \rec\alpha \ld \sigma(\vec\alpha,
\alpha)$ satisfying $\sigma(\vec\alpha, \rec\alpha \ld
\sigma(\vec\alpha, \alpha)) \iso \rec\alpha \ld \sigma(\vec\alpha,
\alpha)$. As before, we need to split occurrences of the variable
$\alpha$ into positive and negative occurrences, and since we would
like to be able to construct nested recursive types, we need to keep
track of positive and negative occurrences of the variables
$\vec\alpha$ in the solution $\rec\alpha \ld \sigma(\vec\alpha,
\alpha)$ as well. So we will suppose that we are given a type
$\vec\alpha, \vec\beta, \alpha,\beta \ts \sigma(\vec\alpha,\vec\beta,
\alpha, \beta)$ in which the variables $\vec\alpha, \alpha$ occur only
negatively and the variables $\vec\beta, \beta$ only positively.

Of course, the proof proceeds as in the case without parameters.
However, one must take care to obtain the right occurrences of
parameters, and so we sketch the proof here.

\begin{lem}\label{lem:parametriced:init:dialg}
  Suppose $\vec\alpha, \vec\beta, \alpha,\beta \ts
  \sigma(\vec\alpha,\vec\beta, \alpha, \beta)$ is a type in which the
  variables $\vec\alpha, \alpha$ occur only negatively and the
  variables $\vec\beta, \beta$ only positively. There exists types
  $\vec\alpha, \vec\beta \ts \tau(\vec \alpha,\vec \beta)$ in which
  $\vec \alpha$ occurs only negatively and $\vec\beta$ only positively
  and $\vec\alpha, \vec\beta \ts \tau'(\vec \alpha,\vec \beta)$ in
  which $\vec \alpha$ occurs only positively and $\vec\beta$ only
  negatively and  terms 
  \[
  \begin{array}{c}
      \interm\co \sigma(\vec\alpha, \vec\beta,
      \tau'(\vec\alpha,\vec\beta), \tau(\vec\alpha, \vec\beta)) \lpop
      \tau(\vec\alpha,\vec\beta) \\
      \out\co \tau'(\vec\alpha, \vec\beta) \lpop \sigma(\vec\beta,
      \vec\alpha, \tau(\vec\alpha,\vec\beta), \tau'(\vec\alpha,
      \vec\beta))
  \end{array}
  \]
  such that for any pair of types $\vec\alpha, \vec\beta \ts\omega,
  \omega'$, and terms 
  \[
  \begin{array}{c}
      g \co \sigma(\vec\alpha, \vec\beta, \omega', \omega) \lpop
      \omega \\
      g' \co \omega' \lpop \sigma(\vec\beta, \vec\alpha, \omega, \omega')
  \end{array}
  \]
  there exists unique $h$, $h'$  making 
  \[\xymatrix{\sigma(\vec\alpha, \vec\beta,
    \tau'(\vec\alpha,\vec\beta), \tau(\vec\alpha,\vec\beta))
    \xypop[d]_{\sigma(\vec\alpha, \vec\beta, h', h)}
    \xypop[r]^-\interm & \tau(\vec\alpha, \vec\beta)  \xypop[d]^h\\
    \sigma(\vec\alpha, \vec\beta, \omega', \omega) \xypop[r]^-g &
    \omega 
  } \qquad
  \xymatrix{\omega' \xypop[r]^{g'} \xypop[d]_{h'} & \sigma(\vec\beta,
    \vec\alpha, \omega, \omega')  
    \xypop[d]^{\sigma(\vec\beta, \vec\alpha, h, h')} \\
    \tau'(\vec\alpha,\vec\beta) 
    \xypop[r]^-{\out} & \sigma(\vec\beta, \vec\alpha, 
    \tau(\vec\alpha,\vec\beta), \tau'(\vec\alpha,\vec\beta))
    }
  \]
  commute up to internal equality.
\end{lem}

\proof
  Define 
  \[
  \begin{array}{rcl}
  \omega(\vec\alpha, \vec\beta, \alpha) & = & \mu\beta\ld
  \sigma(\vec\alpha, \vec\beta, \alpha, \beta) \\
  \tau'(\vec\alpha, \vec\beta) & = & \nu\alpha\ld \sigma(\vec\beta,
  \vec\alpha, \omega(\vec\alpha, \vec\beta, \alpha), \alpha) \\
  \tau(\vec\alpha, \vec\beta) & = & \omega(\vec\alpha, \vec\beta,
  \tau'(\vec\alpha, \vec\beta))
  \end{array}
  \]
Notice that we have swapped the occurrences of $\vec\alpha, \vec\beta$
in $\sigma$  in the definition of $\tau'$, making all occurrences of
$\vec\alpha$  in $\tau'$  positive and all occurrences of $\vec\beta$
in $\tau'$  negative. The rest of the proof proceeds exactly as the
proof of Theorem~\ref{thm:initdialg}.
\qed

\begin{thm}\label{thm:sym:parametriced:init:dialg}
  Suppose $\vec\alpha, \vec\beta, \alpha, \beta\ts \sigma(\vec\alpha,
  \vec\beta, \alpha, \beta)$ is a type as in
  Lemma~\ref{lem:parametriced:init:dialg}. Then there exists a type
  $\rec\alpha\ld \sigma(\vec\alpha, \vec\beta, \alpha, \alpha)$ with
  $\vec\alpha$ occurring only negatively and $\vec\beta$ only
  positively, and an isomorphism
  \[i \co \sigma(\vec\alpha, \vec\beta,
  \rec\alpha\ld \sigma(\vec\beta, \vec\alpha, \alpha, \alpha),
  \rec\alpha\ld \sigma(\vec\alpha, \vec\beta, \alpha, \alpha)) \lpop
  \rec\alpha\ld \sigma(\vec\alpha,\vec\beta, \alpha, \alpha)
  \]
  satisfying the conclusion of
  Lemma~\ref{thm:sym:parametriced:init:dialg} with 
  \begin{displaymath}
    \begin{array}{rcl}
      \tau(\vec\alpha,\vec \beta) & = &
        \rec\alpha\ld \sigma(\vec\alpha, \vec\beta, \alpha,\alpha),\\
      \tau'(\vec\alpha,\vec\beta) & = & 
         \rec\alpha\ld \sigma(\vec \beta, \vec\alpha, \alpha, \alpha),\\
      i & = & \interm,\\
      \out & = & i \inv.
    \end{array}
  \end{displaymath}
\end{thm}

\proof
  Using Theorem~\ref{thm:compactness}, we can prove as in the proof of
  Lemma~\ref{lem:compactness} that the pair
  \[
  \begin{array}{c}
    \out\inv \co \sigma(\vec\alpha,\vec \beta, \tau(\vec\beta,
    \vec\alpha), \tau'(\vec \beta, \vec\alpha)) \lpop \tau'(\vec\beta,
    \vec\alpha) \\
    \interm\inv \co \tau(\vec\beta, \vec\alpha) \lpop \sigma(\vec
    \beta, \vec \alpha, \tau'(\vec\beta, \vec \alpha), \tau(\vec
    \beta, \vec \alpha))
  \end{array}
  \]
  also satisfies the conclusion of
  Lemma~\ref{thm:sym:parametriced:init:dialg}. Proceeding as in the proof
  of Lemma~\ref{lem:sym:dialg} we get an isomorphism $\tau(\vec\alpha,
  \vec\beta) \iso \tau'(\vec\beta, \vec\alpha)$ up to internal
  equality,  which implies the theorem. 
\qed


The mixed induction/coinduction principle of
Theorem~\ref{thm:induction:recursive:types}  can be generalized to
recursive types with parameters as follows. 

\begin{thm} \label{thm:parametrized:recursive}
  Suppose $\vec R_+\admrel{\vec \omega_+}{\vec \omega_+'}$ and
  $\vec R_-\admrel{\vec \omega_-}{\vec \omega_-'}$ are vectors of
  admissible relations, and 
  \[
  \begin{array}{c}
    S_+\admrel {\rec\alpha \ld \sigma(\vec\omega_-, \vec\omega_+,
    \alpha, \alpha)}
    {\rec\alpha \ld \sigma(\vec\omega_-', \vec\omega_+', \alpha,
    \alpha)}  \\ 
    S_-\rel {\rec\alpha \ld \sigma(\vec\omega_+, \vec\omega_-, \alpha,
    \alpha)}
    {\rec\alpha \ld \sigma(\vec\omega_+', \vec\omega_-', \alpha, \alpha)}
  \end{array}
  \]
  are relations. Then the following rule holds:
  \[
  \begin{prooftree}
    (i\inv, i\inv) \co S_- \lpop \sigma(\vec R_+, \vec R_-, S_+, S_-)
    \qquad (i,i)\co \sigma(\vec R_-, \vec R_+, S_-, S_+) \lpop S_+
    \justifies S_- \subset \rec\alpha \ld \sigma(\vec R_+, \vec R_-,
    \alpha, \alpha) \qquad \rec\alpha \ld \sigma(\vec R_-, \vec R_+,
    \alpha, \alpha) \subset S_+
  \end{prooftree}
  \]
\end{thm}

\proof
  The proof proceeds as the proof of
  Theorem~\ref{thm:induction:recursive:types}, and we start by
  considering the case where $S_-$ is admissible. This time the terms
  generating $h,h'$ have types
  \[
  \begin{array}{l}
    k\co \Prod\vec\alpha,\vec\beta \ld\Prod\omega',\omega \ld\\
    \qquad
    (\sigma(\vec\alpha, \vec\beta, \omega', \omega)\lpop 
    \omega)\lpop (\omega' \lpop \sigma(\vec\beta, \vec\alpha,
    \omega,\omega')) \lpop \rec\alpha \ld \sigma(\vec\alpha,
    \vec\beta, \alpha, \alpha) \lpop \omega \\
    k'\co \Prod\vec\alpha,\vec\beta \ld\Prod\omega,\omega' \ld\\
    \qquad
    (\sigma(\vec\alpha, \vec\beta, \omega', \omega)\lpop 
    \omega)\lpop (\omega' \lpop \sigma(\vec\beta, \vec\alpha,
    \omega,\omega')) \lpop \omega' 
    \lpop \rec\alpha \ld \sigma(\vec \beta,\vec \alpha, \alpha, \alpha)
  \end{array}
  \]
  Now, notice first that 
  \begin{eqnarray}
    \label{eq:hisidfirst}
    k\: \vec\omega_+\: \vec\omega_- \: \rec\alpha \ld \sigma(\vec\omega_+,
    \vec\omega_-, \alpha, \alpha) \: \rec\alpha \ld
    \sigma(\vec\omega_-, \vec\omega_+, \alpha, \alpha)\: i\: i\inv = 
    \id_{\rec\alpha \ld \sigma(\vec\omega_+, \vec\omega_-, \alpha, \alpha)}\\
    k'\: \vec\omega_+\: \vec\omega_- \: \rec\alpha \ld \sigma(\vec\omega_+,
    \vec\omega_-, \alpha, \alpha) \: \rec\alpha \ld
    \sigma(\vec\omega_-, \vec\omega_+, \alpha, \alpha)\: i\: i\inv = 
    \id_{\rec\alpha \ld \sigma(\vec\omega_-, \vec\omega_+, \alpha, \alpha)} \\
    k\: \vec\omega_+'\: \vec\omega_-' \: \rec\alpha \ld \sigma(\vec\omega_+',
    \vec\omega_-', \alpha, \alpha) \: \rec\alpha \ld
    \sigma(\vec\omega_-', \vec\omega_+', \alpha, \alpha)\: i\: i\inv = 
    \id_{\rec\alpha \ld \sigma(\vec\omega_+', \vec\omega_-', \alpha, \alpha)}\\
    \label{eq:hisidfinal}
    k'\: \vec\omega_+'\: \vec\omega_-' \: \rec\alpha \ld \sigma(\vec\omega_+',
    \vec\omega_-', \alpha, \alpha) \: \rec\alpha \ld
    \sigma(\vec\omega_-', \vec\omega_+', \alpha, \alpha)\: i\: i\inv = 
    \id_{\rec\alpha \ld \sigma(\vec\omega_-', \vec\omega_+', \alpha, \alpha)}
  \end{eqnarray} 
  as in the proof of Theorem~\ref{thm:induction:recursive:types}.
  
  The theorem will follow from instantiating the parametricity schema
  of $k,k'$ with $\vec R_-$ substituted for $\vec\alpha$, $\vec R_+$
  substituted for $\vec\beta$ and $S_+$ for $\omega$ and $S_-$ for
  $\omega'$. This tells us that if  
  \[
  \begin{array}{c}
    (i\inv, i\inv) \co S_- \lpop \sigma(\vec R_+, \vec R_-, S_+, S_-) \\
    (i,i)\co \sigma(\vec R_-, \vec R_+, S_-, S_+) \lpop S_+
  \end{array}
  \]
  then (using (\ref{eq:hisidfirst})-(\ref{eq:hisidfinal}) above)
  \[
  \begin{array}{c}
    (\id_{\rec\alpha \ld \sigma(\vec\omega_+, \vec\omega_-, \alpha,
    \alpha)}, \id_{\rec\alpha \ld \sigma(\vec\omega_+', 
    \vec\omega_-', \alpha, \alpha)}) \co S_- \lpop \rec\alpha \ld
    \sigma(\vec R_+, \vec R_-, \alpha, \alpha) \\ 
    (\id_{\rec\alpha \ld \sigma(\vec\omega_-, \vec\omega_+, \alpha,
    \alpha)}, \id_{\rec\alpha \ld \sigma(\vec\omega_-', 
    \vec\omega_+', \alpha, \alpha)}) \co \rec\alpha \ld \sigma(\vec
    R_-, \vec R_+, \alpha, \alpha) \lpop S_+ 
  \end{array}
  \]
  which was what we needed to prove.

  For the general case, dropping the assumption that $S_-$  is
  admissible, the proof proceeds exactly as in
  Theorem~\ref{thm:induction:recursive:types}. 
\qed

\section{Conclusion}
\label{sec:conclusion}

We have presented the logic LAPL for reasoning about parametricity in
the domain theoretic case, and we have shown how in this logic
Plotkin's encodings of recursive types can be verified. In later
papers we will present a general notion of model of LAPL, and show how
various earlier suggested domain theoretic models of parametric
polymorphism fit this general notion of model. These models include a
model based on admissible pers over a reflexive domain
\cite{MogelbergR:lapl-tr}, Rosolini and Simpson's
construction in Synthetic Domain Theory \cite{Rosolini:Simpson:04} and
a model based on the language $\lily$~\cite{BiermanGM:opeplp}. 

In all these cases, a central point in verifying that these give rise
to models of LAPL is to show that the various notions of admissible
relations in the specific models satisfy the axioms for admissible
relations presented in this paper. In the case of admissible pers the
admissible relations are given by pointed chain complete subpers, in
Synthetic Domain Theory the admissible relations are given by
subdomain and in the case of $\lily$ these are given by the
$\perp\perp$-closed relations. Of course the axioms presented here
have been constructed to be general enough to fit all these cases. 

As mentioned in the introduction, the logic LAPL can be seen as an
axiomatization of a good category of domains. An interesting question
is whether this is actually a useful axiomatization, particularly
because the solutions to recursive domain equations obtained using
LAPL satisfy universal properties with respect to linear maps of
$\pilly$, and most programming languages that one might want to model
using domains do not correspond to linear calculi. A recent paper by
the second author~\cite{Mogelberg:icalp06} provides evidence of the
usefulness of LAPL by showing how models of it give rise to models of
FPC --- a simply typed lambda calculus with general recursive types
first suggested by Plotkin~\cite{Plotkin:85} (see also
\cite{Fiore:96}) --- and that these models model the expected
reasoning principles for recursive types, reflecting a famous similar
result in classical domain theory.


\section*{Acknowledgments}

We gratefully acknowledge discussions with Milly Maietti,
Gordon Plotkin, John Rey\-nolds, Pino Rosolini and Alex Simpson. We also
thank the anonymous referees for many helpful suggestions.

\end{document}